\documentclass[aps,prd,preprintnumbers,superscriptaddress,showpacs,nofootinbib]{revtex4-2}

\usepackage{amsmath}
\usepackage{amssymb}
\usepackage{graphicx}
\usepackage{hyperref}
\usepackage{xcolor}
\usepackage{makecell}
\usepackage{comment} 
\usepackage{siunitx} 
\usepackage{booktabs} 
\usepackage{lineno}
\usepackage{multirow}
\usepackage[normalem]{ulem} 

\newcommand{\numu}{\ensuremath{\nu_\mu}}
\newcommand{\numubar}{\ensuremath{\bar{\nu}_\mu}}

\begin{document}

\title{Measurement of $\pi^0$ Production in $\bar{\nu}_{\mu}$ Charged-Current Interactions in the NOvA Near Detector}
\preprint{FERMILAB-PUB-25-0813-PPD}

\voffset 0.5in 
\newcommand{\ANL}{Argonne National Laboratory, Argonne, Illinois 60439, 
USA}
\newcommand{\Bandirma}{Bandirma Onyedi Eyl\"ul University, Faculty of 
Engineering and Natural Sciences, Engineering Sciences Department, 
10200, Bandirma, Balıkesir, Turkey}
\newcommand{\ICS}{Institute of Computer Science, The Czech 
Academy of Sciences, 
182 07 Prague, Czech Republic}
\newcommand{\IOP}{Institute of Physics, The Czech 
Academy of Sciences, 
182 21 Prague, Czech Republic}
\newcommand{\Atlantico}{Universidad del Atlantico,
Carrera 30 No.\ 8-49, Puerto Colombia, Atlantico, Colombia}
\newcommand{\BHU}{Department of Physics, Institute of Science, Banaras 
Hindu University, Varanasi, 221 005, India}
\newcommand{\UCLA}{Physics and Astronomy Department, UCLA, Box 951547, Los 
Angeles, California 90095-1547, USA}
\newcommand{\Caltech}{California Institute of 
Technology, Pasadena, California 91125, USA}
\newcommand{\Cochin}{Department of Physics, Cochin University
of Science and Technology, Kochi 682 022, India}
\newcommand{\Charles}
{Charles University, Faculty of Mathematics and Physics,
 Institute of Particle and Nuclear Physics, Prague, Czech Republic}
\newcommand{\Cincinnati}{Department of Physics, University of Cincinnati, 
Cincinnati, Ohio 45221, USA}
\newcommand{\CSU}{Department of Physics, Colorado 
State University, Fort Collins, CO 80523-1875, USA}
\newcommand{\CTU}{Czech Technical University in Prague,
Brehova 7, 115 19 Prague 1, Czech Republic}
\newcommand{\Dallas}{Physics Department, University of Texas at Dallas,
800 W. Campbell Rd. Richardson, Texas 75083-0688, USA}
\newcommand{\DallasU}{University of Dallas, 1845 E 
Northgate Drive, Irving, Texas 75062 USA}
\newcommand{\Delhi}{Department of Physics and Astrophysics, University of 
Delhi, Delhi 110007, India}
\newcommand{\JINR}{Joint Institute for Nuclear Research,  
Dubna, Moscow region 141980, Russia}
\newcommand{\Erciyes}{
Department of Physics, Erciyes University, Kayseri 38030, Turkey}
\newcommand{\FNAL}{Fermi National Accelerator Laboratory, Batavia, 
Illinois 60510, USA}
\newcommand{\FSU}{Florida State University, Tallahassee, Florida 32306, USA}
\newcommand{\UFG}{Instituto de F\'{i}sica, Universidade Federal de 
Goi\'{a}s, Goi\^{a}nia, Goi\'{a}s, 74690-900, Brazil}
\newcommand{\Guwahati}{Department of Physics, IIT Guwahati, Guwahati, 781 
039, India}
\newcommand{\Harvard}{Department of Physics, Harvard University, 
Cambridge, Massachusetts 02138, USA}
\newcommand{\Houston}{Department of Physics, 
University of Houston, Houston, Texas 77204, USA}
\newcommand{\IHyderabad}{Department of Physics, IIT Hyderabad, Hyderabad, 
502 205, India}
\newcommand{\Hyderabad}{School of Physics, University of Hyderabad, 
Hyderabad, 500 046, India}
\newcommand{\IIT}{Illinois Institute of Technology,
Chicago IL 60616, USA}
\newcommand{\Indiana}{Indiana University, Bloomington, Indiana 47405, 
USA}
\newcommand{\INR}{Institute for Nuclear Research of Russia, Academy of 
Sciences 7a, 60th October Anniversary prospect, Moscow 117312, Russia}
\newcommand{\UIowa}{Department of Physics and Astronomy, University of Iowa, 
Iowa City, Iowa 52242, USA}
\newcommand{\ISU}{Department of Physics and Astronomy, Iowa State 
University, Ames, Iowa 50011, USA}
\newcommand{\Irvine}{Department of Physics and Astronomy, 
University of California at Irvine, Irvine, California 92697, USA}
\newcommand{\Jammu}{Department of Physics and Electronics, University of 
Jammu, Jammu Tawi, 180 006, Jammu and Kashmir, India}
\newcommand{\Lebedev}{Nuclear Physics and Astrophysics Division, Lebedev 
Physical 
Institute, Leninsky Prospect 53, 119991 Moscow, Russia}
\newcommand{\Magdalena}{Universidad del Magdalena, Carrera 32 No 22-08 Santa Marta, Colombia}
\newcommand{\MSU}{Department of Physics and Astronomy, Michigan State 
University, East Lansing, Michigan 48824, USA}
\newcommand{\Crookston}{Math, Science and Technology Department, University 
of Minnesota Crookston, Crookston, Minnesota 56716, USA}
\newcommand{\Duluth}{Department of Physics and Astronomy, 
University of Minnesota Duluth, Duluth, Minnesota 55812, USA}
\newcommand{\Minnesota}{School of Physics and Astronomy, University of 
Minnesota Twin Cities, Minneapolis, Minnesota 55455, USA}
\newcommand{\Mississippi}{University of Mississippi, University, Mississippi 38677, USA}
\newcommand{\NISER}{National Institute of Science Education and Research, An OCC of Homi Bhabha
National Institute, Bhubaneswar, Odisha, India}
\newcommand{\OSU}{Department of Physics, Ohio State University, Columbus,
Ohio 43210, USA}
\newcommand{\Oxford}{Subdepartment of Particle Physics, 
University of Oxford, Oxford OX1 3RH, United Kingdom}
\newcommand{\Panjab}{Department of Physics, Panjab University, 
Chandigarh, 160 014, India}
\newcommand{\Pitt}{Department of Physics and Astronomy, 
University of Pittsburgh, Pittsburgh, Pennsylvania 15260, USA}
\newcommand{\QMU}{Particle Physics Research Centre, 
Department of Physics and Astronomy,
Queen Mary University of London,
London E1 4NS, United Kingdom}
\newcommand{\RAL}{Rutherford Appleton Laboratory, Science 
and 
Technology Facilities Council, Didcot, OX11 0QX, United Kingdom}
\newcommand{\SAlabama}{Department of Physics, University of 
South Alabama, Mobile, Alabama 36688, USA} 
\newcommand{\Carolina}{Department of Physics and Astronomy, University of 
South Carolina, Columbia, South Carolina 29208, USA}
\newcommand{\SDakota}{South Dakota School of Mines and Technology, Rapid 
City, South Dakota 57701, USA}
\newcommand{\SMU}{Department of Physics, Southern Methodist University, 
Dallas, Texas 75275, USA}
\newcommand{\Stanford}{Department of Physics, Stanford University, 
Stanford, California 94305, USA}
\newcommand{\Sussex}{Department of Physics and Astronomy, University of 
Sussex, Falmer, Brighton BN1 9QH, United Kingdom}
\newcommand{\Syracuse}{Department of Physics, Syracuse University,
Syracuse NY 13210, USA}
\newcommand{\Tennessee}{Department of Physics and Astronomy, 
University of Tennessee, Knoxville, Tennessee 37996, USA}
\newcommand{\Texas}{Department of Physics, University of Texas at Austin, 
Austin, Texas 78712, USA}
\newcommand{\Tufts}{Department of Physics and Astronomy, Tufts University, Medford, 
Massachusetts 02155, USA}
\newcommand{\UCL}{Physics and Astronomy Department, University College 
London, 
Gower Street, London WC1E 6BT, United Kingdom}
\newcommand{\Virginia}{Department of Physics, University of Virginia, 
Charlottesville, Virginia 22904, USA}
\newcommand{\WSU}{Department of Mathematics, Statistics, and Physics,
 Wichita State University, 
Wichita, Kansas 67260, USA}
\newcommand{\WandM}{Department of Physics, William \& Mary, 
Williamsburg, Virginia 23187, USA}
\newcommand{\Wisconsin}{Department of Physics, University of 
Wisconsin-Madison, Madison, Wisconsin 53706, USA}
\newcommand{\deceased}{Deceased.}
\affiliation{\ANL}
\affiliation{\Atlantico}
\affiliation{\Bandirma}
\affiliation{\BHU}
\affiliation{\Caltech}
\affiliation{\Charles}
\affiliation{\Cincinnati}
\affiliation{\Cochin}
\affiliation{\CSU}
\affiliation{\CTU}
\affiliation{\Delhi}
\affiliation{\Erciyes}
\affiliation{\FNAL}
\affiliation{\FSU}
\affiliation{\UFG}
\affiliation{\Guwahati}
\affiliation{\Houston}
\affiliation{\Hyderabad}
\affiliation{\IHyderabad}
\affiliation{\IIT}
\affiliation{\Indiana}
\affiliation{\ICS}
\affiliation{\INR}
\affiliation{\IOP}
\affiliation{\UIowa}
\affiliation{\ISU}
\affiliation{\Irvine}
\affiliation{\JINR}
\affiliation{\Magdalena}
\affiliation{\MSU}
\affiliation{\Duluth}
\affiliation{\Minnesota}
\affiliation{\Mississippi}
\affiliation{\OSU}
\affiliation{\NISER}
\affiliation{\Panjab}
\affiliation{\Pitt}
\affiliation{\QMU}
\affiliation{\SAlabama}
\affiliation{\Carolina}
\affiliation{\SMU}
\affiliation{\Sussex}
\affiliation{\Syracuse}
\affiliation{\Texas}
\affiliation{\Tufts}
\affiliation{\UCL}
\affiliation{\Virginia}
\affiliation{\WSU}
\affiliation{\WandM}
\affiliation{\Wisconsin}

\author{S.~Abubakar}
\affiliation{\Erciyes}

\author{M.~A.~Acero}
\affiliation{\Atlantico}

\author{B.~Acharya}
\affiliation{\Mississippi}

\author{P.~Adamson}
\affiliation{\FNAL}









\author{N.~Anfimov}
\affiliation{\JINR}


\author{A.~Antoshkin}
\affiliation{\JINR}


\author{E.~Arrieta-Diaz}
\affiliation{\Magdalena}

\author{L.~Asquith}
\affiliation{\Sussex}


\author{A.~Aurisano}
\affiliation{\Cincinnati}



\author{A.~Back}
\affiliation{\Indiana}
\affiliation{\ISU}



% ...

\collaboration{The NOvA Collaboration}
\noaffiliation




\date{\today}

\begin{abstract}
We present a high-statistics measurement of muon antineutrino-induced charged-current neutral pion production on a hydrocarbon target using the NOvA Near Detector. The differential cross sections as functions of the momenta and angles of the outgoing pion and muon, the squared four-momentum transfer, and the invariant mass of the hadronic system at an average neutrino energy of 2~GeV are measured and compared with predictions from various neutrino interaction models. The results agree with the GENIE prediction but suggest that other models underestimate the cross section in the $\Delta$(1232) resonance region. These results represent the most precise measurement of antineutrino-induced neutral pion production to date. 
\end{abstract}

\maketitle

\section{Introduction}
\label{sec:introduction}

Precise knowledge of neutrino and antineutrino cross sections on a variety of nuclear targets in the few GeV range will be essential to achieving the goals of future oscillation measurements. Cross section uncertainties, including the modeling of effects due to the nuclear environment, may limit the ability to accurately calculate the neutrino energy spectrum and may ultimately limit precision on oscillation parameters. Neutrino (and antineutrino) induced neutral pion ($\pi^0$) production is of particular importance for neutrino oscillation experiments, as the final state $\pi^0$ decays electromagnetically to two photons which could be misidentified as a signal event for electron neutrino appearance. Although neutral-current (NC) $\pi^0$ production is the dominant background, it is challenging to measure directly, and the lack of information about the outgoing lepton limits the ability to constrain the cross section model parameters. Quantitative information on the charged-current (CC) cross section's dependence on pion and lepton kinematics provides valuable input for related NC $\pi^0$ production. 

While there are several recent precise measurements of $\nu$-induced $\pi^0$ production on hydrocarbon~\cite{MiniBooNE:2010cxl,MINERvA:2017okh,NOvA:2023uxq}, there is only one previous measurement of antineutrino-induced CC $\pi^0$ production ($\numubar$CC$\pi^0$)~\cite{MINERvA:2015slz} at average energy 3.6~GeV. We present here a measurement of the differential dependence of the cross section on the muon  and $\pi^0$ momentum and angle in muon-antineutrino scattering on a hydrocarbon target ($\numubar + {\rm CH} \rightarrow \mu^+ + \pi^0 + X $)  over the energy range 1--5~GeV with a peak near 2~GeV. We also present the cross section dependence on derived variables, the squared four-momentum transfer, $Q^2$ and $W_{\rm EXP}$, which is related to the invariant mass of the hadronic system (defined in Sec.~\ref{sec:kinematics}). This result represents an increase in statistics by a factor of six compared to the previous measurement on the same nuclear target~\cite{MINERvA:2015slz}.

Antineutrino-induced $\pi^0$ production in the few GeV region arises from nucleon resonances and non-resonant processes. Charge exchange of a charged pion ($\pi^{\pm}\rightarrow \pi^0$) in the nucleus, through intranuclear final-state interactions (FSI), also contributes to the production cross section. We define signal $\pi^0$'s to be those produced through a primary antineutrino-nucleus interaction or intranuclear FSI, e.g., $\pi^\pm$ charge exchange ($p\pi^-\rightarrow n\pi^0$, $n\pi^+\rightarrow p\pi^0$) within the target nucleus. A $\pi^0$ produced in the primary interaction but then absorbed in the nucleus is excluded from the signal definition. Those produced through downstream secondary interactions in the detector or through $\nu_\mu$ primary interactions are treated as backgrounds. To ensure adequate signal efficiency and purity, the signal definition restricts the phase space of muon momentum, $p_\mu$, to $0.5 \leq p_\mu < 2.5$~GeV/c and muon angle, $\theta_\mu$, measured with respect to the neutrino beam direction, to $\theta_\mu < 60^\circ$.

\section{The NOvA Experiment}
The NOvA (NuMI Off-axis $\nu_e$ Appearance) experiment is a long-baseline neutrino oscillation experiment primarily designed to measure $\numu/\numubar$ disappearance and $\nu_e/\bar{\nu}_e$ appearance. It uses the Neutrinos at the Main Injector (NuMI) beam~\cite{Adamson:2015dkw} at Fermilab. Interactions are measured in two functionally identical liquid scintillator detectors, the Near Detector (ND), located about 1 km from the NuMI target, and the Far Detector, located 810 km downstream in Ash River, MN. The high-statistics ND data are used to provide constraints on the event rate for oscillation analyses and to measure neutrino-nucleus cross sections, such as the measurements presented in this paper. 

\subsection{The NuMI Beam}
\label{subsec:numi_beam}

\begin{figure}[!htb]
\includegraphics[height=5.8cm]{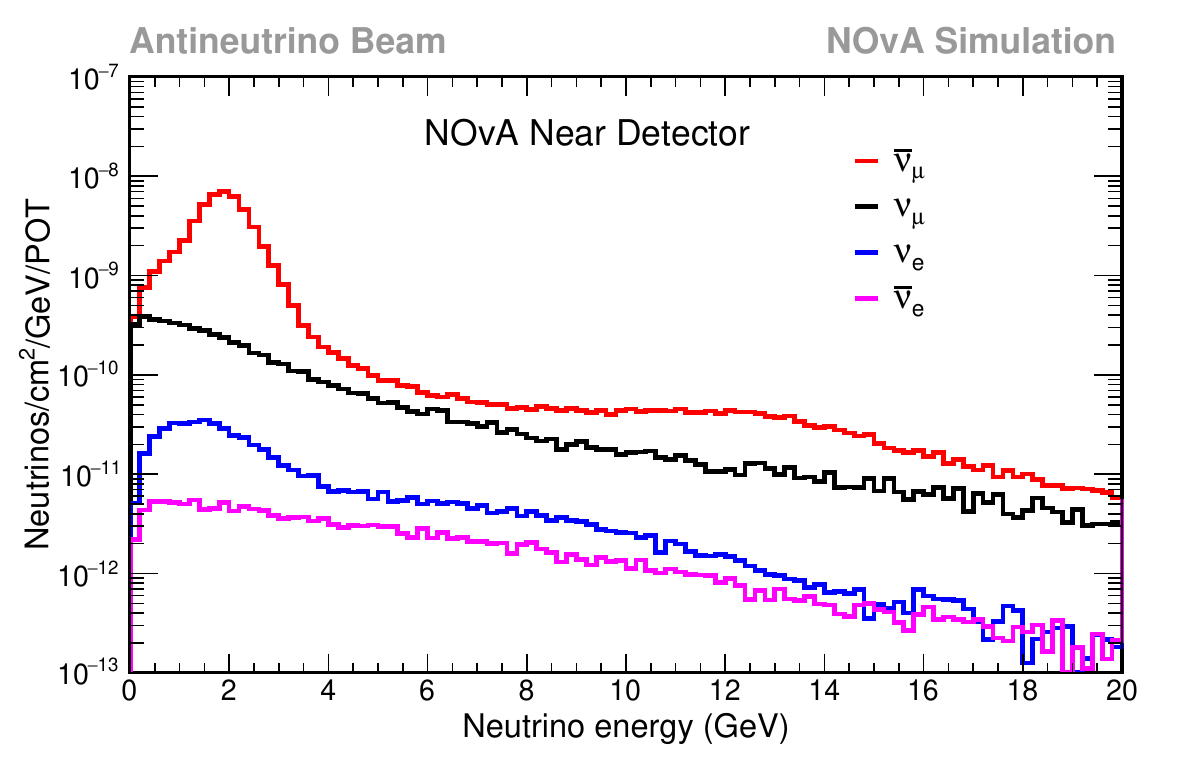}
\caption{Neutrino flux components at the NOvA near detector in antineutrino running mode. The muon antineutrino fraction is about 92\%.}
\label{fig:flux_mode}
\end{figure}

The NuMI beamline uses 120 GeV protons extracted from the Main Injector which strike a graphite target producing pions and kaons. The secondaries are focused using two magnetic horns and directed towards a 675 m decay pipe where they decay to produce primarily muon neutrinos and muons. The polarity of the focusing horns can be reversed to produce a beam that is primarily antineutrinos. Both detectors are situated approximately 14.6 mrad off the beam axis which results in a high-purity narrowband beam centered around 2 GeV. The antineutrino beam flux is shown in Fig.~\ref{fig:flux_mode} along with $\nu_\mu$, $\nu_e$, and $\bar{\nu}_{e}$ beam components. In the 1--5 GeV energy region, the wrong-sign component ($\nu_\mu$ component in $\bar{\nu}_{\mu}$) is $6.6\%$, and the $\nu_e + \bar{\nu}_e$ component is less than $1\%$. The data were accumulated during an antineutrino run from June 29, 2016 to February 26, 2019, corresponding to an exposure of $11.38\times 10^{20}$ Protons-On-Target (POT).

\subsection{The NOvA Near Detector}

The ND is a tracking calorimeter composed of liquid scintillator-filled PVC cells.  It is divided into a fully active region that has a cross sectional area of (3.9~$\times$~3.9)~m$^2$ and extends \SI{12.8}{\meter} along the beam direction. Immediately downstream is a 3.2-m-long  muon catcher. The fully active region consists of alternating vertical and horizontal planes, segmented into 3.9 m long cells with a rectangular cross section of \SI{6.6}{\centi \meter} (0.15 radiation lengths) along the beam direction by \SI{3.9}{\centi \meter}  transverse to the beam. Each cell is filled with a blend of 95\% mineral oil and 5\% pseudocumene with trace concentrations of wavelength shifting fluors~\cite{Mufson:2015kga}. The resulting composition by mass is 67\% carbon, 16\% chlorine, 11\% hydrogen, 3\% titanium, and 3\% oxygen along with other trace elements. The downstream muon catcher alternates 10-cm-thick steel plates with orthogonal scintillator plane pairs to improve muon containment. It spans the full detector width and 2/3 of its height, allowing muon containment for muons up to $\sim$2.5~GeV. The ND is 290 tons in total, of which 130 tons is scintillator, 78 tons steel in the muon catcher, and the rest primarily PVC.

\subsection{Simulation}
\label{sec:simulation}

Simulation is used throughout this analysis to calculate selection efficiencies and purities, estimate detector resolution effects, and assess systematic uncertainties. The stages of simulation include modeling the neutrino beam flux, simulating neutrino interactions within the detector, tracking the propagation of final-state particles, and modeling the detector's response. 

The NuMI flux is simulated using detailed beamline geometry and transport implemented in Geant4~(v9.2.p03)~\cite{GEANT4:2002zbu}. Outgoing hadron spectra produced from protons incident on the target are adjusted using PPFX (Package to Predict the FluX)~\cite{MINERvA:2016iqn} to correct for hadron production mismodeling by using data from external hadron production measurements~\cite{NA49:2006oyk,MIPP:2014shj,NA61SHINE:2011dsu,Barton:1982dg}.

\sloppy Neutrino interactions are simulated using a custom version of GENIE v3.0.6\cite{Andreopoulos:2015wxa,genie_manual_v3}, configured with the \texttt{G18\_10j\_00\_000} model. The initial-state nucleon is simulated using a local Fermi gas (LFG) model with a spherically symmetric, position dependent momentum distribution. Quasielastic (QE) scattering is modeled by the Valencia model \cite{Gran:2013kda} using a $z$-expansion formalism \cite{Meyer:2016oeg}, which includes long range nucleon-nucleon (RPA) correlations and multi-nucleon (e.g., 2p2h) interactions. Resonance (RES) production is modeled via the Berger-Sehgal model~\cite{Berger:2007rq}, which incorporates lepton mass effects into the original Rein-Sehgal model~\cite{Rein:1980wg}. GENIE's deep inelastic scattering (DIS) model, loosely defined here as non-resonant inelastic scattering, uses an effective leading-order framework with Bodek--Yang modifications~\cite{Bodek:2002ps}. The model spans a wide range in $Q^2$, covering both the perturbative QCD regime ($Q^2>4$~GeV$^2$) and extending to lower-$Q^2$ inelastic scattering, often referred to as shallow inelastic scattering (SIS), which describes the region where resonance processes transition to DIS. At low $Q^2$, the Bodek-Yang model includes corrections for higher-twist~\cite{Yang:1998zb} and target-mass~\cite{Georgi:1976ve, Barbieri:1976bj, Barbieri:1976rd} effects using a modified scaling variable and $Q^2$-dependent corrections to the parton distribution functions which are tuned to achieve agreement with global data. Additionally, for $W<1.7$~GeV, the inelastic cross section is decomposed into channels by hadronic final state multiplicity and tuned using a data-driven model~\cite{Yang:2009zx} to represent the non-resonant contribution in the resonance region.

Hadrons produced in neutrino interactions in nuclei can re-interact as they exit the nucleus through various channels, including nucleon knock-out, elastic scattering, charge exchange, and absorption. These final-state interactions (FSI) are simulated in GENIE using the $hN$ semiclassical intranuclear cascade model~\cite{Salcedo:1987md}, which propagates hadrons in finite steps through a nuclear density profile, with interaction probabilities governed by free hadron--nucleon cross sections. In the NOvA implementation, the pion-fate parameters (mean free path and charge exchange, absorption, and elastic scattering fractions) were adjusted to improve agreement with pion–carbon data (see Table II of Ref.~\cite{PinzonGuerra:2018rju}).

The model used in this analysis, called ``NOvA Tune v2", is the same model as in our recent 3-flavor oscillation paper~\cite{NOvA:2023iam}. It includes custom tuning applied to the default GENIE model to adjust the Valencia MEC model to match a subset of NOvA ND data as described in Ref.~\cite{MartinezCasales:2023bkf}. The adjustments to pion fate parameters described above are also part of this custom tune.

The final-state particles emerging from the nucleus are propagated through the detector using a detailed Geant4 (v4.10)~\cite{GEANT4:2002zbu} simulation of the NOvA ND and surrounding materials. The detector response is simulated using measured detector and optical parameters, including scintillator light yield, fiber attenuation, and channel-by-channel calibration constants from cosmic-ray muons together with a detailed model of the electronics response~\cite{Mufson:2015kga,Aurisano:2015oxj,Anfimov:2020okt}.

\subsection{Event and kinematic variable reconstruction}
\label{sec:kinematics}

Reconstruction of neutrino interactions, referred to as ``events,'' starts from the recorded energy depositions in NOvA cells, referred to as ``hits.'' Each hit is associated with a time, 2D position, and energy deposition. Hits are calibrated using minimum ionizing regions of cosmic ray muons that stop in the detector. Cell-to-cell differences are removed using cosmic muon samples that traverse the detector. Hits close in time and space are clustered and a three-dimensional interaction vertex is identified using the straight-line features of the hit patterns. From this vertex, particle trajectories are reconstructed in both detector views and combined to form 3D objects called ``prongs,'' representing individual particles. Separately, a tracking algorithm based on a Kalman filter~\cite{Kalman:1960mft,Fruhwirth:1987fm} is applied to the event to better reconstruct long, track-like particles such as muons. Hit patterns that are identified as track-like by this algorithm are referred to as “tracks”.

To isolate a sample of $\numubar$CC$\pi^0$ we require a final state muon along with at least two photon candidates, which are used to reconstruct the $\pi^0$. The muon-candidate momentum is determined by its track length with a typical resolution of 3\%. The muon angular resolution is better than 3$^{\circ}$ over the angular range of the sample. The momentum and angle of the $\pi^0$ are directly reconstructed from the momentum vectors of the two candidate photon prongs, $\vec{p}_1$ and  $\vec{p}_2$, according to
\begin{equation}
  \vec{p}_{\pi^0}=\vec{p}_1+\vec{p}_2\,.
\end{equation}
The photon-candidate momentum vectors are reconstructed using the prong direction and the calorimetric energy, corrected by a scale factor for dead material and threshold effects. The $\pi^0$ momentum resolution is about 14\% and the angular resolution averaged over the sample is about 6$^\circ$. To construct the squared four-momentum transfer, $Q^2$, which is given by
\begin{equation}
  Q^2=2E_\nu(E_\mu-|p_\mu| \cos{\theta_\mu})-m^2_\mu\,,
  \label{eq:q2}
\end{equation}
we need an estimate of the neutrino energy, $E_\nu$.  We construct $E_\nu=E_\mu+ E_{\rm HAD}$ from the muon energy and the hadronic energy, $E_{\rm HAD}$, which is the sum of the calorimetric energy from all hits that are not associated with the muon track, as in Ref.~\cite{NOvA:2021eqi}. In this prescription, the $\pi^0$ is treated as part of the hadronic system rather than separately reconstructed. The reconstructed $Q^2$ has resolution of about 16\%.

$W_{\rm EXP}$ is related to the invariant mass of the hadronic system, but ignores the initial momentum of the struck nucleon, which enters into the true $W$. Following Ref.~\cite{MINERvA:2016sfc}, we define the measured invariant mass, $W_{\rm EXP}$, as
\begin{equation}
  W_{\rm EXP}^2=M^2+2M(E_\nu-E_\mu)-Q^2\,,
  \label{eq:w}
\end{equation}
where $M$ is the proton mass and $Q^2$ is defined in Eq.~\ref{eq:q2}. We use $W_{\rm EXP}$ to distinguish among contributions from the $\Delta(1232)$, higher-mass $N^*$ resonances, and DIS processes. The resolution of $W_{\rm EXP}$ is approximately 7\%.

The selected sample is binned into nine bins of $\pi^{0}$ momentum between 0 and 2.0~GeV/c and eleven bins of $\pi^{0}$ angle between 0 and 180~degrees. For the muon kinematic distributions fourteen momentum bins between 0.5 and 2.5~GeV/c and twelve angular bins between 0 and 60~degrees are used. The distributions in $Q^2$ are binned into nine intervals between 0 and 2.0~GeV$^2$, and those in $W_{\rm EXP}$ are binned into eleven intervals between 1.0 and 3.0~GeV.

\section{Event selection}
\label{sec:eventsel}

To reject poorly reconstructed events, candidate events must have a minimum of 20 hits and must include at least one reconstructed track that crosses more than four contiguous planes. Preselected events are also required to have a vertex within the fiducial volume, defined as a 2.7~m~$\times$~2.7~m$~\times$~9.0~m region beginning 1~m downstream of the detector front face. In addition, containment requirements are applied to all reconstructed prongs and tracks. Prongs are required to stop within the detector active region, while tracks are allowed to stop in the muon catcher.

The candidate primary muon is identified using a Boosted Decision Tree (BDT) developed in a previous analysis, called MuonID~\cite{NOvA:2021eqi}. Candidate tracks are required to satisfy MuonID~$\geq 0.27$. (This threshold has been re-optimized for muon–antineutrino interactions compared with 0.24 from Ref.~\cite{NOvA:2021eqi}). In addition to the MuonID requirement, the muon selection limits 
the reconstructed muon momentum and angle: $0.5 < p_\mu \leq 2.5$~GeV/c and $\theta_\mu < 60^\circ$ (as mentioned in Sec.~\ref{sec:introduction}). 

\begin{figure}[!htb]
  \centering
  \includegraphics[width=0.42\textwidth]{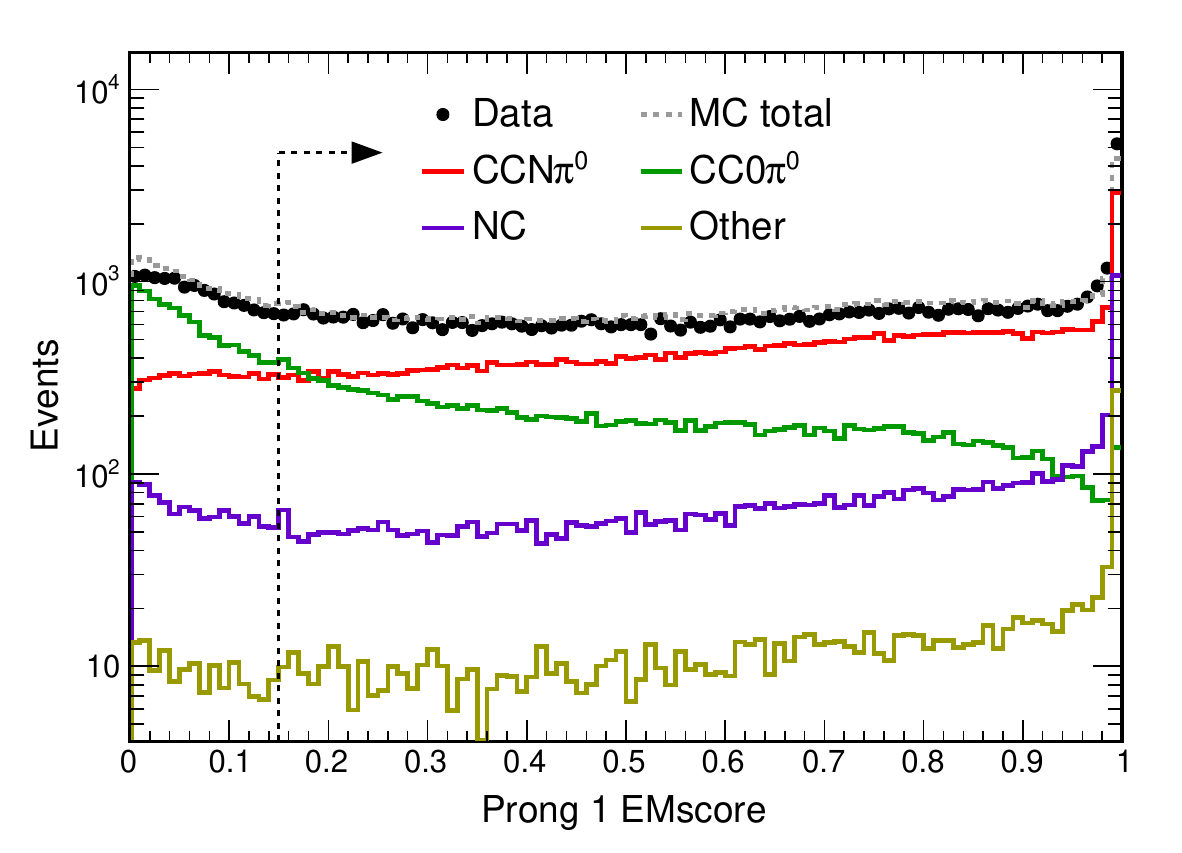}
  \includegraphics[width=0.42\textwidth]{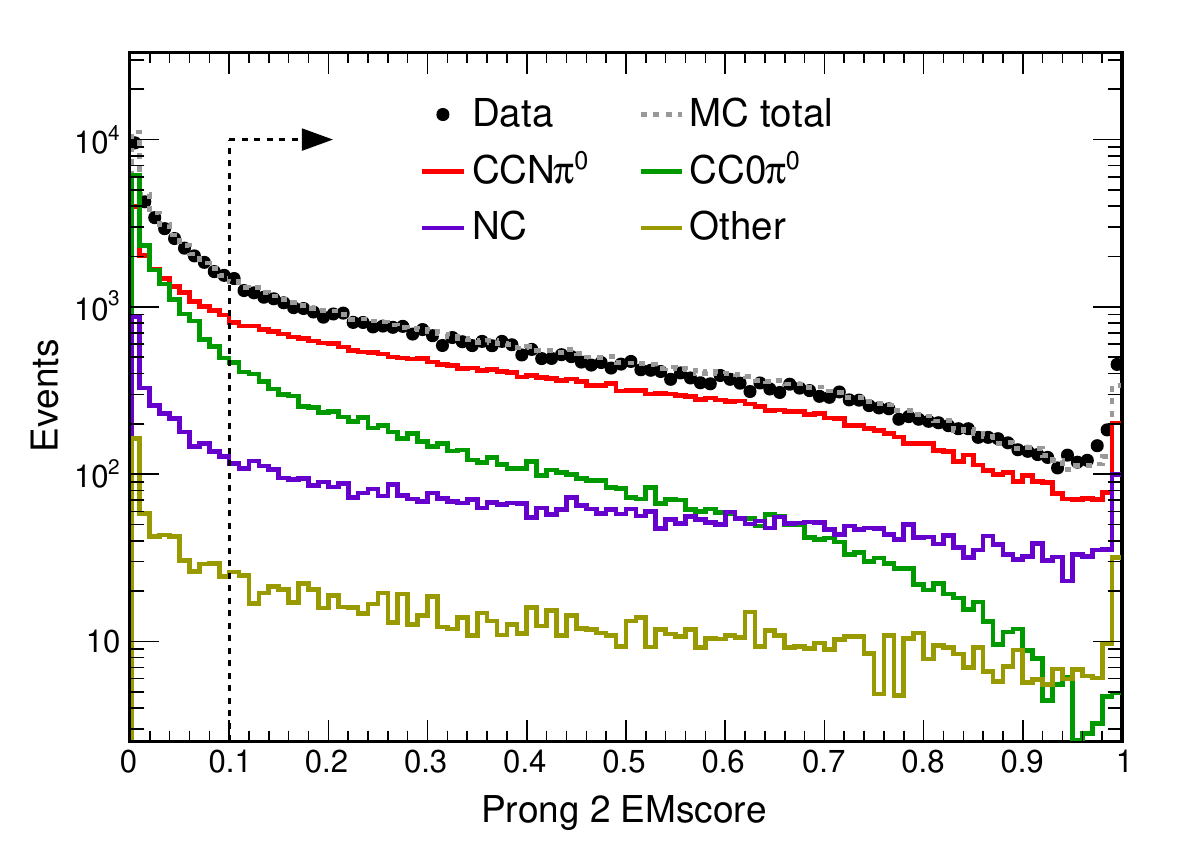}
  \caption{Distributions of EMscore for leading (left) and subleading (right) prongs. Event quality preselection, muon selection, prong multiplicity, and minimum number of prong hit cuts have been applied. The black points represent the data, while the dashed gray line shows the total MC prediction. The vertical dashed lines indicate the signal region selection thresholds. The CCN$\pi^0$ event category includes the signal, while CC0${\pi^0}$, NC, and Other are background event categories.}
  \label{fig:EMscores}
\end{figure}

\begin{figure}[!htb]
  \centering
  \includegraphics[height=7cm]{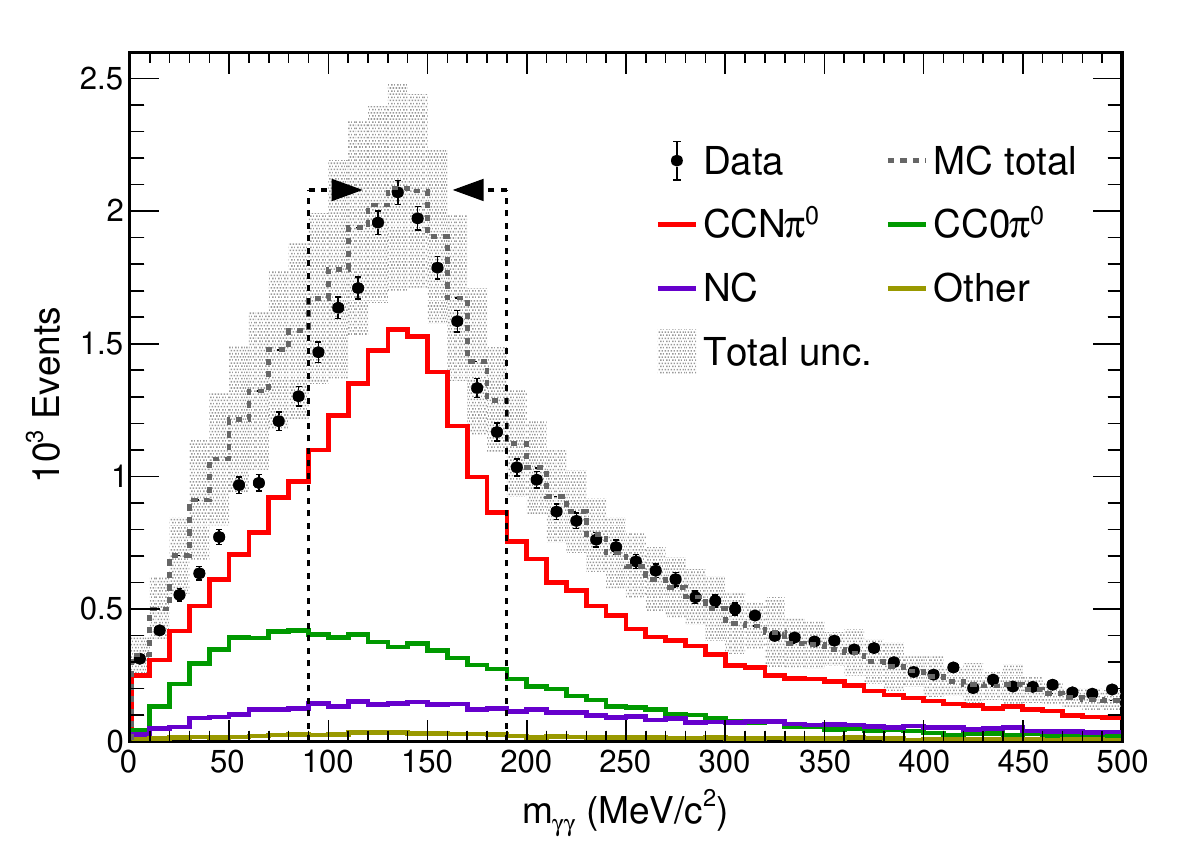}
  \caption{Reconstructed $\pi^{0}\rightarrow\gamma\gamma$ mass distribution after applying the preselection, muon selection, prong multiplicity, prong hit count, and EMscore requirements. Black points represent data with statistical uncertainties, while the shaded gray band denotes the total uncertainty, including both systematic and simulation statistical components. The vertical dashed lines separate the signal region from two background-dominated (low-mass and high-mass) regions used to constrain the backgrounds.}
  \label{fig:pi0mass}
\end{figure}

After identifying the muon candidate, and associating it with a reconstructed prong (called the ``muon prong''), the remaining prongs are examined to select electromagnetic-like (EM-like) candidates used in the reconstruction of the $\pi^0$. Events must contain a total of at least three reconstructed prongs (prong multiplicity cut): one identified as a muon candidate and two EM-like prongs, consistent with photons from $\pi^0\rightarrow\gamma\gamma$, which are used to reconstruct the $\pi^0$. We employ a convolutional neural network (CNN) trained on simulated samples of individual particles ($\gamma$, electron, $\pi^\pm$, and proton) propagating in the NOvA Near Detector. The training uses uniformly distributed kinetic energies and angles spanning the phase space of final-state particles produced in antineutrino interactions. Only fully contained prongs with more than four hits are included in the training samples. A binary classifier is constructed to distinguish EM-like ($\gamma$ or $e$) prongs from non-EM-like ($p$, $\pi^\pm$) prongs. Figure~\ref{fig:EMscores} shows the CNN output ``EMscore'' for the leading (highest EMscore) and subleading (second-highest EMscore) prongs in events passing preselection, muon selection, prong multiplicity, and minimum prong hit count requirements. The black points represent the data, while the dashed gray line shows the total Monte Carlo (MC) prediction. In Fig.~\ref{fig:EMscores}, events with at least one $\pi^0$ (CCN$\pi^0$) include the signal category, while the remaining curves include events without a $\pi^0$ (CC0${\pi^0}$), NC events, and Other (discussed in detail below).
The leading prong is required to have at least six hits and EMscore $\geq$ 0.15, while the subleading prong must have at least four hits and EMscore $\geq$ 0.10. The minimum hit thresholds improve the CNN's ability to classify energy deposition patterns. A less stringent hit requirement is applied to the subleading prong, since the presence of a leading EM-like prong increases the likelihood of a second. The two selected EM-like prongs are combined to form a diphoton invariant mass, shown in Fig.~\ref{fig:pi0mass}. To suppress backgrounds (discussed below), we retain only events with reconstructed masses within $\pm$1$\sigma$ of the fitted $\pi^{0}$ peak (90 MeV/$c^2$ $\leq m_{\gamma\gamma} < 190$ MeV/$c^2$). Events outside this window are used to constrain backgrounds, as described in Sec.~\ref{sec:sidebands}. A discrepancy between data and simulation is observed in the low invariant mass region which is consistent with overestimation of the CC0${\pi^0}$ background (discussed below) which peaks in this region.

There are two irreducible backgrounds that contain both a final-state muon and $\pi^0$. The first arises from CC primary $\pi^{0}$ production induced by the $\nu_\mu$ beam component in antineutrino mode, and the second corresponds to $\nu_{\mu}/\bar{\nu}_{\mu}$ CC secondary $\pi^0$ production, where the $\pi^0$ is produced in a secondary interaction outside the target nucleus. These contributions are predicted using the GENIE neutrino interaction simulation, combined with the flux prediction described in Sec.~\ref{subsec:numi_beam}, and their associated uncertainties are evaluated through model and flux variations as discussed in Sec.~\ref{sec:systematics}. The remaining dominant backgrounds arise from CC0${\pi^0}$ and from NC interactions. CC0${\pi^0}$ and from NC interactions. Finally, ``Other'' backgrounds are predominantly events that are outside of the true phase space of the measurement but leak into our sample by misreconstruction or smearing effects, and $\nu_e/\bar{\nu}_e$ CC interactions. These backgrounds are determined from simulation and are negligible.

The NC background dominates for low-momentum muon candidate tracks where, typically, a charged pion is misidentified as the muon. The minimum muon momentum requirement $p_\mu \geq 0.5$~GeV/c is effective at reducing this background. The diphoton invariant mass cut suppresses both the CC0${\pi^0}$ component, which peaks at low invariant mass, as well as the NC background which more uniformly populates the invariant mass distribution.

Table~\ref{tab:nevents_cuts} summarizes the numbers of signal and background events passing each selection cut, along with the signal selection efficiency and purity. The table also shows the fractions of the initial MC and data samples surviving each cut, allowing a direct comparison of the cut-by-cut behavior in data and simulation prior to the template fit. The muon selection requirements reduce the signal by about 10\% while rejecting about half the background. The primary effect is on the NC background, in which the misidentified muon track typically has low momentum or fails the MuonID requirement. Prong multiplicity and hit count both have sizable effects on the background, with the biggest effect on the CC0${\pi^0}$ background, which contains more events with poorly reconstructed, low-quality prongs that have very few hits. In the signal sample, the prong multiplicity cut rejects events for which one of the photons is not reconstructed (e.g., the photons do not have sufficient angular separation to be resolved, or one does not convert in the detector).  The prong EMscore requirement mainly serves to reject the dominant CC0$\pi^0$, which arises from final-state charged pions and nucleons that are incorrectly identified as EM-like prongs. After all selection cuts, the selected data sample consists of 16,687 selected events with signal purity of 42.3\%.

\begin{table}[htbp]
  \begin{center}
    \caption{Number of selected events (normalized to $11.38\times10^{20}$ POT) passing each selection cut before the template fit. Columns 2-5 show the MC signal and background yields, along with the signal selection efficiency and purity. The last two columns show the fractions of the MC and data samples surviving each cut.} 
    \resizebox{0.75\textwidth}{!}{
      \begin{tabular}{lrrcccc}
        \hline\hline
        \multicolumn{1}{c}{Selection cut} & \multicolumn{1}{c}{Signal} & \multicolumn{1}{c}{Background} & \multicolumn{1}{c}{Efficiency} & \multicolumn{1}{c}{Purity} & \multicolumn{1}{c}{MC fraction} & \multicolumn{1}{c}{Data fraction} \\
        \hline
        Preselection & 92,401 & 1,739,734 & 1.000 & 0.050 & 1.000 & 1.000\\
        Muon selection & 82,132 & 812,303 & 0.889 & 0.092 & 0.488 & 0.469 \\
        Prong multiplicity & 51,055 & 188,768 & 0.553 & 0.213 & 0.131 & 0.107 \\
        Prong hit count & 22,578 & 57,423 & 0.244 & 0.282 & 0.044 & 0.044 \\
        Prong EMscore & 15,715 & 29,013 & 0.170 & 0.351 & 0.024 & 0.025 \\
        $\pi^{0}$ mass & 7,567 & 10,308 & 0.082 & 0.423 & 0.010 & 0.010 \\
        \hline\hline
      \end{tabular}
      }
      \label{tab:nevents_cuts}
  \end{center}
\end{table}

\section{Sideband Regions and Template Fit}
\label{sec:sidebands}

We consider four categories of events, which are fit to determine the selected sample composition as described below. The ``signal-like" events (CCN$\pi^0$) contain both a true muon and at least one $\pi^0$. This category consists of 59.7\% signal (from $\numubar$CC$\pi^0$), and the remainder is from the two indistinguishable backgrounds: $\nu_\mu$-induced primary interactions (11.5\%) and $\nu_{\mu}/\bar{\nu}_{\mu}$ CC secondary $\pi^0$ production (28.8\%). The additional three categories are the CC0$\pi^0$, NC, and Other background classifications described in Sec.\ref{sec:eventsel}. 

To constrain the CC0$\pi^0$ and NC backgrounds, we construct four background-dominated sideband regions. Each background component has a shape that is different from that of the signal-like CCN$\pi^0$ category, which allows a simultaneous fit to determine its normalization as described below.

The CC0${\pi^0}$ background dominates in regions with low prong EMscore as shown in Fig.~\ref{fig:EMscores}. Due to the strong correlation between the prong EMscores, we construct a sideband with a low Prong 1 EMscore of [0.05, 0.15) and without an additional requirement on the Prong 2 EMscore. The events with Prong 1 EMscore $<$ 0.05 typically have lower reconstructed energy and broader angular distributions than those in the signal region. They are excluded due to these significant kinematic differences. To increase statistics in the CC0${\pi^0}$  background sample, we define an additional sideband that includes events with signal-like Prong 1 EMscore [0.15, 0.80) but low reconstructed $\pi^0$ invariant mass ($m_{\gamma\gamma}<$~90 MeV/$c^2$).

\begin{figure}[tb]
  \centering
  \includegraphics[height=7cm]{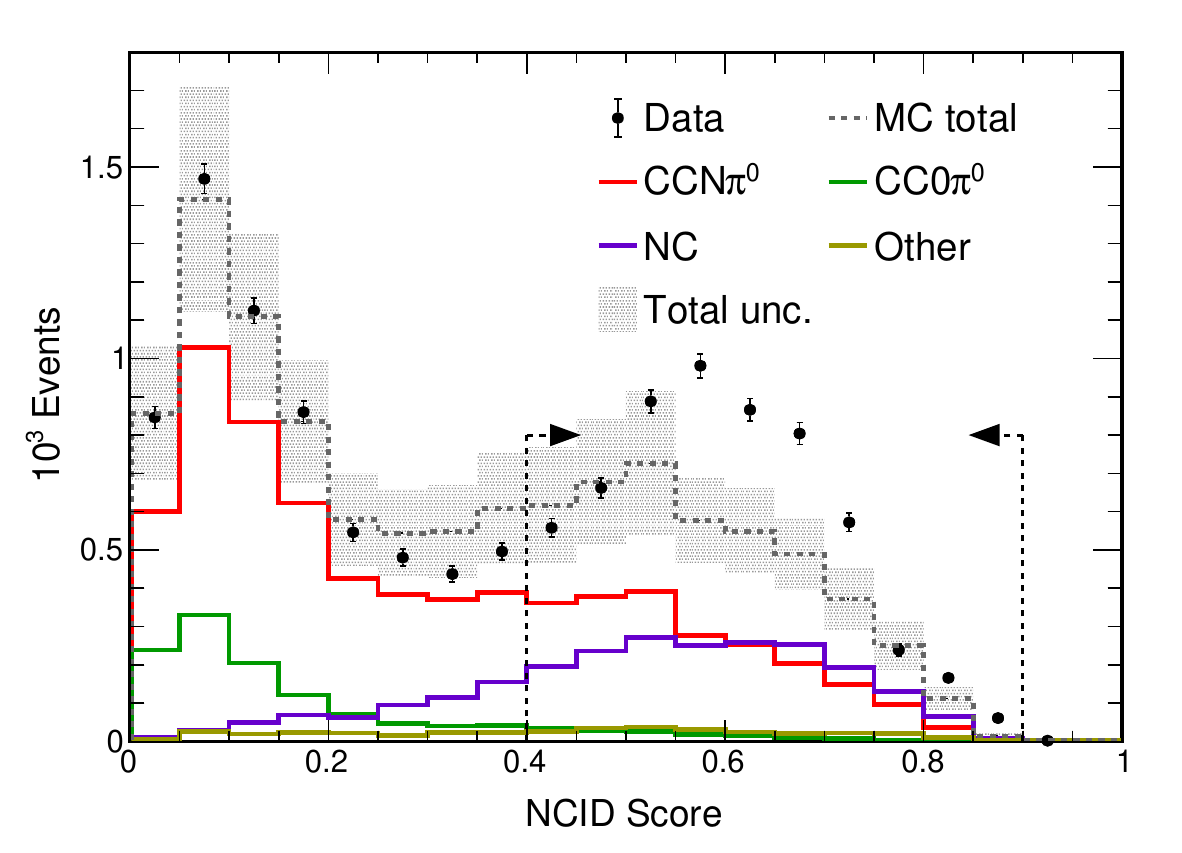}
  \caption{
    NCID score distribution after applying the preselection, muon selection, prong multiplicity, prong hit count, EMscore, $m_{\gamma\gamma}\geq$ 190~MeV/$c^2$, and NProngs $>$ 3 cuts. Black points represent data with statistical uncertainties, while the shaded gray band denotes the total uncertainty, including both systematic and simulation statistical components. The vertical dashed lines indicate the region used for the template fit.}
  \label{fig:ncid_fullscore}
\end{figure}

NC interactions contribute a subdominant background that is broadly distributed in $m_{\gamma\gamma}$ (see Fig.~\ref{fig:pi0mass}). We use a multi-class BDT, referred to as NCID, to separate NC events from both signal and CC0$\pi^0$ background in the region $m_{\gamma\gamma} \geq$ 190 MeV/$c^2$ where the backgrounds are comparable in size. The NC background events contain a particle (usually a charged pion) that is misdentified as a muon. Typically, these events will have less energy in the reconstructed hadronic system than a true charged-current event, which will erroneously not include the energy from the particle that is misidentified as the muon. They will also on average have greater reconstructed missing transverse momentum, and uncharacteristic particle angular distributions, due to the incorrectly labeled particle. The BDT is trained on variables that are sensitive to these differences, including the opening angle between the reconstructed muon and $\pi^0$ in the center-of-mass frame, missing transverse momentum, extra reconstructed energy (outside of the muon and $\pi^0$ prong candidates), and the $\pi^0$ energy fraction. The NCID score distribution is shown in Fig.~\ref{fig:ncid_fullscore}. The NC-enhanced sideband is defined by selecting events with $m_{\gamma\gamma} \geq$ 190 MeV/$c^2$ and 0.4 $\leq$ NCID $<$ 0.9. A discrepancy between data and simulation is observed in the high NCID region where the NC background peaks, indicative of mismodeling of this background (This is remedied by the data-driven background constraints described below).

To enhance shape discrimination, signal and sideband regions are further separated based on the number of reconstructed final state prongs. Subsamples with three (NProngs = 3) and greater than three (NProngs $>$ 3) prongs correspond to different average hadronic invariant mass ($W$) and are sensitive to different primary processes, with the delta resonance dominating the NProngs = 3 subsample. This separation results in two CCN$\pi^0$ signal regions, four CC0$\pi^0$ sideband regions, and two NC sideband regions. The low EMscore NProngs $>3$ sample is dominated by poorly reconstructed events in all categories and it is therefore excluded. The NC NProngs = 3 sideband is also discarded due to low statistical precision.

\begin{table}[htbp]
  \centering
  \caption{Summary of template definitions, including targeted event category and selection criteria, after applying the preselection, muon selection, prong multiplicity, and prong hit count cuts.}
  \begin{tabular}{ccl}
    \hline\hline
    \multicolumn{1}{c}{Name} & 
    \multicolumn{1}{c}{~~~~~Target~~~~~} & 
    \multicolumn{1}{c}{Definition}
    \\
    \hline
    T1 & CCN$\pi^{0}$ & Prong 1 EMscore $\geq$ 0.15; Prong 2 EMscore $\geq$ 0.1; $m_{\gamma\gamma} \in$ [90, 190)~MeV/$c^2$; NProngs $=$ 3 \\
    T2 & CCN$\pi^{0}$ & Prong 1 EMscore $\geq$ 0.15; Prong 2 EMscore $\geq$ 0.1; $m_{\gamma\gamma} \in$ [90, 190)~MeV/$c^2$; NProngs $>$ 3 \\
    T3 & CC0$\pi^{0}$ & Prong 1 EMscore $\in$ [0.15, 0.8); $m_{\gamma\gamma} <$90~MeV/$c^2$; NProngs $=$ 3 \\
    T4 & CC0$\pi^{0}$ & Prong 1 EMscore $\in$ [0.15, 0.8); $m_{\gamma\gamma}<$90~MeV/$c^2$; NProngs $>$ 3 \\
    T5 & CC0$\pi^{0}$ & Prong 1 EMscore $\in$ [0.05, 0.15); NProngs $=$ 3 \\
    T6 & NC & Prong 1 EMscore $>$ 0.15; Prong 2 EMscore $>$ 0.1; $m_{\gamma\gamma}\geq$ 190~MeV/$c^2$; NCID $\in$ [0.4, 0.9); NProngs $>$3 \\
    \hline\hline
  \end{tabular}
  \label{tab:selected_templates}
\end{table}

\begin{figure}[htb]
  \centering
  \includegraphics[width=0.3\textwidth]{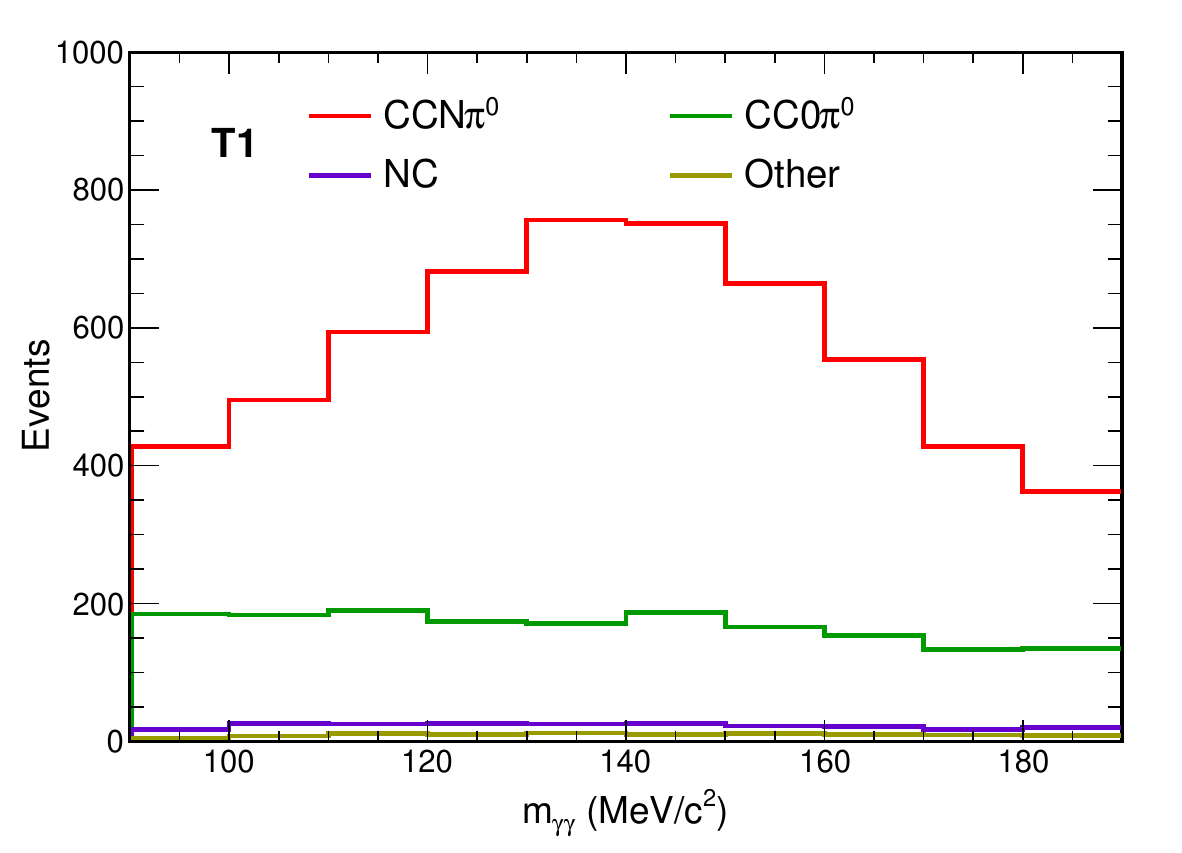}
  \includegraphics[width=0.3\textwidth]{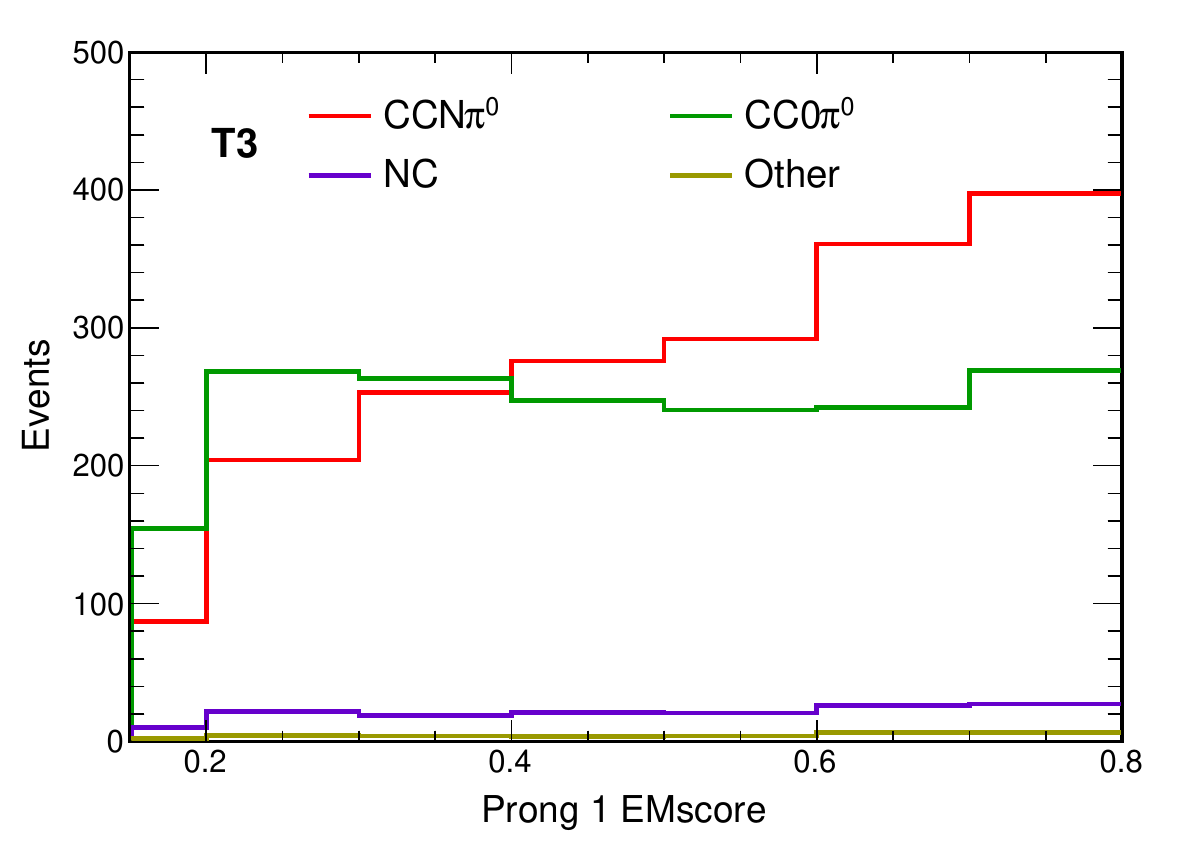}
  \includegraphics[width=0.3\textwidth]{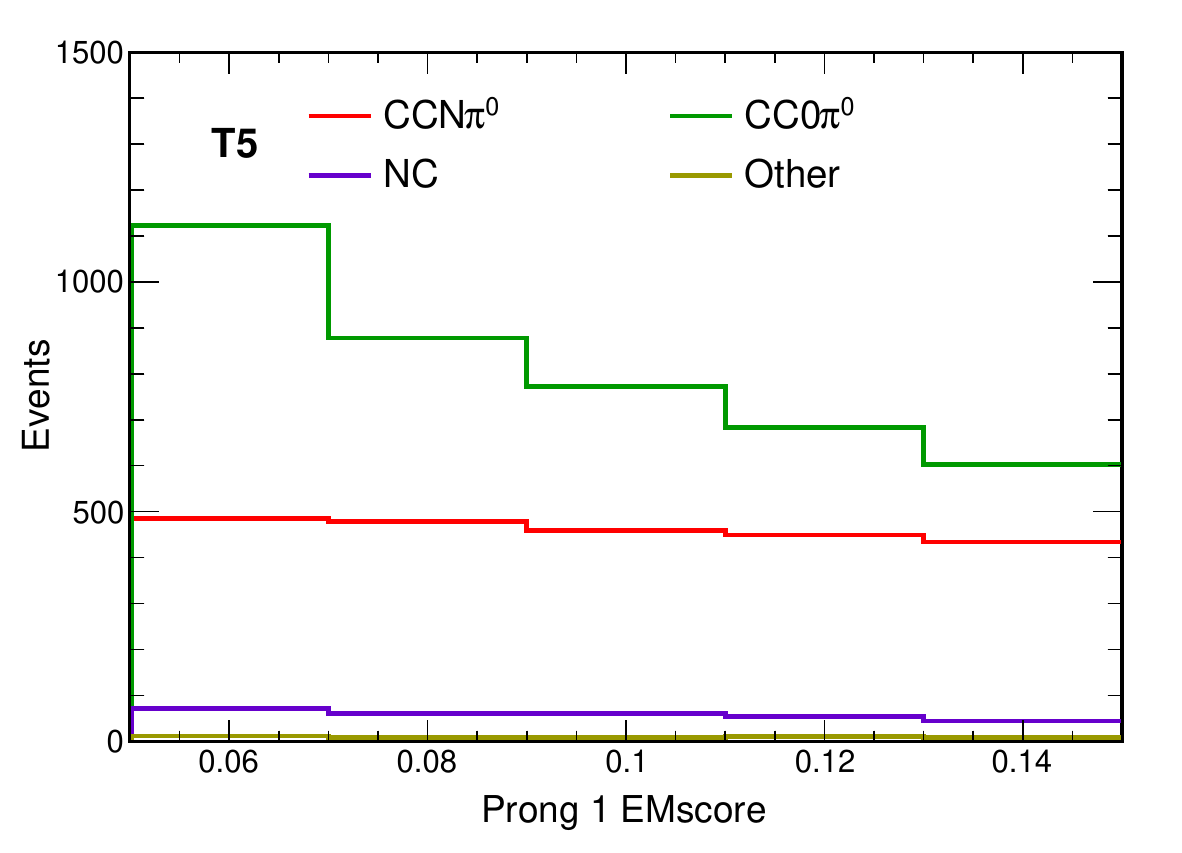}
  \includegraphics[width=0.3\textwidth]{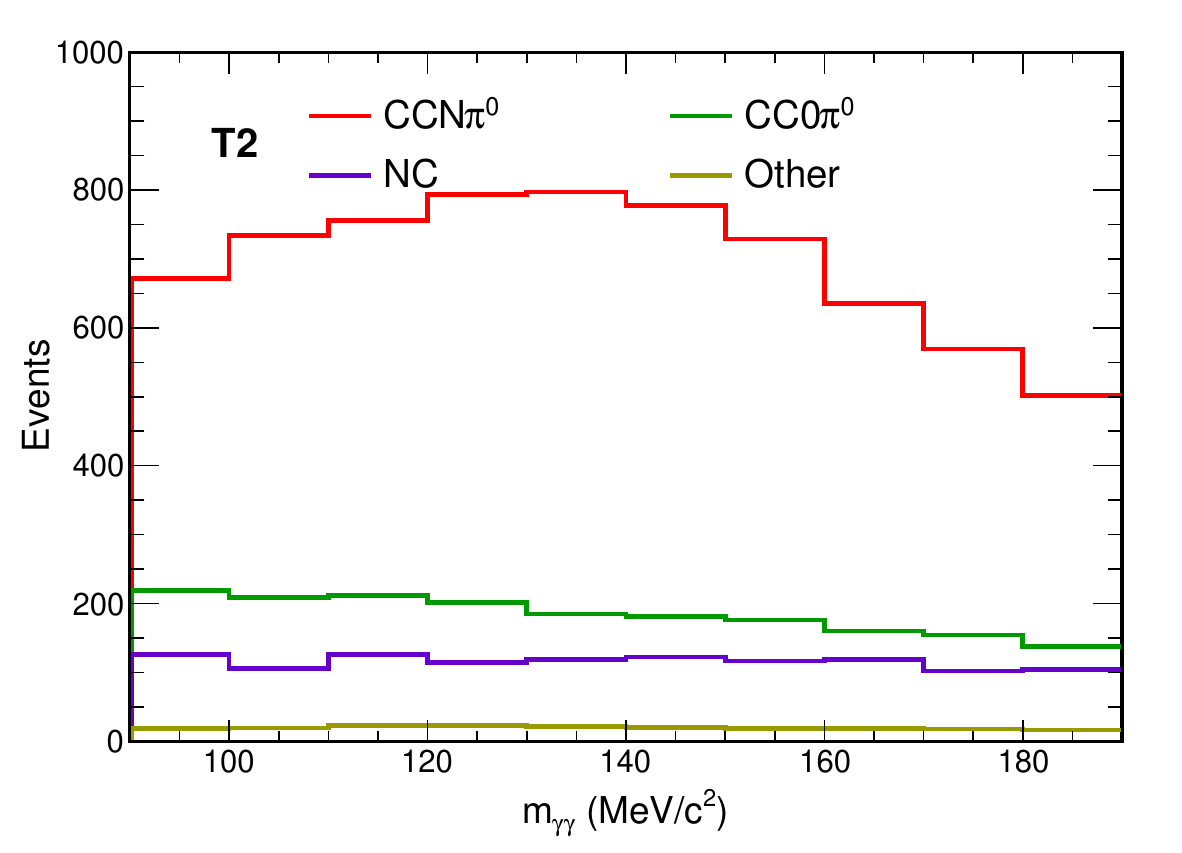}
  \includegraphics[width=0.3\textwidth]{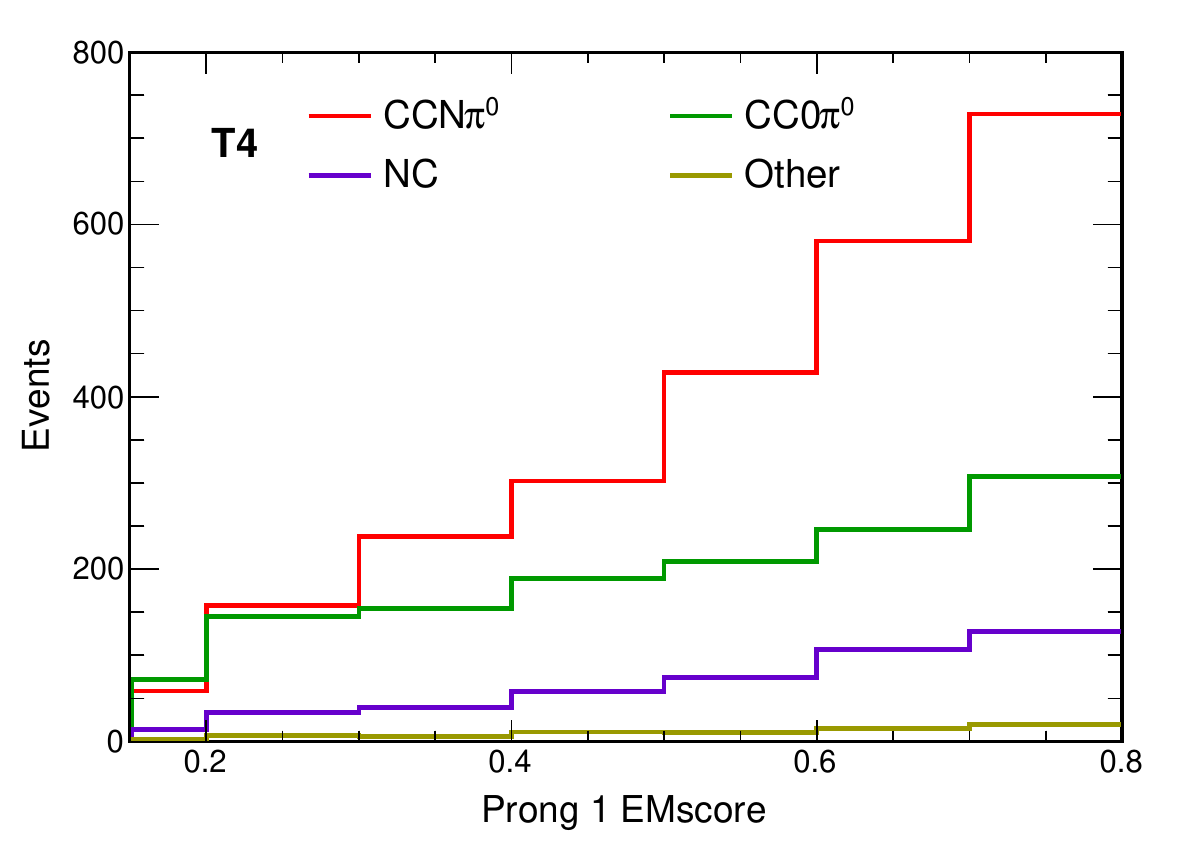}
  \includegraphics[width=0.3\textwidth]{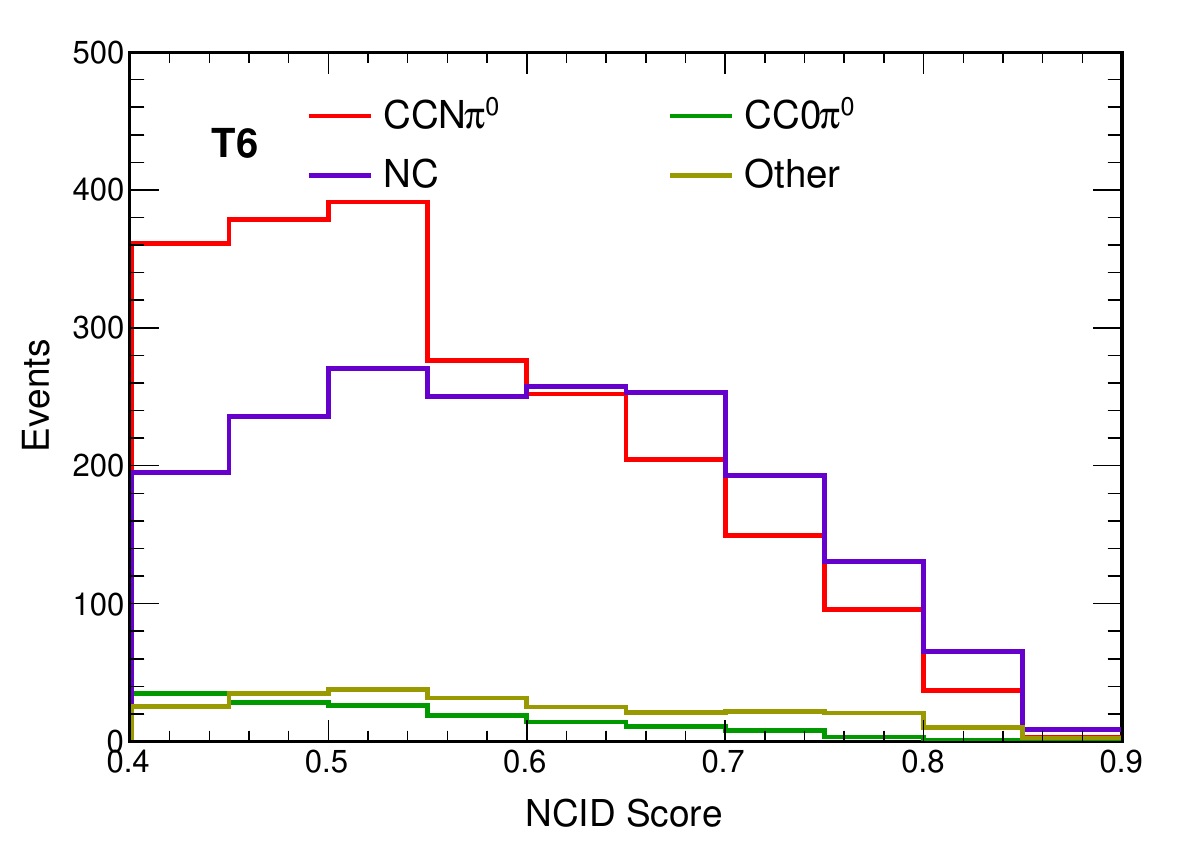}
  \caption{Signal and sideband samples used in the template fit, integrated over all kinematic bins.  Top row: NProngs $=$ 3; bottom row: NProngs $>$ 3. Left: distributions in signal region; middle and top right: sideband distributions for $\rm{CC}0\pi^0$; bottom right: NC sideband.}
  \label{fig:template}
\end{figure}

The two signal templates (T1 and T2) and four sideband templates (T3--T6) are shown for the full sample in Fig.~\ref{fig:template} and summarized in Tab.~\ref{tab:selected_templates}. These distributions are shown at the pre-fit level to illustrate the composition and shape differences of the signal and background templates used in the fit. For each observable of interest ($p_{\pi^0}$, $\theta_{\pi^0}$, $p_\mu$, $\theta_{\mu}$, $Q^2$, and $W_{\rm EXP}$) a separate fit is performed to extract the level of the three corresponding components (CCN$\pi^0$, CC0$\pi^0$, and NC) for each kinematic bin. The template distributions are binned in the respective discriminating variables, indexed by $j$, and then further divided into kinematic bins for the observable of interest, indexed by $k$. The prediction in each bin, $\mu_{jk}$, is 
\begin{equation}
  \mu_{jk} = a_{k} (N_{\mathrm{CCN\pi^0}})_{jk} + b_{k}(N_{\mathrm{CC\,0\pi^0}})_{jk} + c_{k} (N_{\mathrm{NC}})_{jk}+(N_{\mathrm{Other}})_{jk}\,,
  \label{eq:template}
\end{equation}
where $N_{\mathrm{CCN\pi^0}}$ is the signal-like component (CCN$\pi^0$), and $N_{\mathrm{CC0\pi^0}}$, $N_{\mathrm{NC}}$, and $N_{\mathrm{Other}}$ are from CC0$\pi^0$, NC, and Other backgrounds, respectively. In order to perform the fit, $\mu_{jk}$ is arranged into ``template'' bins which run over all of the $j\times k$ bins. Here, multiple template bins share the same normalization parameters --- $a_k$, $b_k$, and $c_k$ --- which correspond to CCN$\pi^0$, CC0$\pi^0$, and NC components in kinematic bin $k$, respectively. Contributions from $\nu_\mu$CC$\pi^0$ and $\nu_{\mu}/\bar{\nu}_{\mu}$ CC secondary $\pi^0$ production, which share similar reconstructed final-state topologies and exhibit comparable features in the template observables, are included in the $N_{\mathrm{CCN\pi^0}}$ component. $N_{\mathrm{Other}}$ is negligible, and along with bins with fewer than ten events, is not adjusted. The normalization parameters are obtained by minimizing the global chi-square
\begin{equation}
  \chi^2 = (x_\alpha-\mu_\alpha)^T V_{\alpha\beta}^{-1}(x_\beta-\mu_\beta)\,,
\end{equation}
where $x_\alpha$($x_\beta$) is the observed data yield in template bin $\alpha$($\beta$) and $V_{\alpha\beta}$ is the full covariance matrix that includes both statistical (diagonal) and systematic uncertainties, as described in Sec.~\ref{sec:systematics}. This procedure relies on simulation to model only the shape within a bin, while the normalization is determined by the fit. Therefore, in constructing $V_{\alpha\beta}$, we normalize each systematic effect to the nominal shape distribution to isolate the component of the uncertainty that affects only the shape. In each kinematic bin, the six templates are combined to construct the full covariance matrix, capturing the correlations among template bins.

\begin{figure}[h!]
  \centering
  \includegraphics[width=0.45\textwidth]{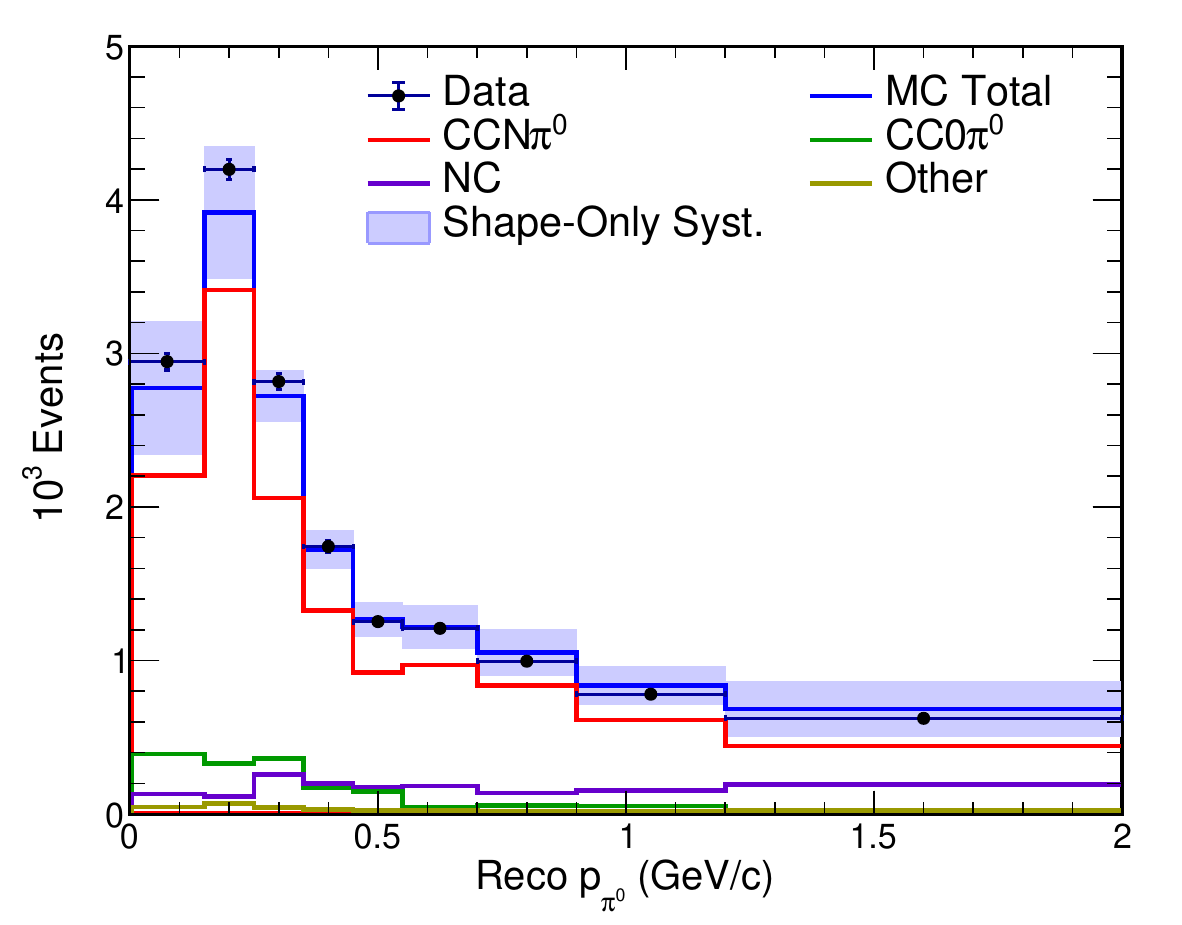}
  \includegraphics[width=0.45\textwidth]{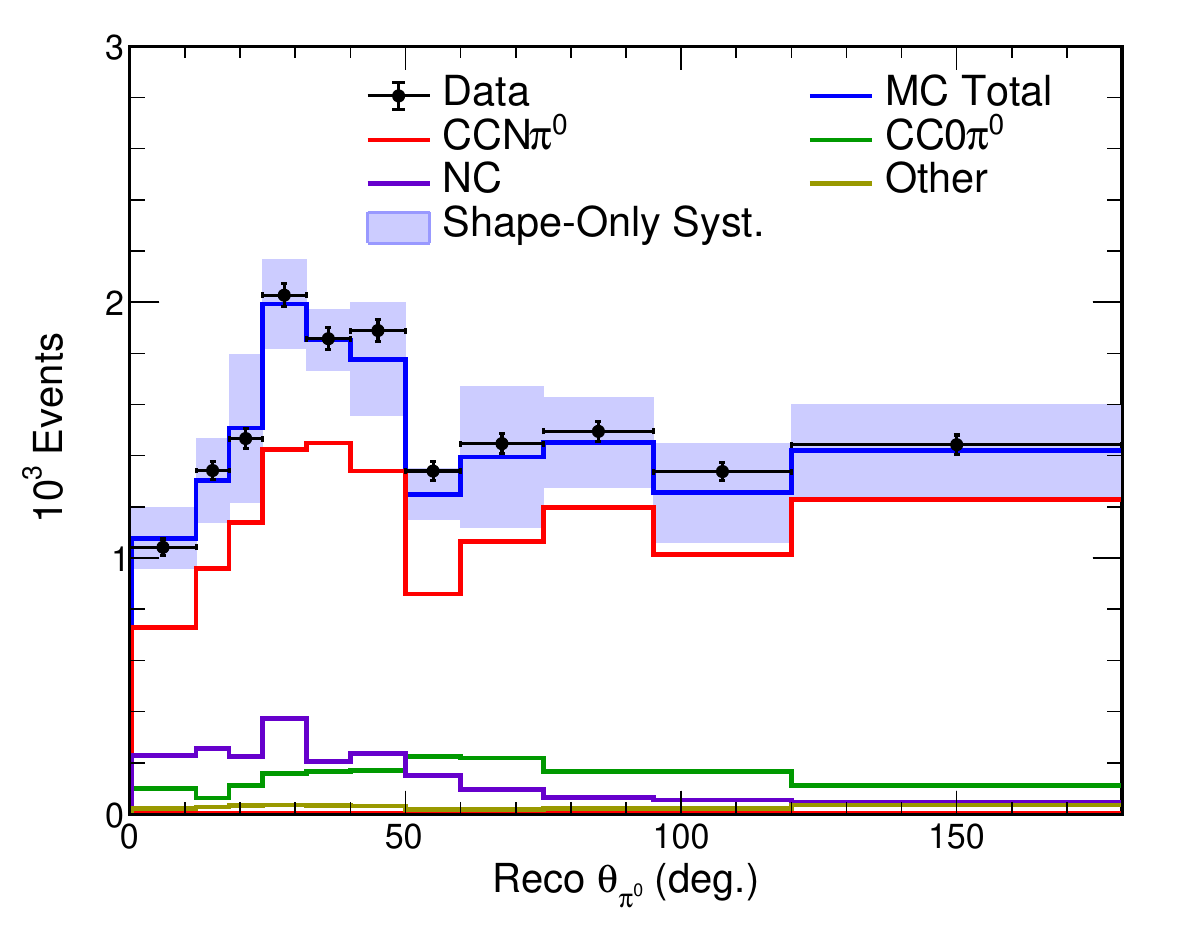}
  \caption{Reconstructed $\pi^0$  kinematic distributions with template fit weights applied to CCN$\pi^0$, CC0$\pi^0$, and NC components. The error band shows only the uncertainties on the shape.}
  \label{fig:pi0_events_postfit}
\end{figure}

\begin{figure}[h!]
  \centering
  \includegraphics[width=0.45\textwidth]{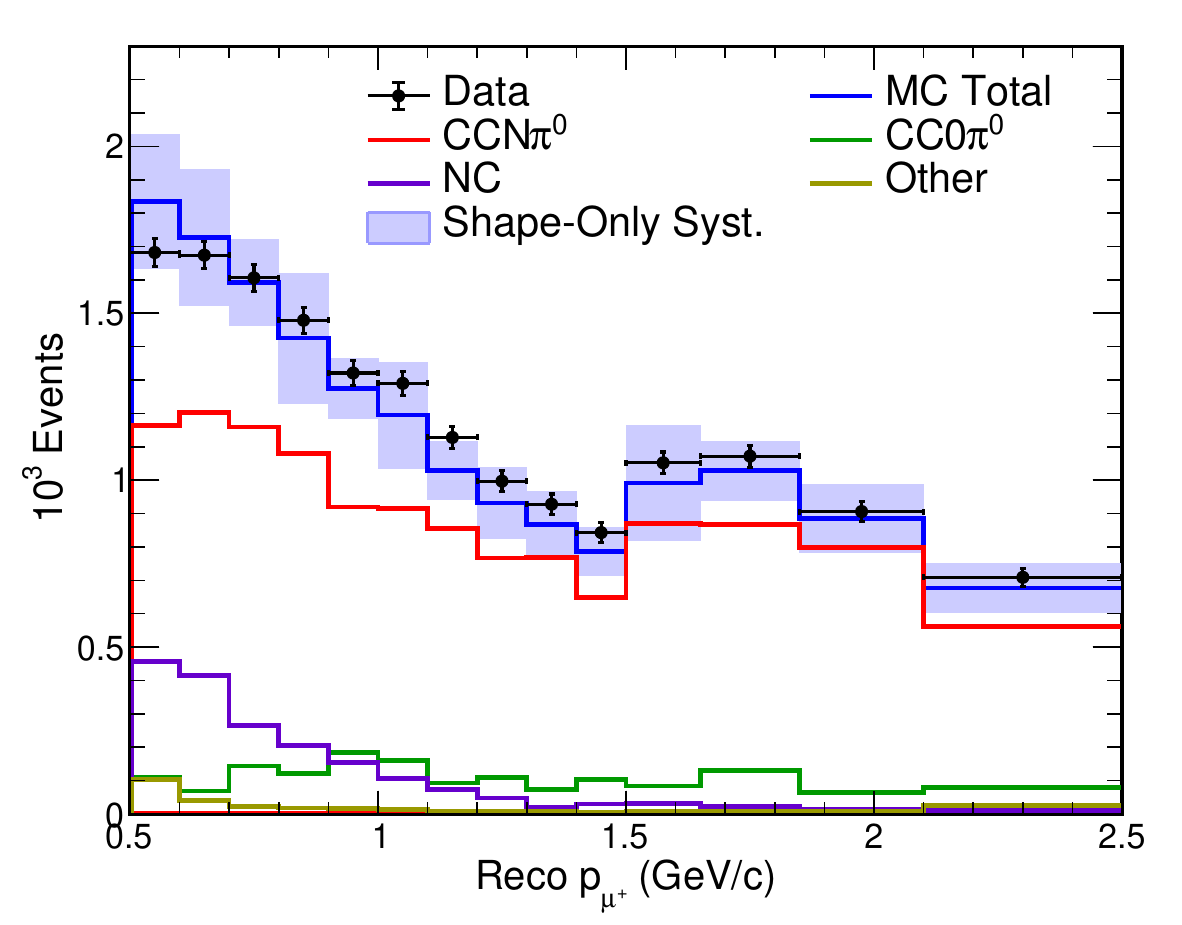}
  \includegraphics[width=0.45\textwidth]{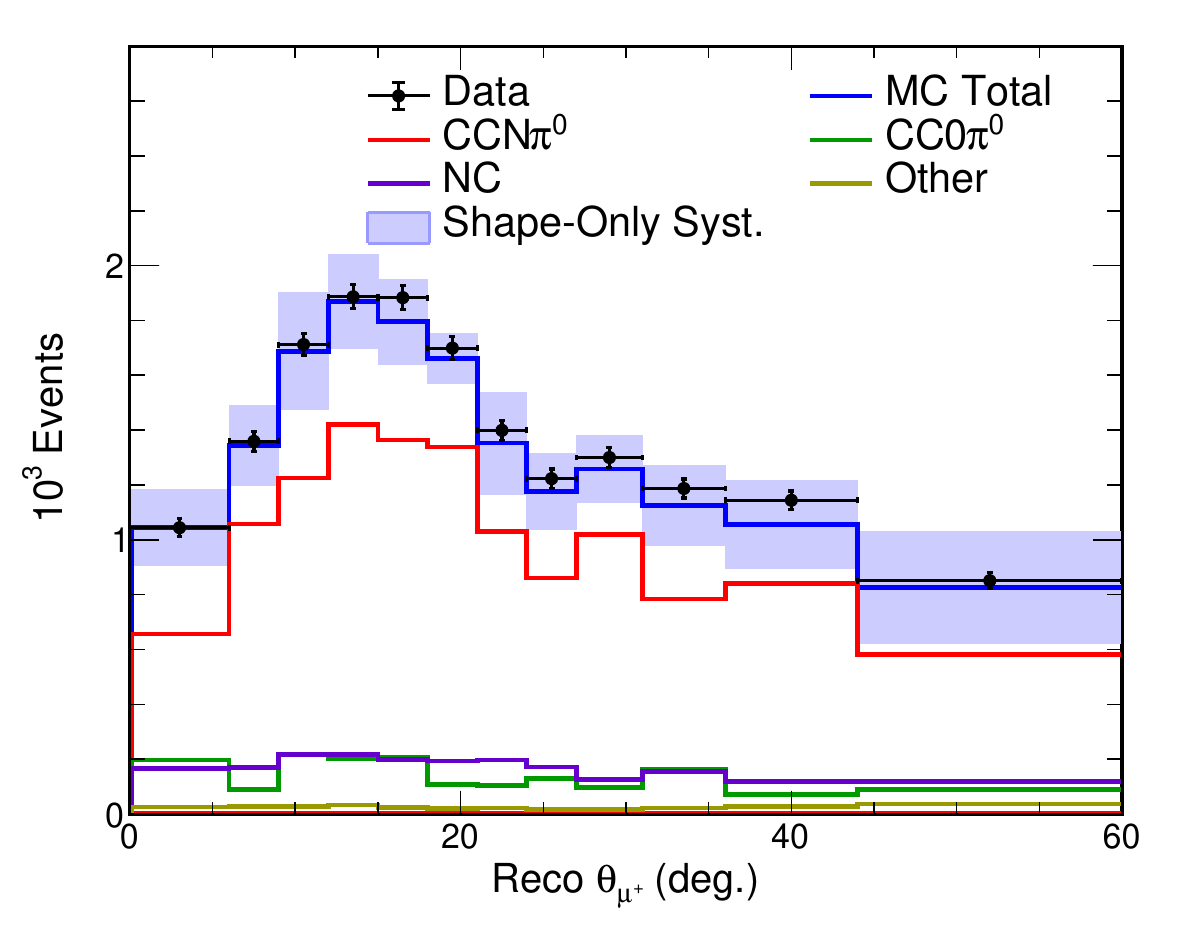}
  \caption{Reconstructed muon kinematic distributions with template fit weights applied to CCN$\pi^0$, CC0$\pi^0$, and NC components. The error band shows only the uncertainties on the shape.}
  \label{fig:muon_events_postfit}
\end{figure}

The sample composition before the fit (71.6\%~CCN$\pi^0$, 20.0\%~CC0$\pi^0$, 6.8\%~NC, 1.5\%~Other) is adjusted to 77.9\%~CCN$\pi^0$, 10.6\%~CC0$\pi^0$, 9.8\%~NC, 1.7\%~Other after the fit. The main effect of the fit is to reduce the size of the CC0$\pi^0$ background by about 55.5\% and to increase the NC background by 30.2\% compared with their respective nominal values, which represent $1.8\sigma$ and $0.8\sigma$ systematic shifts for CC0$\pi^0$ and NC backgrounds. The CCN$\pi^0$ sample is reduced by about 1.2\% relative to its nominal value. Figure~\ref{fig:pi0_events_postfit} shows the reconstructed $\pi^0$  kinematic distributions compared with the simulation. Template fit normalization parameters have been applied to CCN$\pi^0$, CC0$\pi^0$, and NC components. The same comparisons are shown for the reconstructed muon momentum and angle distributions in Fig.~\ref{fig:muon_events_postfit}. After the fit, the $\chi^2/\mathrm{NDF}$ averaged over the observables of interest is found to be 1.17, compared with the pre-fit value of 3.57.


\section{Cross Section extraction}
\label{sec:crosssec}

The single differential cross section in bin $i$ for kinematic variable $x$ can be expressed as
\begin{equation}
  \left(\frac{d\sigma}{dx}\right)_i
  =\frac{\sum_jU_{ij}^{-1}\left(N^{\mathrm{data}}_j(x)-N^{\mathrm{bkg}}_j(x)\right)}
  {N_t\,\Phi\,\epsilon_i(x)\,\Delta {(x)_i}}\,,
\end{equation}
where $N_{t}$ is the number of nucleons in the target fiducial region, $\Phi$ is the integrated flux, $\epsilon_i$ is the efficiency and $\Delta(x)_i$ is the bin width. The estimated number of signal events in bin $j$, obtained by subtracting the background estimate $N^{\mathrm{bkg}}_j$ from the selected data $N^{\mathrm{data}}_j$, is unfolded as discussed below. The response matrix $U_{ij}$ predicts the expected number of true events in bin $i$ given the number of reconstructed events measured in bin $j$. The total background in each bin is found by summing the background contributions after applying the template fit parameters $a$, $b$, and $c$ as defined in Eq.~\ref{eq:template}. Recall that the charged-current neutrino-induced and secondary $\pi^0$ production backgrounds are combined with the signal in the template and are scaled by the $a$ fit parameters.

Reconstructed event counts are affected by detector effects, including smearing due to finite resolution,  which can cause bin migration and distort the measured distribution shapes from those of the underlying true distributions. The D’Agostini iterative unfolding algorithm~\cite{DAgostini:1994fjx}, implemented in the RooUnfold framework in ROOT~\cite{RooUnfold,Brenner:2019lmf}, with two unfolding iterations, is applied to correct for these effects in the cross section extraction. We determine the optimal number of iterations using the results from $10^5$ toy experiments by studying the mean and RMS of $\chi^2/\mathrm{NDF}$ between extracted and input cross sections. In each toy experiment, Poisson fluctuations are applied to the measured distribution to simulate statistical variations in the data. Systematic uncertainties and bin resolution effects are incorporated by varying the response matrix accordingly. In addition, unfolding studies were carried out with systematically shifted fake data sets, which were consistent with expectations and confirmed the robustness of the procedure. This treatment accounts for both statistical and response-related effects. The observed bin shifts were less than the size of the statistical uncertainties, leading us to conclude that any resulting bias is small. We include an additional systematic uncertainty to account for potential effects arising from the choice of the number of unfolding iterations. It is obtained by comparing the average effect on the extracted cross section when using the neighboring iteration numbers and an alternative unfolding method, singular value decomposition (SVD) unfolding~\cite{Hocker:1995kb} as implemented in RooUnfold~\cite{RooUnfold}, to the nominal with two iterations. The resulting uncertainty is small, approximately the size of the statistical error in each bin.

\section{Systematic Uncertainties}
\label{sec:systematics}

The high-statistics data collected with the NOvA ND make this measurement primarily limited by systematic uncertainties. These are assessed by re-extracting cross sections from simulation samples where parameters --- such as those modeling neutrino flux, neutrino-nucleus interactions, neutron production, secondary $\pi^0$ production, and detector response --- are systematically varied. The difference between the cross section obtained from each varied sample and that from the nominal simulation is taken as the associated uncertainty, accounting for the effects of systematic variations on background composition, selection efficiency, and event reconstruction.

For uncertainties influenced by multiple correlated parameters, such as neutrino cross-section modeling and flux, many ($\ge100$) ``universes" are generated by randomly varying each parameter following a normal distribution within its 1$\sigma$ uncertainty. Events in each universe are reweighted according to the combined effect of these parameter shifts, and a cross section is extracted. The $\pm1\sigma$ spread of extracted cross section values across all simulations is taken as the associated systematic uncertainty. We refer to this as the multi-universe method below.

\begin{figure}[h!]
  \centering
  \includegraphics[width=0.45\textwidth]{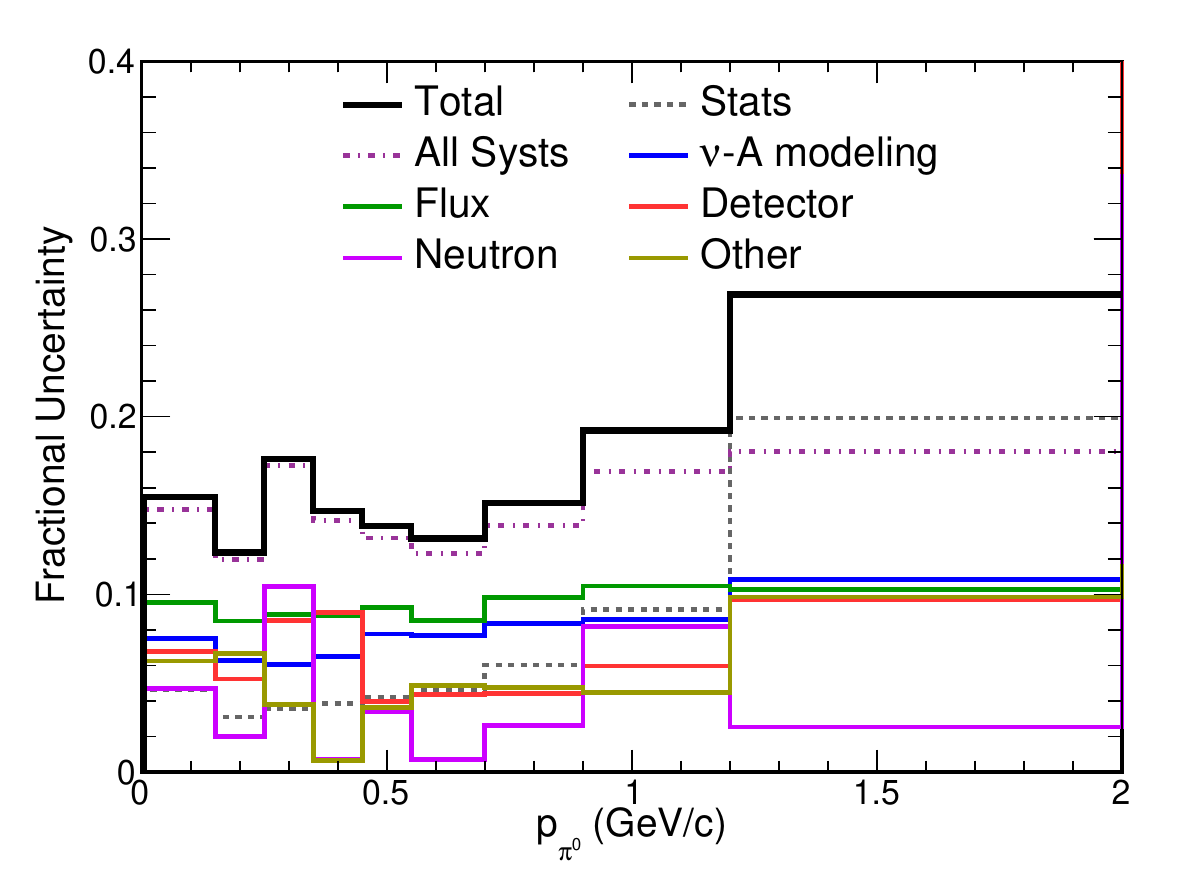}
  \includegraphics[width=0.45\textwidth]{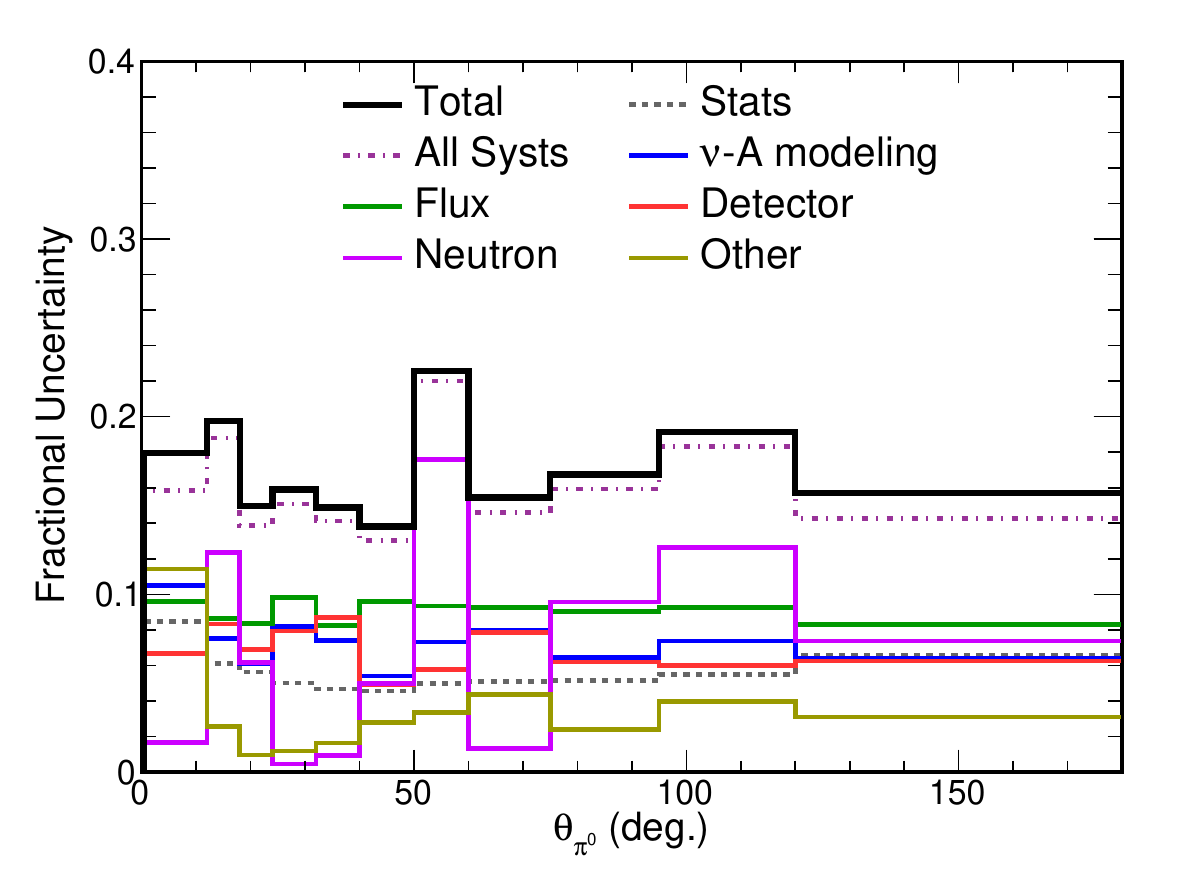}
  \caption{Breakdown of uncertainties on cross section for $\pi^0$ momentum (left) and angle (right). The solid black histogram shows the total uncertainty.} 
  \label{fig:pi0syst}
\end{figure}

\begin{figure}[h!]
  \centering
  \includegraphics[width=0.45\textwidth]{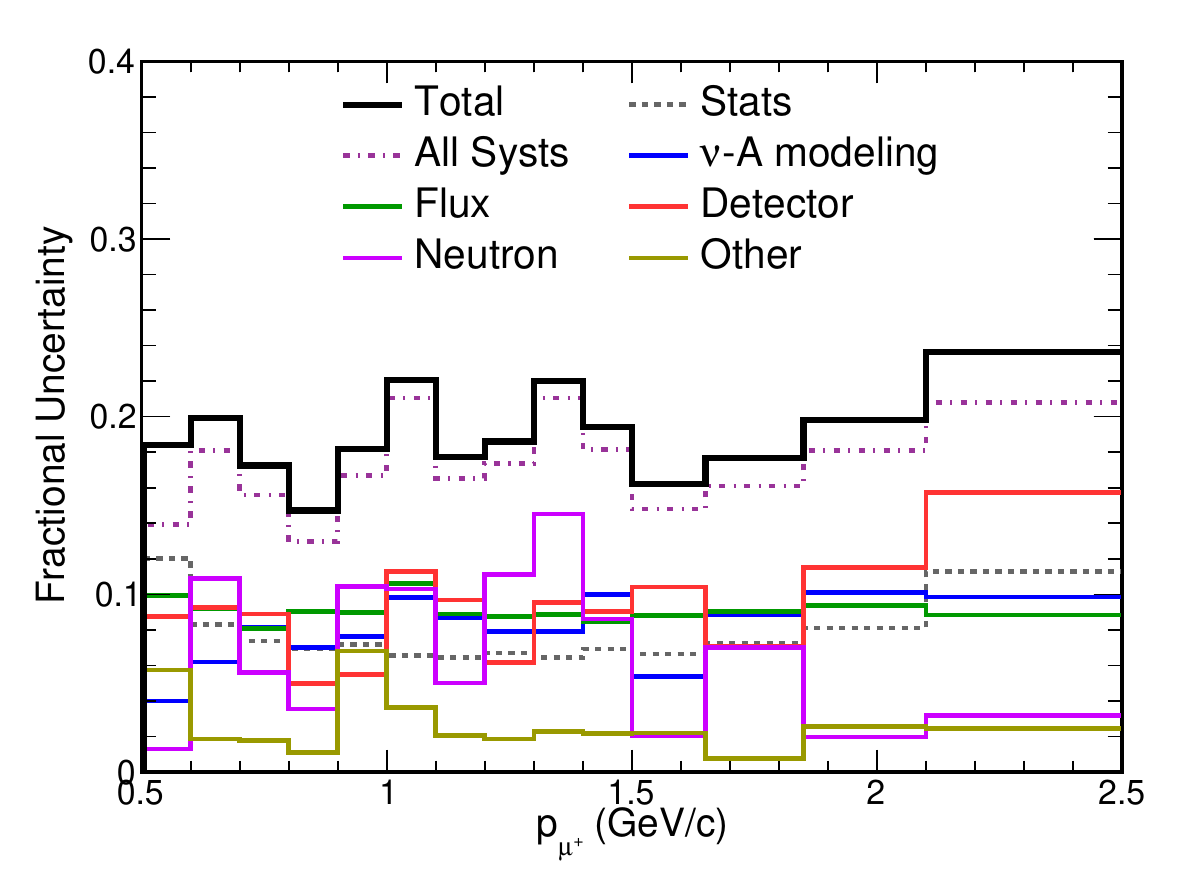}
  \includegraphics[width=0.45\textwidth]{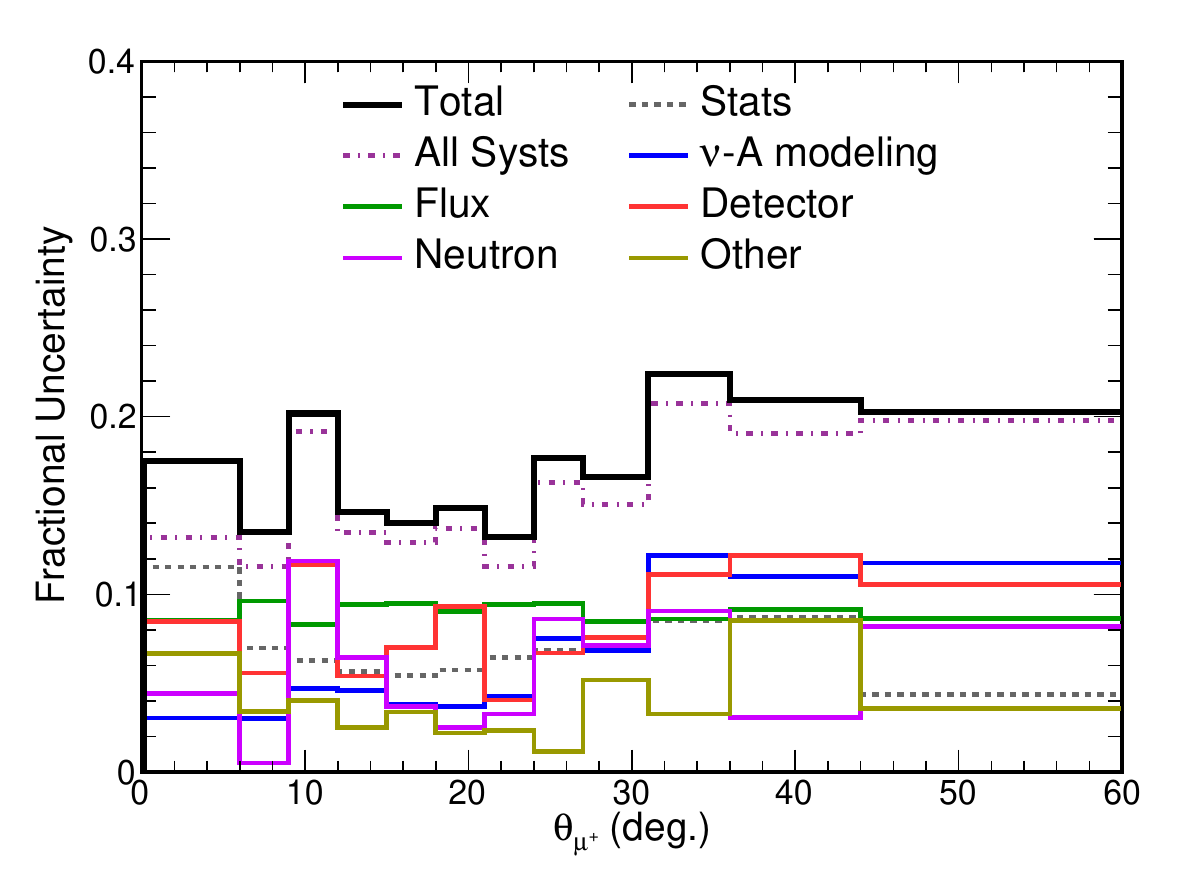}
  \caption{Breakdown of uncertainties on cross section for muon momentum (left) and angle (right). The solid black histogram shows the total uncertainty. }
  \label{fig:musyst}
\end{figure}

Flux uncertainties arise from hadron production and NuMI beamline transport modeling. The hadron production uncertainty is evaluated by simultaneously varying the parameters fit to external production data in the PPFX framework~\cite{MINERvA:2016iqn} within their uncertainties in the multi-universe method. Transport uncertainties, such as those on horn current and skin depth, proton beam spot size and position on target, and horn alignment, are also included, but have a much smaller impact. The total $\bar{\nu}_{\mu}$ flux uncertainty (shown in Figs.~\ref{fig:pi0syst} and \ref{fig:musyst}) around the NOvA flux peak is about 8\%. The flux uncertainty in the wrong-sign ($\nu_\mu$) component of the beam is approximately 13\%. Since neutrino-induced events are treated as background in the signal extraction, this uncertainty only contributes at the 1–2\% level to the final cross section.

The reference GENIE NOvA-tuned cross-section model is used to correct for efficiency and backgrounds, as well as in unfolding. GENIE provides an uncertainty range for each model parameter that can be varied and propagated through the analysis to obtain its resulting systematic effect on the measurement. We evaluate the cross section model uncertainty using the multi-universe method by simultaneously varying the cross section model parameters within their respective $\pm 1\sigma$ ranges. Uncertainties in our NOvA custom-tuned 2p2h and FSI model parameters are also included in the multi-universe variations. The parameters with the largest uncertainties are from the resonance axial (MaCCRES) and vector (MvCCRES) masses, DIS nonresonant pion production, and FSI model uncertainties (formation zone). The average fractional uncertainties for signal events are 5--8\% from CC resonance knobs, 2--8\% from $hN$ FSI knobs, and roughly 4\% from DIS knobs. 

Detector response uncertainties result primarily from energy calibration, light production and transport modeling, and detector aging. The largest effect comes from calibration of the visible hadronic energy scale, which is studied by comparing proton candidate prong energy deposition as a function of length between data and simulation. A $\pm$5\% uncertainty is assigned based on the observed difference between data and simulation. A calibration shape uncertainty is also included to account for nonuniformity in the response as a function of distance from the readout. The muon energy scale uncertainty is 0.7\% for muons stopping in the fully active detector volume, and 0.5\% for muons stopping in the muon catcher. The scintillator light level model includes the scintillation light yield component described by the Birks model \cite{Birk,Chou} as well as Cherenkov light production. Model parameters related to each component are varied to cover the differences between data and simulation for stopping muon and proton data samples. The resulting uncertainties are found to be $\pm$5\% for the scintillation yield and $\pm$6.2\% for Cherenkov light. Detector aging is also treated as an uncertainty and modeled as a linear decrease in the light model parameterization over time, with an observed annual 4.5\% downward drift. An upward drift in the overall calibration scale is applied to compensate for aging effects and is also treated as the corresponding uncertainty.

Neutrons are common in the final states of $\bar{\nu}_{\mu}$ CC interactions. The neutron systematic uncertainties are evaluated based on neutron–carbon inelastic reactions defined by the MENATE$\_$R model~\cite{Kohley:2012awa} within Geant4. MENATE$\_$R is a revised version of the MENATE model \cite{Roeder2008} for neutron propagation, and we use the difference between MENATE$\_$R and the default Geant4 model to evaluate the uncertainty on the neutron response.

Other uncertainties may arise from secondary $\pi^{0}$ production, where a showering hadron or interacting charged pion produces a $\pi^0$ in the detector medium. Charged-pion events undergoing charge exchange are responsible for about 86$\%$ of all secondary $\pi^0$ events. Uncertainties associated with these processes are assessed using a multi-universe approach, where Geant4 particle interaction weights are varied based on hadron scattering cross-section uncertainties. Initial fit results revealed that the standard $1\sigma$ error bands for $\pi^\pm$ charge exchange were insufficient to capture the spread seen in external data. To ensure conservative coverage, the uncertainties for these components were increased to $2\sigma$. In addition, the systematic uncertainty arising from the unfolding procedure, as discussed in Sec.~\ref{sec:crosssec}, is included within the ``Other" uncertainty category.

Figs.~\ref{fig:pi0syst} and \ref{fig:musyst} present the measurement uncertainties, separated by source, as a function of the $\pi^0$ and muon momentum and angle. The integrated fractional uncertainties from the various sources, averaged over the reported kinematic variables, are summarized in Tab.~\ref{tab:uncertainties}. As discussed above, the major contributions are from the flux, neutrino-nucleon interaction modeling (GENIE), and detector response. The results are dominated by systematic uncertainties.


\begin{table}[htbp]
  \begin{center}
    \caption{Relative uncertainties from each source, averaged over the reported kinematic variables. All values are expressed in percent.}
    \begin{tabular}{lr}
      \hline\hline
      \multicolumn{1}{c}{Uncertainty Source} & \multicolumn{1}{c}{Relative Error (\%)} \\
      \hline
      Stats & 5.5 \\
      $\nu$-A Modeling & 6.5 \\
      Flux & 8.9 \\
      Detector & 7.4 \\
      Neutron & 5.3 \\
      Other & 2.7 \\
      Total Syst. & 14.9 \\
      Total & 16.2 \\
      \hline\hline
    \end{tabular}
    \label{tab:uncertainties}
  \end{center}
\end{table}

\section{Results}

In this section, we present the differential cross-section distributions as functions of the momentum and angle of the $\pi^0$ and muon, as well as of $Q^2$ and $W_{\rm EXP}$. Comparisons with some commonly studied generators are provided in Sec.~\ref{sec:gencomparisons}.

\subsection{$\pi^0$ Kinematics}

\begin{figure}[h!]
  \centering
  \includegraphics[width=0.45\textwidth]{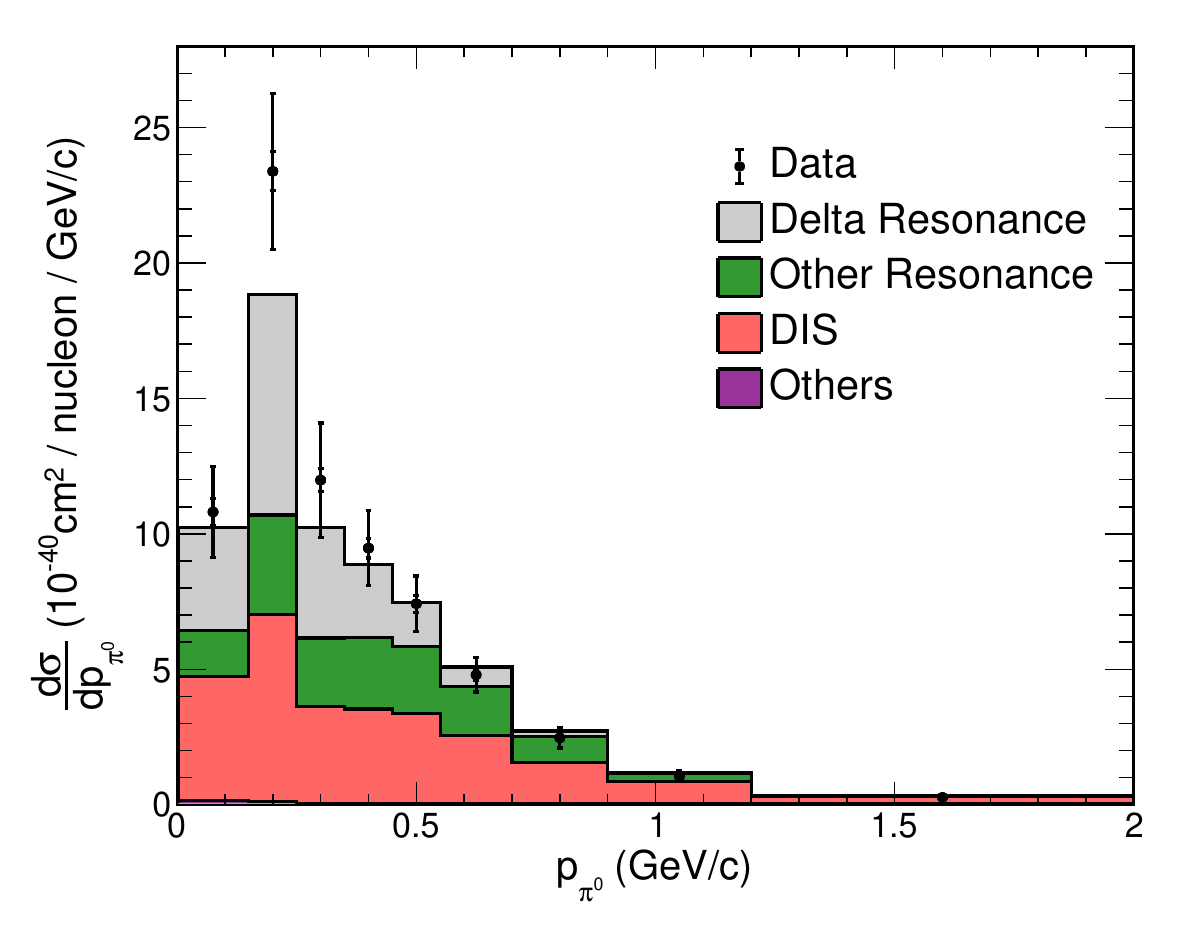}
  \includegraphics[width=0.45\textwidth]{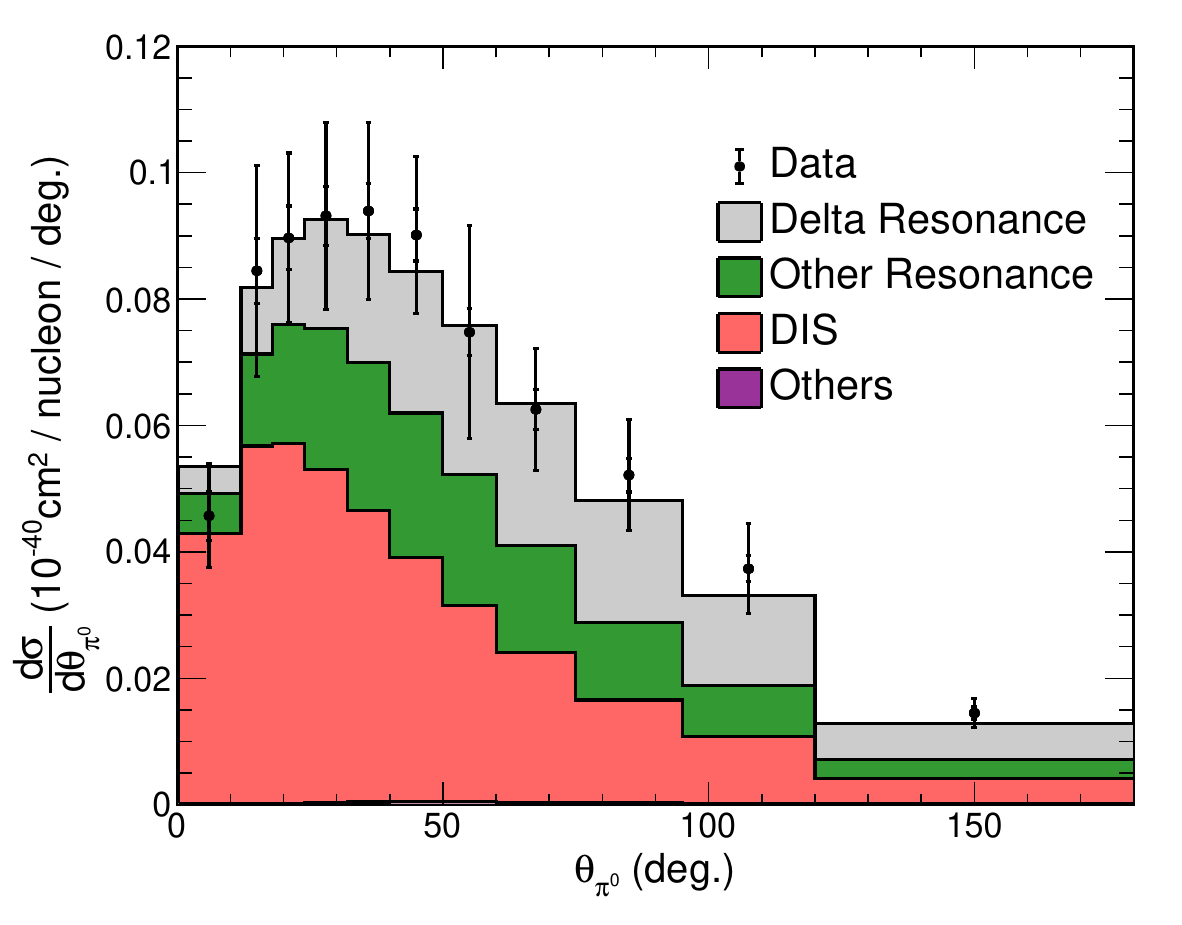}
  \caption{Differential distribution in $\pi^0$ momentum (left) and angle (right), showing the breakdown by primary process for the GENIE v3.0.6 model with NOvA Tune v2.}
  \label{fig:pi0xsecprocess}
\end{figure}

The measured cross sections as functions of $\pi^0$ momentum and angle are shown in Fig.~\ref{fig:pi0xsecprocess}, along with the breakdown by primary interaction process predicted by the GENIE v3.0.6 model with the NOvA Tune v2 configuration. Using GENIE labeling, approximately 29\% of the sample arises from the delta resonance, 24\% from higher order resonance production, and 47\% from DIS. As described in Sec.~\ref{sec:simulation}, the DIS category extends to low $Q^2$ and includes a non-resonant contribution in the $\Delta$ resonance region.

The cross section peaks in both the momentum and angle of the pion, at 0.2~GeV/c and 27$^{\circ}$, respectively. The shapes are similar to those reported in Ref.~\cite{MINERvA:2015slz}, which is at higher average energy (3.6~GeV) and on the same target, but the distributions peak at slightly lower values of momentum and angle. While no single bin in the $\pi^0$ momentum distribution is significantly discrepant, NO$\nu$A-tuned GENIE underestimates the data in the region below 0.4~GeV, where delta resonance production is relatively more important. This is most pronounced in the 0.2~GeV/c peak bin where the model underestimates the data by about 20\%. The angular distribution on the right in Fig.~\ref{fig:pi0xsecprocess} shows good shape agreement with simulation. 

\begin{figure}[h!]
  \centering
  \includegraphics[width=0.45\textwidth]{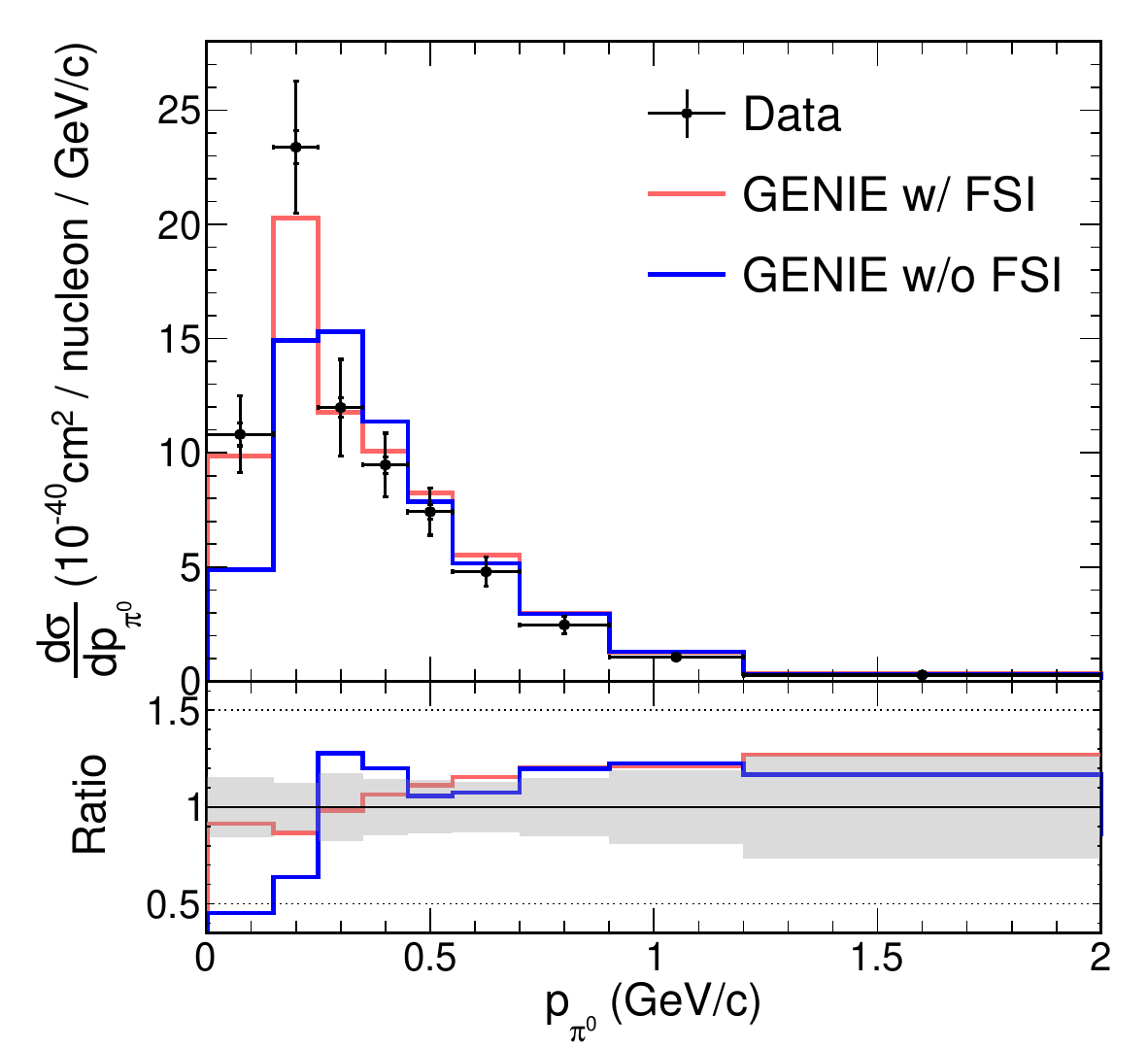}
  \includegraphics[width=0.45\textwidth]{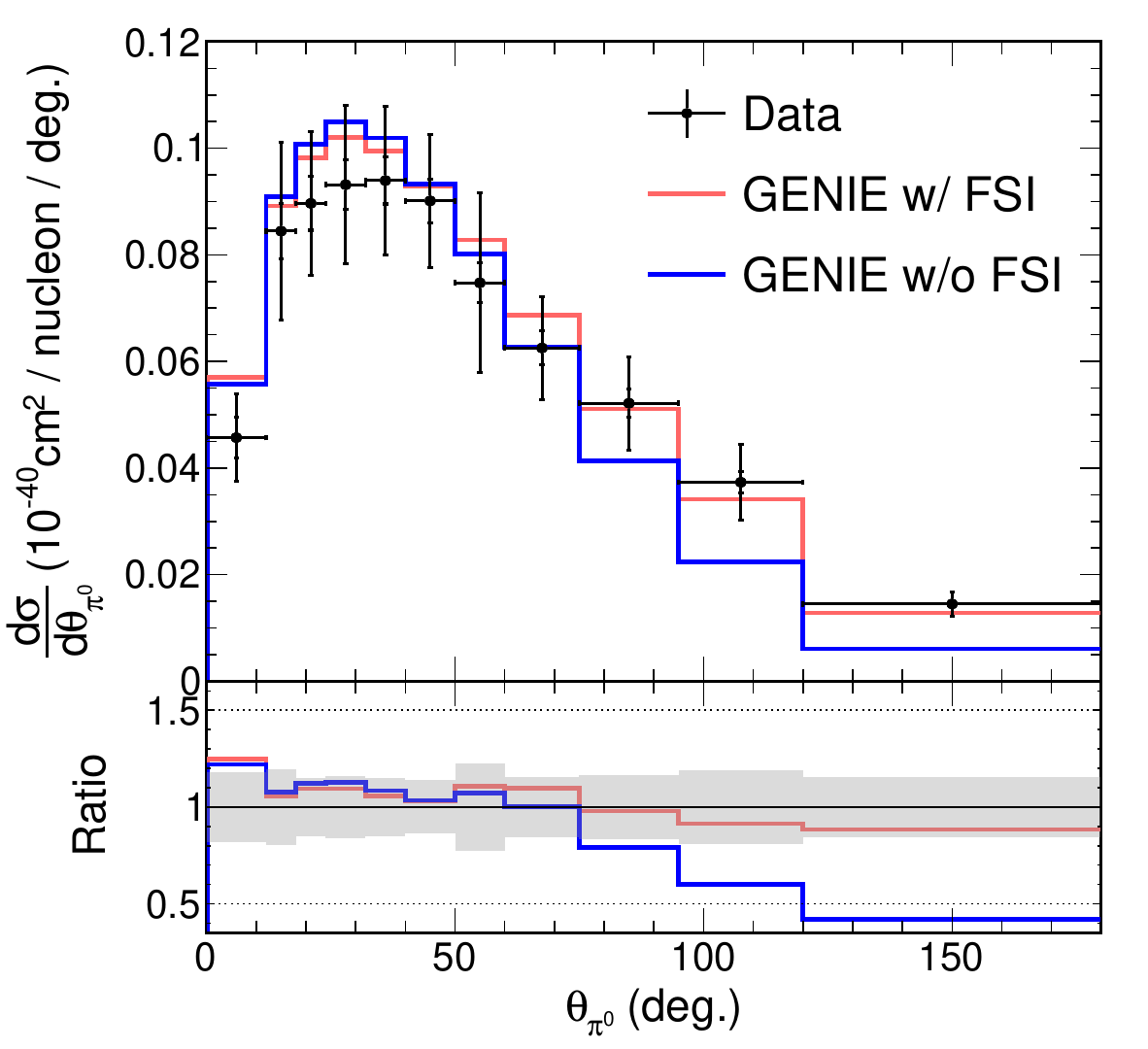}
  \caption{Differential distribution in $\pi^0$ momentum (left) and angle (right) showing GENIE v3.0.6 (untuned) model with FSI on/off. The lower plot panels show the ratio of each model to the data. The grey band is the measurement uncertainty.}
  \label{fig:pi0xsecfsi}
\end{figure}

Figure~\ref{fig:pi0xsecfsi} compares the measured $\pi^0$ kinematic variable distributions for the GENIE 3.0.6 (untuned) model with and without FSI effects included. Intranuclear scattering reduces the average $\pi^0$ momentum and broadens the $\pi^0$ angular distribution. The modest increase in cross section seen results from tradeoffs between charged-pion charge-exchange, which increases the number of final state $\pi^0$'s, and $\pi^0$ absorption. Including FSI dramatically improves agreement with the $\pi^0$ momentum distribution, most noticeably in the low pion momentum region. The $\chi^2$/NDF calculated over the $\pi^0$ kinematic bins between our measurement and the GENIE 3.0.6 model increases from 0.8 to 2.9 without FSI effects included.

\subsection{Muon Kinematics}

\begin{figure}[h!]
  \centering
  \includegraphics[width=0.45\textwidth]{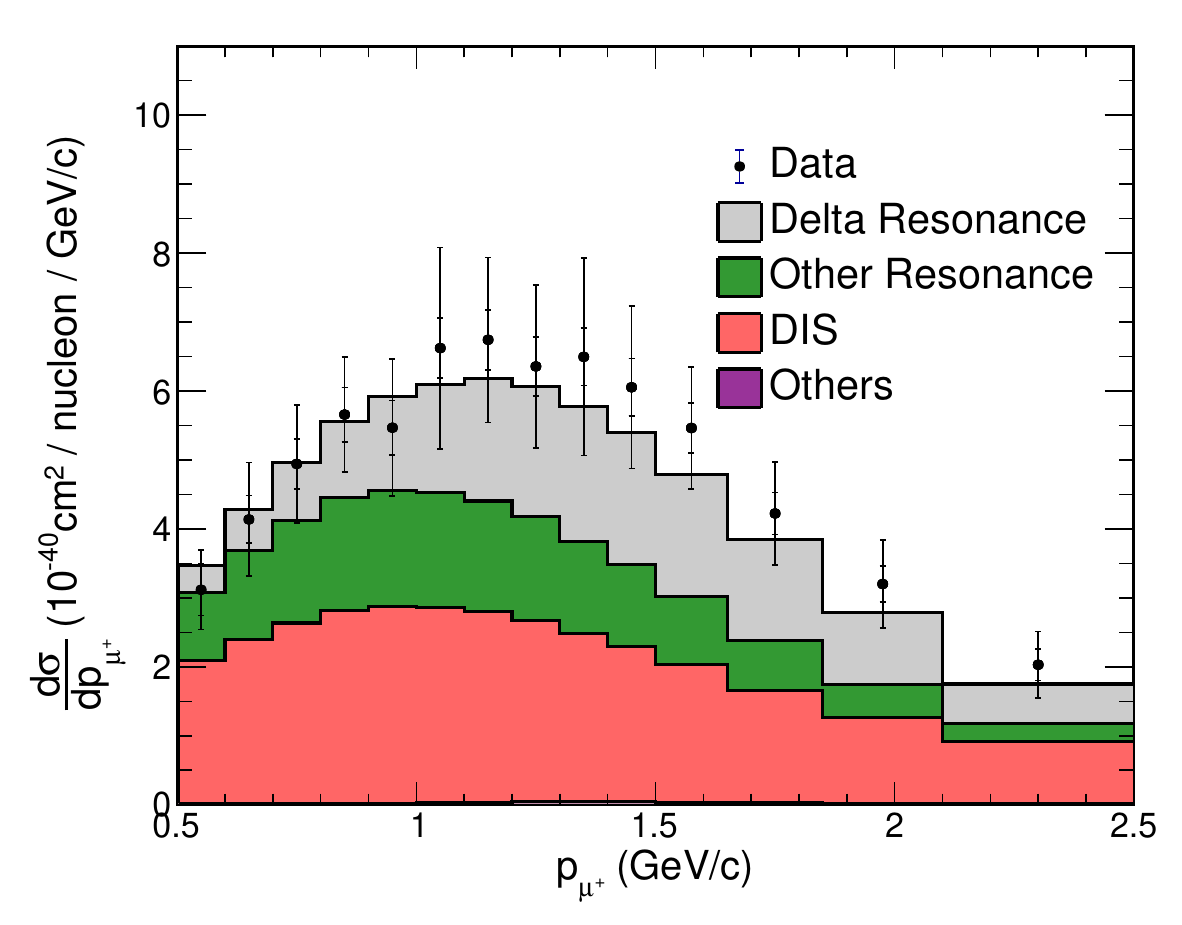}
  \includegraphics[width=0.45\textwidth]{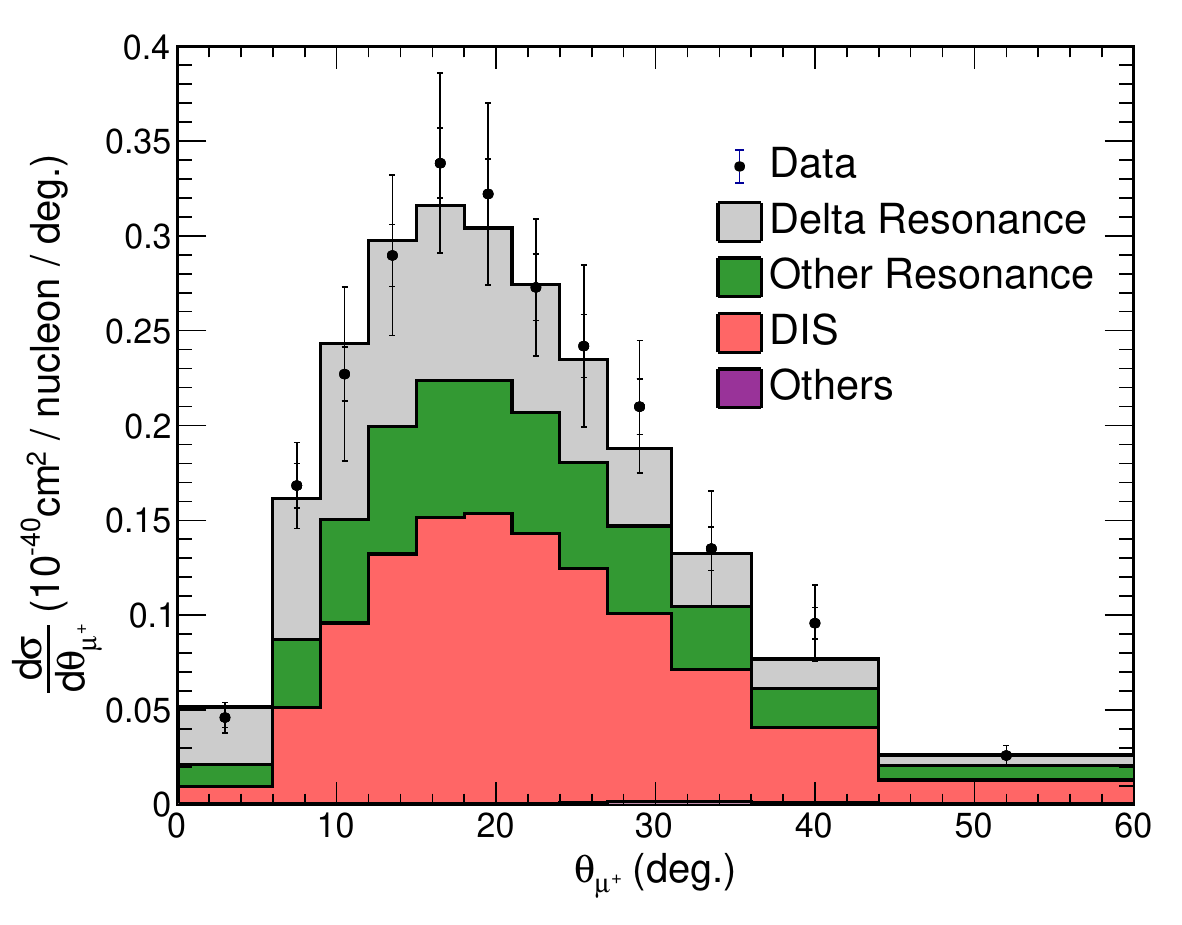}
  \caption{Differential distribution in muon momentum (left) and angle (right), showing the breakdown by primary process for the GENIE v3.0.6 model with NOvA Tune v2.}
  \label{fig:muoninteractions}
\end{figure}

Fig.~\ref{fig:muoninteractions} shows the measured distributions of muon momentum and angle compared with the NOvA-tuned GENIE prediction broken down by primary process. In muon momentum, the data below $\sim$1~GeV/c are in good agreement with the model curve, while above that the data lie systematically above the curve, indicating a preference for a slightly harder muon momentum distribution than predicted. Consistent with the effects seen in the $\pi^0$ momentum distribution, the biggest discrepancies are in regions where the contributions from the delta resonance are greatest. The angular distribution of the muon, which is more uniformly populated by all primary processes, is in good agreement with the shape of the model.

\subsection{$Q^2$ and $W_{\rm EXP}$}

\begin{figure}[h!]
  \centering
  \includegraphics[width=0.45\textwidth]{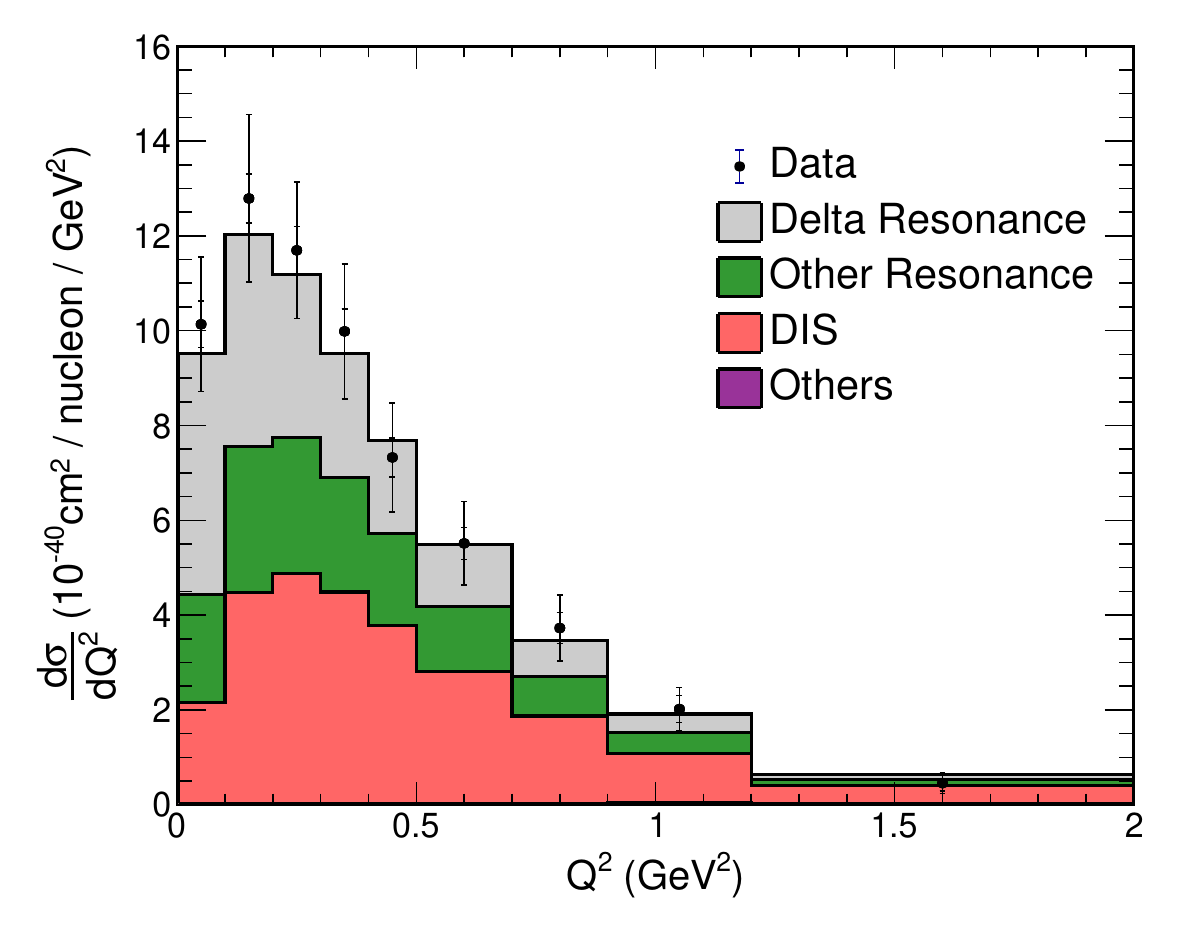}
  \includegraphics[width=0.45\textwidth]{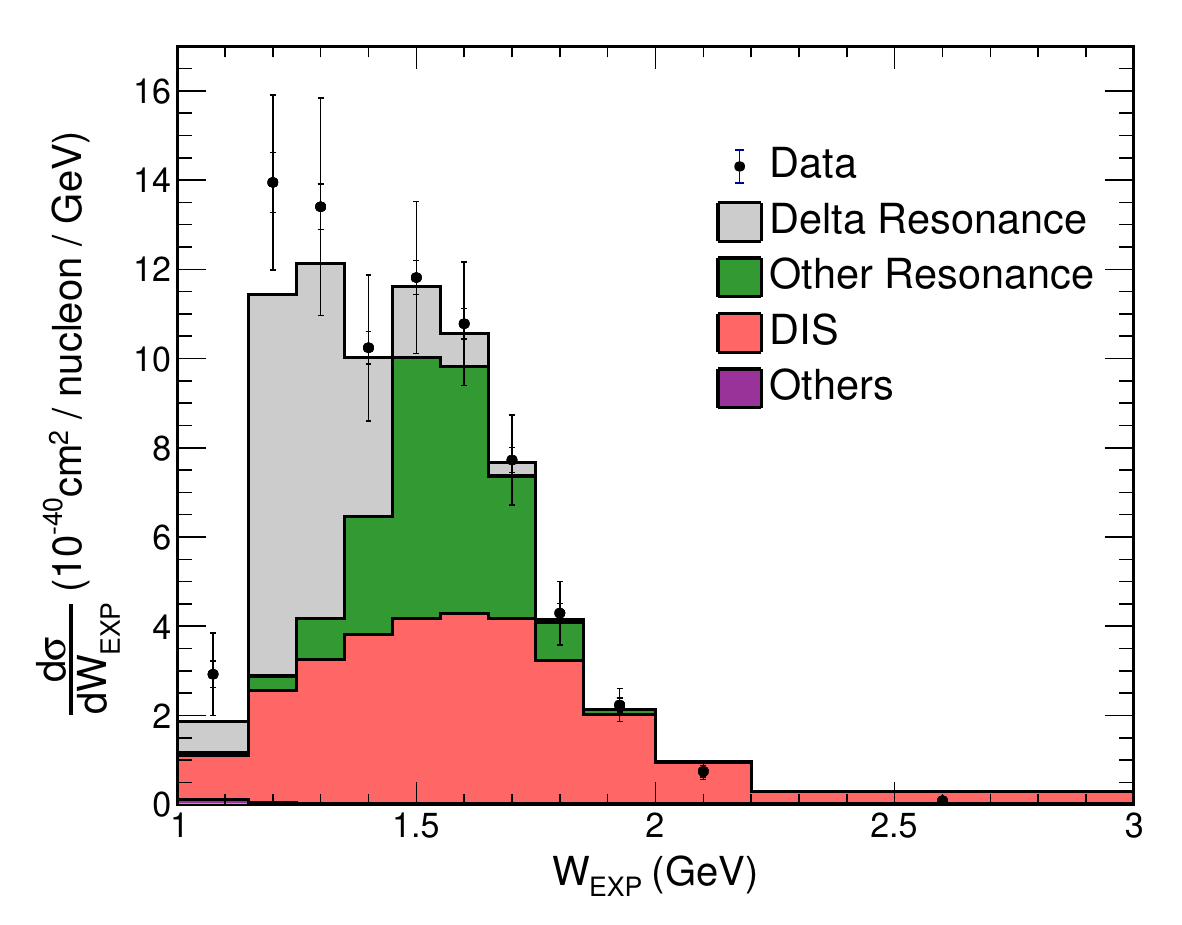}
  \caption{Differential distribution in $Q^2$ (left) and $W_{\rm EXP}$ (right), showing the breakdown by primary process for the GENIE v3.0.6 model with NOvA Tune v2.}
  \label{fig:q2andwprocess}
\end{figure}

Fig.~\ref{fig:q2andwprocess} shows the measured distributions in $Q^2$ and $W_{\rm EXP}$, defined in Eqs.~\ref{eq:q2} and \ref{eq:w}, respectively. The $Q^2$ shape is in good agreement with NOvA-tuned GENIE over the full measurement range. Ref.~\cite{MINERvA:2016sfc} reported a mild suppression  at the lowest $Q^2$ ($<$ 0.2~GeV$^2$) compared with the GENIE v2.6.2 model prediction. Although our measurement is similar in other features, we do not see evidence of unmodeled low $Q^2$ suppression with our version of GENIE.

The $W_{\rm EXP}$ distribution shows two distinct regions. In the region below $\sim$1.4~GeV, $\pi^0$ production from the $\Delta(1232)$ resonance decay ($\Delta^0 (1232) \rightarrow n +\pi^0$) dominates, while the non-resonant contribution is small ($<20$\%). Above $\sim$1.4~GeV non-resonant processes and higher order resonances become important. In the delta-dominated region, the data are systematically above the NO$\nu$A-tuned GENIE curve, while at higher $W_{\rm EXP}$ they align well with the model. The shape of the measured cross section for $W_{\rm EXP}$ corroborates the above observations from the kinematic distributions and constitutes our strongest evidence for underestimation of $\Delta^0$-resonance production.

\subsection{Comparison with Generator Models}
\label{sec:gencomparisons}

\begin{figure}[!htbp]
  \centering
  \includegraphics[width=0.45\textwidth]{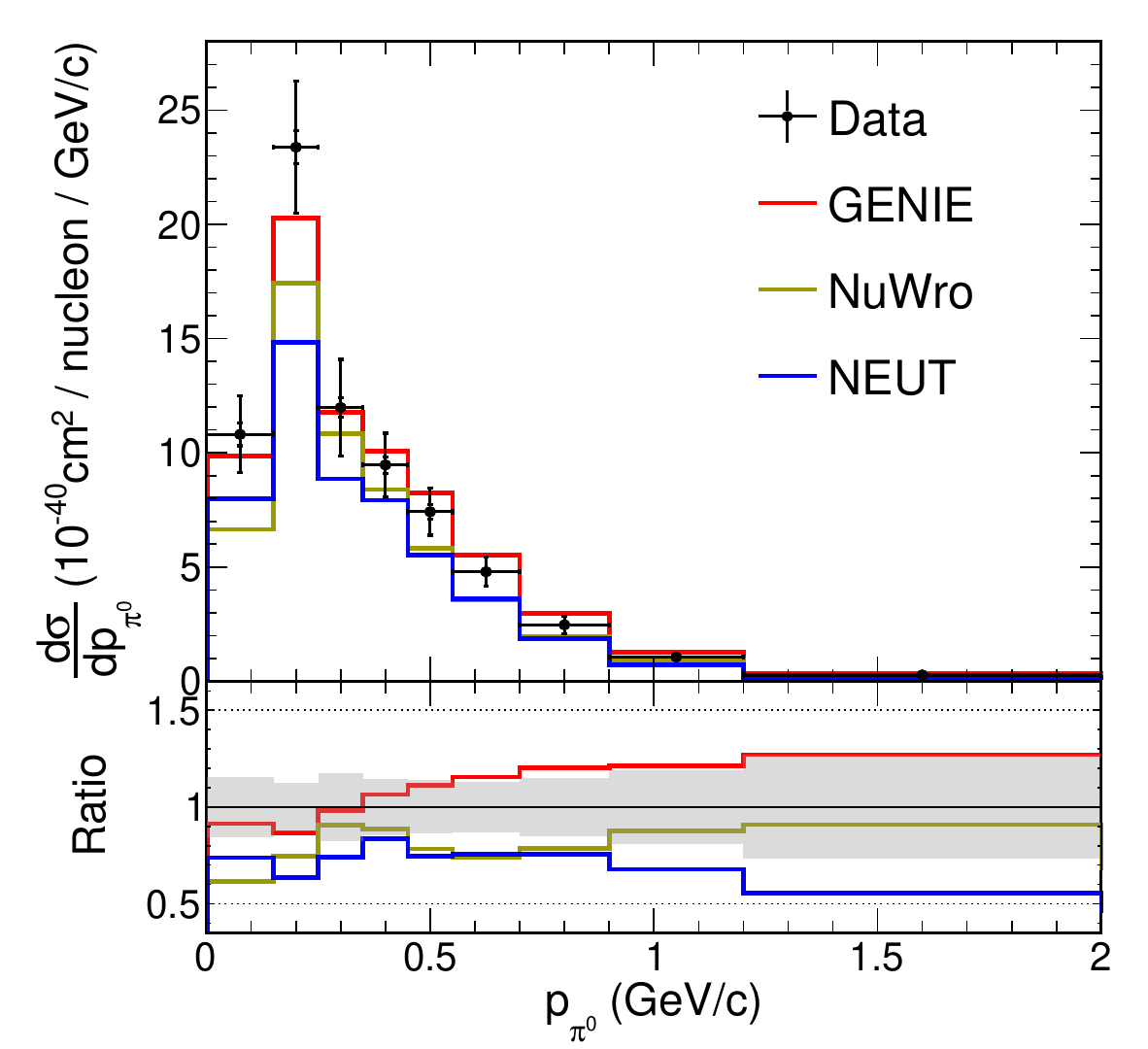}
  \includegraphics[width=0.45\textwidth]{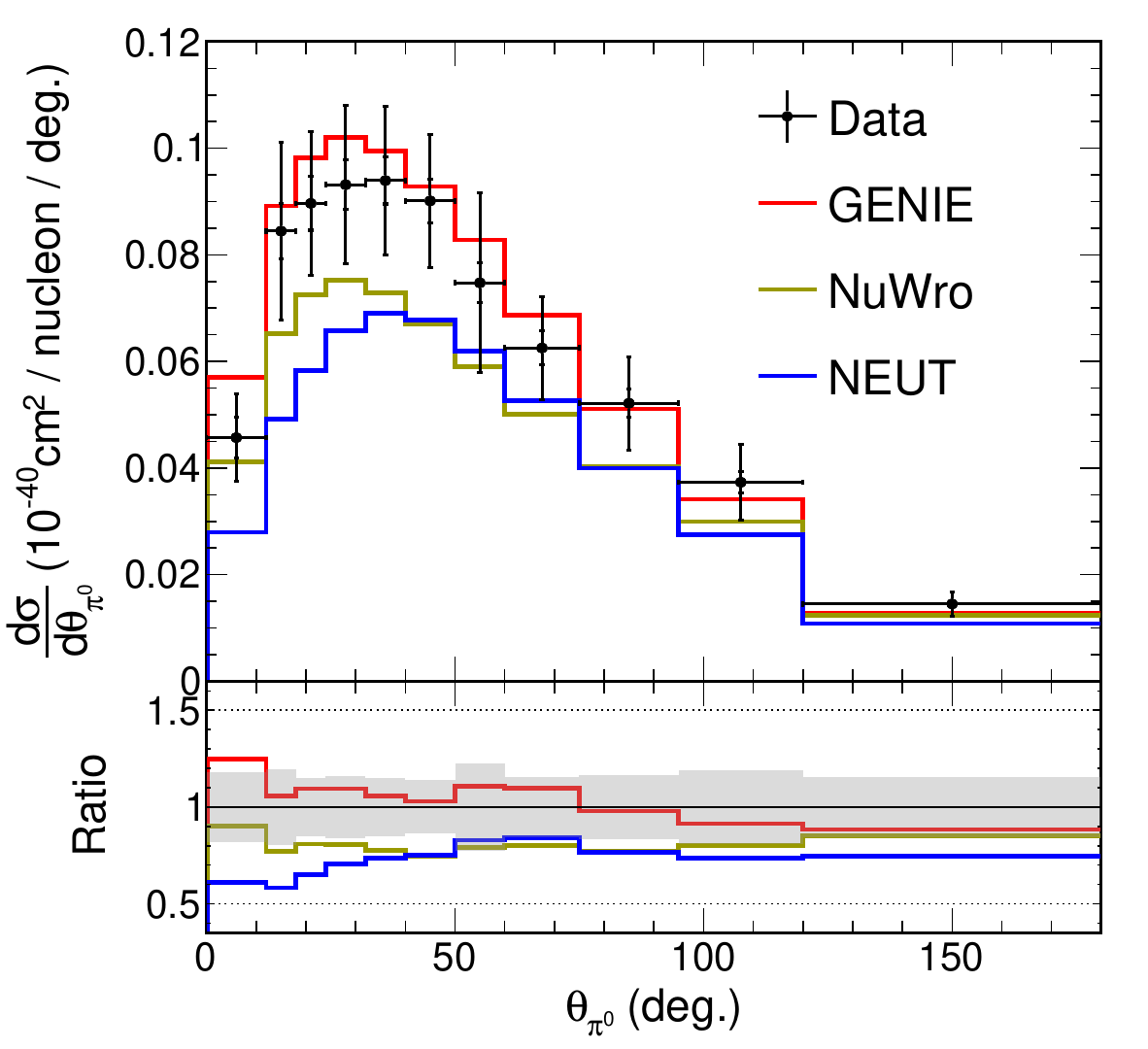}
  \caption{Comparison of various neutrino interaction generators in $\pi^0$ momentum (left) and angle (right). The lower panels show the ratio of each model to the data. The grey band is the measurement uncertainty.}
  \label{fig:pi0xsec}
\end{figure}

\begin{figure}[!htbp]
  \centering
  \includegraphics[width=0.45\textwidth]{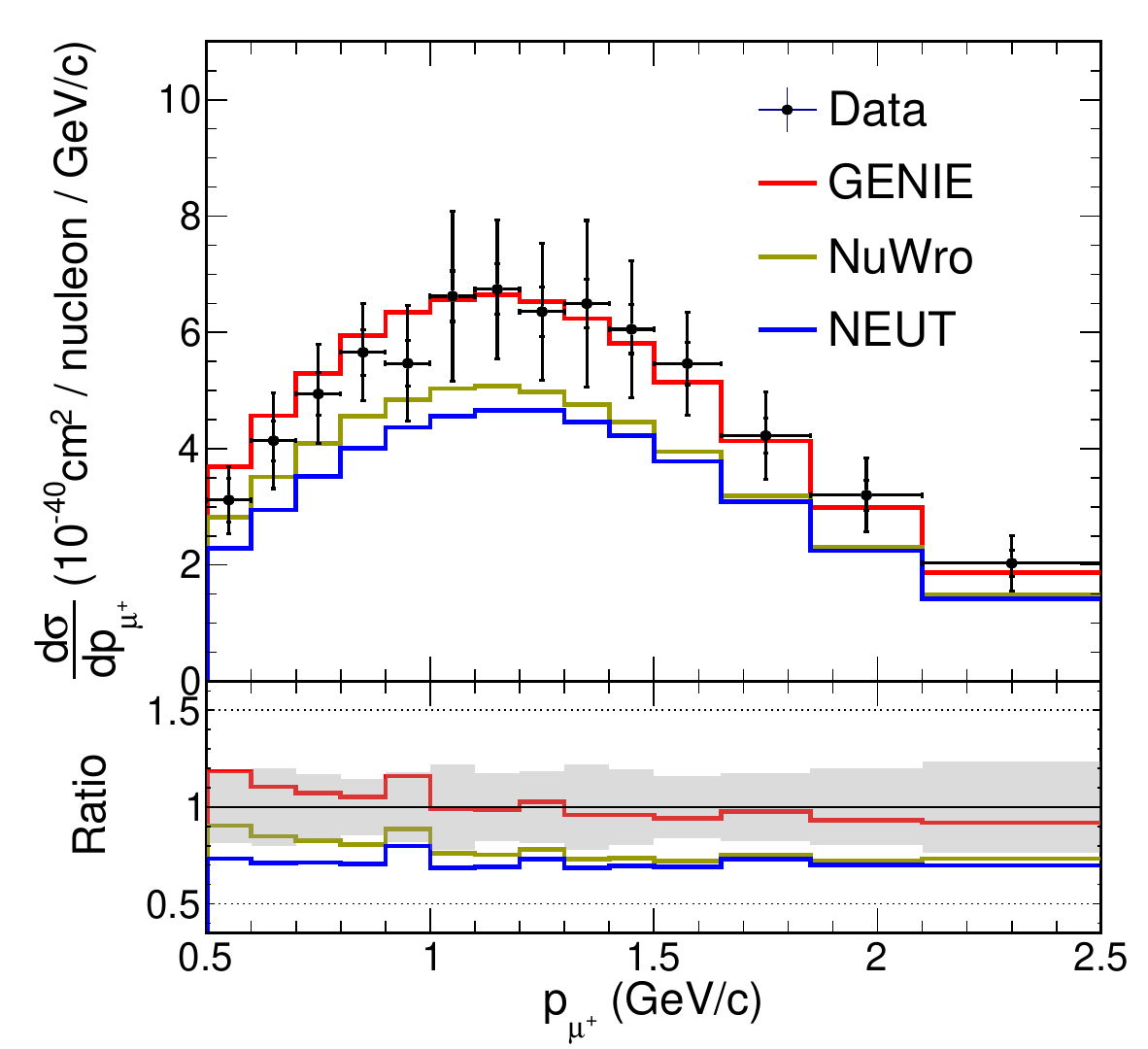}
  \includegraphics[width=0.45\textwidth]{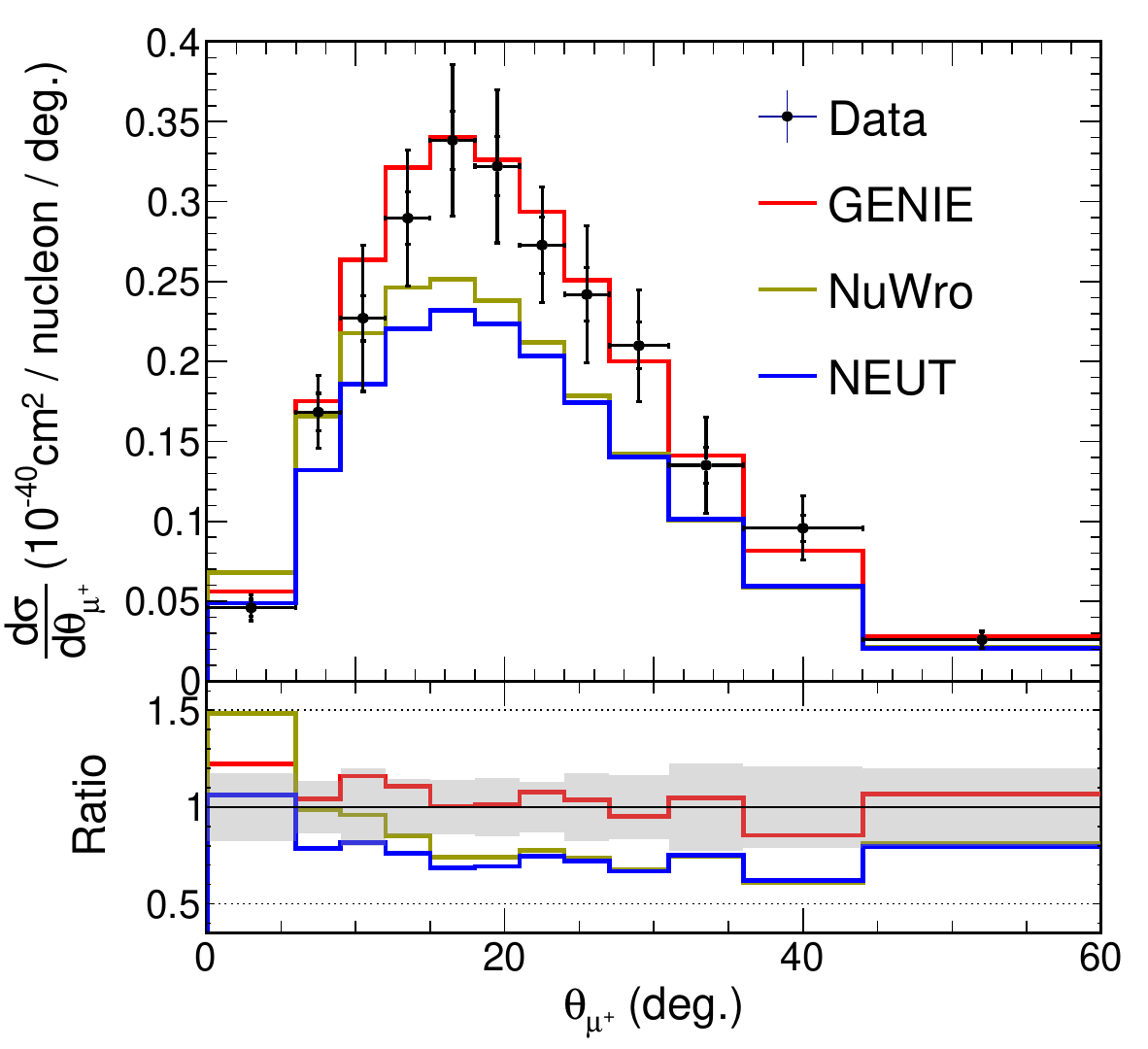}
  \caption{ Comparison of various neutrino interaction generators in muon momentum (left) and angle (right). The lower panels show the ratio of each model to the data. The grey band is the measurement uncertainty.}
  \label{fig:muonxsec}
\end{figure}

\begin{figure}[!htbp]
  \centering
  \includegraphics[width=0.45\textwidth]{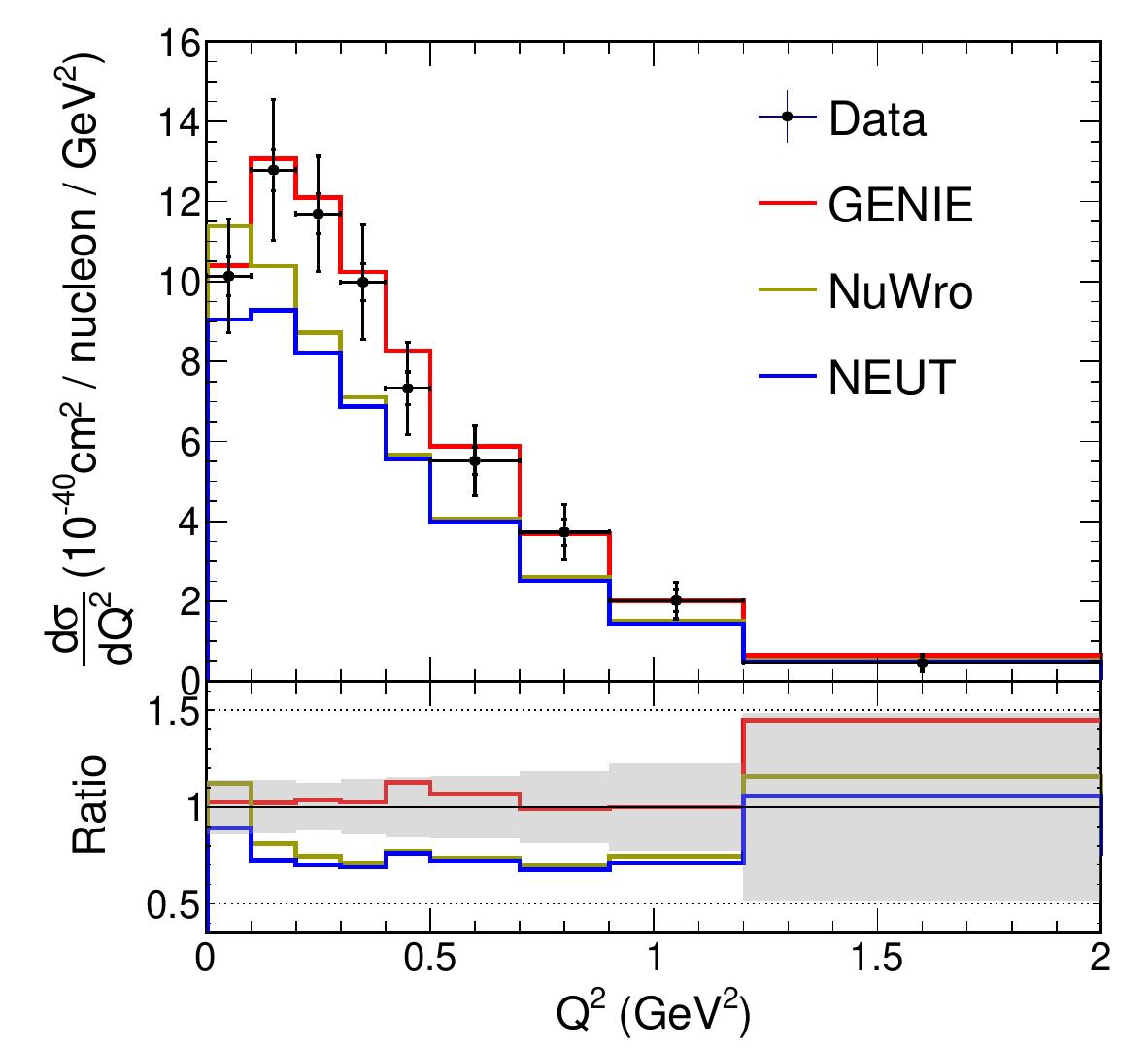}
  \includegraphics[width=0.45\textwidth]{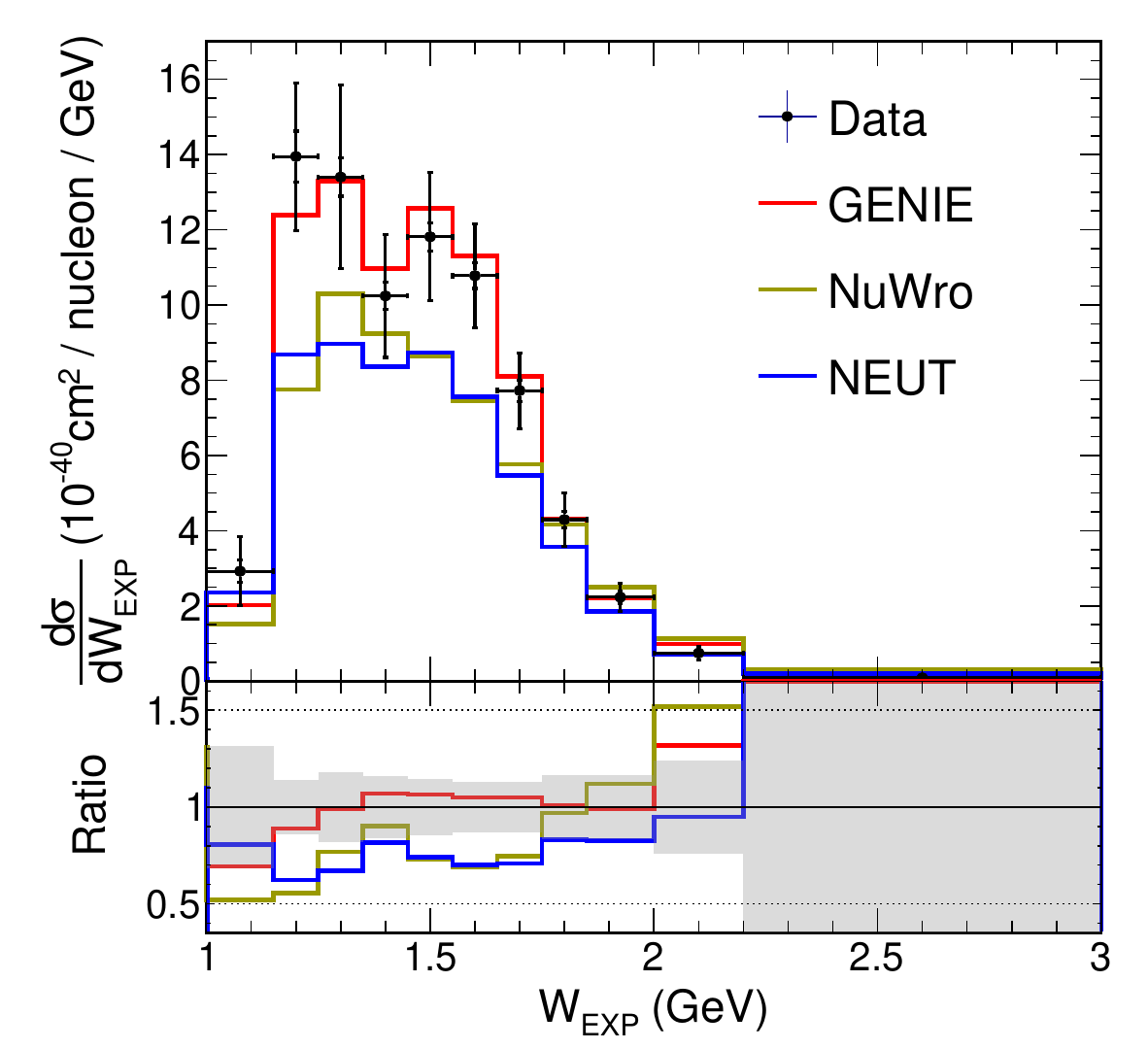}
  \caption{Comparison of various neutrino interaction generators in $Q^2$ (left) and $W_{\rm EXP}$ (right). The lower plot panels show the ratio of each model to the data. The grey band is the measurement uncertainty.}
  \label{fig:q2andw}
\end{figure}

In this section data are compared with GENIE (v3.0.6, using the \texttt{G18\_10j\_00\_000} model configuration), NuWro (21.09.02)~\cite{Golan:2012rfa,Golan:2012wx}, and NEUT (5.7.0)~\cite{Hayato:2021heg}. The $\pi^0$ momentum and angle distributions are shown in  Fig.~\ref{fig:pi0xsec} along with the model to data ratio for each in the frame below the curves. All of the models underestimate the cross section at the $\pi^0$ momentum peak by approximately 10--30\%. GENIE models the shape and normalization of the angular distribution well, while NuWro and NEUT are somewhat below the data for the entire angular range. The discrepancies are largest in the peak, with better agreement above the peak region. Among the models, GENIE best describes the data overall. The other models have similar levels of discrepancy in describing the $\pi^0$ kinematic distributions with each substantially under-predicting the cross section over most of the range of the data. Similar modeling trends are seen in the muon kinematic variables shown in Fig.~\ref{fig:muonxsec}, with GENIE showing the best agreement, while other models are systematically below the data with similar levels of disagreement.

Comparisons of $Q^2$ and $W_{\rm EXP}$ with generators are shown in Fig.~\ref{fig:q2andw}. Here again GENIE shows the best agreement. In Fig.~\ref{fig:q2andwprocess} we noted a deficit in the $\Delta$ dominated region for the NOvA-tuned GENIE model that is less noticeable here. The FSI tuning, which uses external data, is responsible for the additional suppression in the NOvA-tuned version. The adjustments to the MEC component have a negligible effect for this process. GENIE models the region above 1.4~GeV well while NuWro and NEUT remain in disagreement.

The $\chi^2$/NDF values comparing all of the models with data for each measured distribution are given  in Tab.~\ref{tab:xsec_model_compare}. By comparing the computed $\chi^2$/NDF values among the models, we again arrive at the conclusion that GENIE provides the best overall description of the data, with nearly all values $<1$ indicating that uncertainties are slightly overestimated. For NEUT and NuWro the $\chi^2$/NDF values are similar for all distributions with the exception of $W_{\rm EXP}$ where the NuWro value is at its largest. 

The treatment of the resonance region varies widely among the generators studied here. While GENIE, NEUT, and NuWro all use Berger-Seghal as the basis of the model, different values of $M_A$ and differences in form factors  contribute to  differences in modeling the $\Delta(1232)$ resonance. For example, the default value for $M_A^{RES}$ in GENIE is 1.12~GeV, while NEUT and NuWro use somewhat lower values of 0.95~GeV and 0.94~GeV, respectively. The differences in these values can only account for a small fraction of modeling differences in this region ($\sim$5\%). GENIE, NEUT, and NuWro all use the Bodek-Yang model for DIS. In both GENIE and NuWro, this model is extrapolated into the $\Delta$ region to account for non-resonant contributions. NEUT and GENIE include the Berger-Seghal higher order resonances with a prescription for combining them with the DIS component in the transition region where they overlap.

Pion FSI modeling represents another source of differences that can significantly affect the pion kinematic variables. GENIE, NEUT, and NuWro all use semi-classical intranuclear cascade models, as described in Sec.~\ref{sec:simulation}. However, details such as re-interaction probabilities and the resulting particle kinematics are implemented differently in each case~\cite{PinzonGuerra:2018rju, Golan:2012wx}.

\begin{table}[htbp]
  \begin{center}
    \caption{Comparison of $\chi^{2}/\mathrm{NDF}$ between data and models from GENIE (v3.0.6)~\cite{Andreopoulos:2015wxa,genie_manual_v3}, NuWro (21.09.02)~\cite{Golan:2012rfa,Golan:2012wx}, and NEUT (5.7.0)~\cite{Hayato:2021heg}. The $\chi^{2}/\mathrm{NDF}$ values include contributions from uncertainties in data and models.}
    \resizebox{0.35\textwidth}{!}{
      \begin{tabular}{lccc}
        \hline\hline
        \multicolumn{1}{c}{} & \multicolumn{1}{c}{GENIE} & \multicolumn{1}{c}{NuWro} & \multicolumn{1}{c}{NEUT} \\
        \hline
        $p_{\pi^0}$ & 0.77 & 2.04 & 3.06  \\
        $\theta_{\pi^0}$ & 0.39 & 1.52 & 2.95  \\
        $p_{\mu}$ & 0.20 & 1.42 & 2.39 \\
        $\theta_{\mu}$  & 0.33 & 2.46 & 2.59  \\
        $W_{\rm EXP}$ & 1.01 & 3.26 & 2.39  \\
        $Q^2$ & 0.19 & 1.98 & 2.52 \\
        \hline\hline
      \end{tabular}
      }
      \label{tab:xsec_model_compare}
  \end{center}
\end{table}

\section{Conclusion}

We have presented a high-statistics measurement of antineutrino-induced neutral-pion production on a primarily carbon target at average neutrino energy of 2~GeV. The single differential dependences on muon and pion kinematic variables along with $Q^2$ and $W_{\rm EXP}$ were reported. GENIE (v3.0.6) produces the best description of data. The data favor a larger $\Delta$ resonance contribution, indicated by a mild systematic underestimation of data in $\Delta$ dominated regions of phase space by GENIE and larger discrepancies seen with respect to the other models. Neut and NuWro both underestimate the data over much of the phase space of the measured kinematic variables. The measured dependences on lepton and neutral pion kinematic variables will help constrain neutrino interaction models for both signal and background processes in neutrino oscillation measurements.
The data associated with the measured cross sections and their systematic uncertainties, including the full covariance matrices, are available at~\cite{nova_data_release}.

\acknowledgments

This document was prepared by the NOvA collaboration using the resources of the Fermi National Accelerator Laboratory (Fermilab), a U.S. Department of Energy, Office of Science, HEP User Facility. Fermilab is managed by Fermi Forward Discovery Group, LLC, acting under Contract No. 89243024CSC000002.  This work was supported by the U.S. Department of Energy; the U.S. National Science Foundation; the Department of Science and Technology, India; the European Research Council; the MSMT CR, GA UK, Czech Republic; the RAS, the Ministry of Science and Higher Education, and RFBR, Russia; CNPq and FAPEG, Brazil; UKRI, STFC and the Royal Society, United Kingdom; and the state and University of Minnesota.  We are grateful for the contributions of the staffs of the University of Minnesota at the Ash River Laboratory, and of Fermilab. For the purpose of open access, the author has applied a Creative Commons Attribution (CC BY) license to any Author Accepted Manuscript version arising.

\bibliographystyle{apsrev4-2}
\bibliography{references}

\begin{thebibliography}{46}%
\makeatletter
\providecommand \@ifxundefined [1]{%
 \@ifx{#1\undefined}
}%
\providecommand \@ifnum [1]{%
 \ifnum #1\expandafter \@firstoftwo
 \else \expandafter \@secondoftwo
 \fi
}%
\providecommand \@ifx [1]{%
 \ifx #1\expandafter \@firstoftwo
 \else \expandafter \@secondoftwo
 \fi
}%
\providecommand \natexlab [1]{#1}%
\providecommand \enquote  [1]{``#1''}%
\providecommand \bibnamefont  [1]{#1}%
\providecommand \bibfnamefont [1]{#1}%
\providecommand \citenamefont [1]{#1}%
\providecommand \href@noop [0]{\@secondoftwo}%
\providecommand \href [0]{\begingroup \@sanitize@url \@href}%
\providecommand \@href[1]{\@@startlink{#1}\@@href}%
\providecommand \@@href[1]{\endgroup#1\@@endlink}%
\providecommand \@sanitize@url [0]{\catcode `\\12\catcode `\$12\catcode
  `\&12\catcode `\#12\catcode `\^12\catcode `\_12\catcode `\%12\relax}%
\providecommand \@@startlink[1]{}%
\providecommand \@@endlink[0]{}%
\providecommand \url  [0]{\begingroup\@sanitize@url \@url }%
\providecommand \@url [1]{\endgroup\@href {#1}{\urlprefix }}%
\providecommand \urlprefix  [0]{URL }%
\providecommand \Eprint [0]{\href }%
\providecommand \doibase [0]{https://doi.org/}%
\providecommand \selectlanguage [0]{\@gobble}%
\providecommand \bibinfo  [0]{\@secondoftwo}%
\providecommand \bibfield  [0]{\@secondoftwo}%
\providecommand \translation [1]{[#1]}%
\providecommand \BibitemOpen [0]{}%
\providecommand \bibitemStop [0]{}%
\providecommand \bibitemNoStop [0]{.\EOS\space}%
\providecommand \EOS [0]{\spacefactor3000\relax}%
\providecommand \BibitemShut  [1]{\csname bibitem#1\endcsname}%
\let\auto@bib@innerbib\@empty
\bibitem [{\citenamefont {Aguilar-Arevalo}\ \emph {et~al.}(2011)\citenamefont
  {Aguilar-Arevalo} \emph {et~al.}}]{MiniBooNE:2010cxl}%
  \BibitemOpen
  \bibfield  {author} {\bibinfo {author} {\bibfnamefont {A.~A.}\ \bibnamefont
  {Aguilar-Arevalo}} \emph {et~al.} (\bibinfo {collaboration} {MiniBooNE}),\
  }\href {https://doi.org/10.1103/PhysRevD.83.052009} {\bibfield  {journal}
  {\bibinfo  {journal} {Phys. Rev. D}\ }\textbf {\bibinfo {volume} {83}},\
  \bibinfo {pages} {052009} (\bibinfo {year} {2011})},\ \Eprint
  {https://arxiv.org/abs/1010.3264} {arXiv:1010.3264 [hep-ex]} \BibitemShut
  {NoStop}%
\bibitem [{\citenamefont {Altinok}\ \emph {et~al.}(2017)\citenamefont {Altinok}
  \emph {et~al.}}]{MINERvA:2017okh}%
  \BibitemOpen
  \bibfield  {author} {\bibinfo {author} {\bibfnamefont {O.}~\bibnamefont
  {Altinok}} \emph {et~al.} (\bibinfo {collaboration} {MINERvA}),\ }\href
  {https://doi.org/10.1103/PhysRevD.96.072003} {\bibfield  {journal} {\bibinfo
  {journal} {Phys. Rev. D}\ }\textbf {\bibinfo {volume} {96}},\ \bibinfo
  {pages} {072003} (\bibinfo {year} {2017})},\ \Eprint
  {https://arxiv.org/abs/1708.03723} {arXiv:1708.03723 [hep-ex]} \BibitemShut
  {NoStop}%
\bibitem [{\citenamefont {Acero}\ \emph
  {et~al.}(2023{\natexlab{a}})\citenamefont {Acero} \emph
  {et~al.}}]{NOvA:2023uxq}%
  \BibitemOpen
  \bibfield  {author} {\bibinfo {author} {\bibfnamefont {M.~A.}\ \bibnamefont
  {Acero}} \emph {et~al.} (\bibinfo {collaboration} {NOvA}),\ }\href
  {https://doi.org/10.1103/PhysRevD.107.112008} {\bibfield  {journal} {\bibinfo
   {journal} {Phys. Rev. D}\ }\textbf {\bibinfo {volume} {107}},\ \bibinfo
  {pages} {112008} (\bibinfo {year} {2023}{\natexlab{a}})},\ \Eprint
  {https://arxiv.org/abs/2306.04028} {arXiv:2306.04028 [hep-ex]} \BibitemShut
  {NoStop}%
\bibitem [{\citenamefont {Le}\ \emph {et~al.}(2015)\citenamefont {Le} \emph
  {et~al.}}]{MINERvA:2015slz}%
  \BibitemOpen
  \bibfield  {author} {\bibinfo {author} {\bibfnamefont {T.}~\bibnamefont {Le}}
  \emph {et~al.} (\bibinfo {collaboration} {MINERvA}),\ }\href
  {https://doi.org/10.1016/j.physletb.2015.07.039} {\bibfield  {journal}
  {\bibinfo  {journal} {Phys. Lett. B}\ }\textbf {\bibinfo {volume} {749}},\
  \bibinfo {pages} {130} (\bibinfo {year} {2015})},\ \Eprint
  {https://arxiv.org/abs/1503.02107} {arXiv:1503.02107 [hep-ex]} \BibitemShut
  {NoStop}%
\bibitem [{\citenamefont {Adamson}\ \emph {et~al.}(2016)\citenamefont {Adamson}
  \emph {et~al.}}]{Adamson:2015dkw}%
  \BibitemOpen
  \bibfield  {author} {\bibinfo {author} {\bibfnamefont {P.}~\bibnamefont
  {Adamson}} \emph {et~al.},\ }\href
  {https://doi.org/10.1016/j.nima.2015.08.063} {\bibfield  {journal} {\bibinfo
  {journal} {Nucl. Instrum. Meth. A}\ }\textbf {\bibinfo {volume} {806}},\
  \bibinfo {pages} {279} (\bibinfo {year} {2016})},\ \Eprint
  {https://arxiv.org/abs/1507.06690} {arXiv:1507.06690 [physics.acc-ph]}
  \BibitemShut {NoStop}%
\bibitem [{\citenamefont {Mufson}\ \emph {et~al.}(2015)\citenamefont {Mufson}
  \emph {et~al.}}]{Mufson:2015kga}%
  \BibitemOpen
  \bibfield  {author} {\bibinfo {author} {\bibfnamefont {S.}~\bibnamefont
  {Mufson}} \emph {et~al.},\ }\href
  {https://doi.org/10.1016/j.nima.2015.07.026} {\bibfield  {journal} {\bibinfo
  {journal} {Nucl. Instrum. Meth. A}\ }\textbf {\bibinfo {volume} {799}},\
  \bibinfo {pages} {1} (\bibinfo {year} {2015})},\ \Eprint
  {https://arxiv.org/abs/1504.04035} {arXiv:1504.04035 [physics.ins-det]}
  \BibitemShut {NoStop}%
\bibitem [{\citenamefont {Agostinelli}\ \emph {et~al.}(2003)\citenamefont
  {Agostinelli} \emph {et~al.}}]{GEANT4:2002zbu}%
  \BibitemOpen
  \bibfield  {author} {\bibinfo {author} {\bibfnamefont {S.}~\bibnamefont
  {Agostinelli}} \emph {et~al.} (\bibinfo {collaboration} {GEANT4}),\ }\href
  {https://doi.org/10.1016/S0168-9002(03)01368-8} {\bibfield  {journal}
  {\bibinfo  {journal} {Nucl. Instrum. Meth. A}\ }\textbf {\bibinfo {volume}
  {506}},\ \bibinfo {pages} {250} (\bibinfo {year} {2003})}\BibitemShut
  {NoStop}%
\bibitem [{\citenamefont {Aliaga}\ \emph {et~al.}(2016)\citenamefont {Aliaga}
  \emph {et~al.}}]{MINERvA:2016iqn}%
  \BibitemOpen
  \bibfield  {author} {\bibinfo {author} {\bibfnamefont {L.}~\bibnamefont
  {Aliaga}} \emph {et~al.} (\bibinfo {collaboration} {MINERvA}),\ }\href
  {https://doi.org/10.1103/PhysRevD.94.092005} {\bibfield  {journal} {\bibinfo
  {journal} {Phys. Rev. D}\ }\textbf {\bibinfo {volume} {94}},\ \bibinfo
  {pages} {092005} (\bibinfo {year} {2016})},\ \bibinfo {note} {[Addendum:
  Phys. Rev. D 95, 039903 (2017)]},\ \Eprint {https://arxiv.org/abs/1607.00704}
  {arXiv:1607.00704 [hep-ex]} \BibitemShut {NoStop}%
\bibitem [{\citenamefont {Alt}\ \emph {et~al.}(2007)\citenamefont {Alt} \emph
  {et~al.}}]{NA49:2006oyk}%
  \BibitemOpen
  \bibfield  {author} {\bibinfo {author} {\bibfnamefont {C.}~\bibnamefont
  {Alt}} \emph {et~al.} (\bibinfo {collaboration} {NA49}),\ }\href
  {https://doi.org/10.1140/epjc/s10052-006-0165-7} {\bibfield  {journal}
  {\bibinfo  {journal} {Eur. Phys. J. C}\ }\textbf {\bibinfo {volume} {49}},\
  \bibinfo {pages} {897} (\bibinfo {year} {2007})},\ \Eprint
  {https://arxiv.org/abs/hep-ex/0606028} {arXiv:hep-ex/0606028} \BibitemShut
  {NoStop}%
\bibitem [{\citenamefont {Paley}\ \emph {et~al.}(2014)\citenamefont {Paley}
  \emph {et~al.}}]{MIPP:2014shj}%
  \BibitemOpen
  \bibfield  {author} {\bibinfo {author} {\bibfnamefont {J.~M.}\ \bibnamefont
  {Paley}} \emph {et~al.} (\bibinfo {collaboration} {MIPP}),\ }\href
  {https://doi.org/10.1103/PhysRevD.90.032001} {\bibfield  {journal} {\bibinfo
  {journal} {Phys. Rev. D}\ }\textbf {\bibinfo {volume} {90}},\ \bibinfo
  {pages} {032001} (\bibinfo {year} {2014})},\ \Eprint
  {https://arxiv.org/abs/1404.5882} {arXiv:1404.5882 [hep-ex]} \BibitemShut
  {NoStop}%
\bibitem [{\citenamefont {Abgrall}\ \emph {et~al.}(2011)\citenamefont {Abgrall}
  \emph {et~al.}}]{NA61SHINE:2011dsu}%
  \BibitemOpen
  \bibfield  {author} {\bibinfo {author} {\bibfnamefont {N.}~\bibnamefont
  {Abgrall}} \emph {et~al.} (\bibinfo {collaboration} {NA61/SHINE}),\ }\href
  {https://doi.org/10.1103/PhysRevC.84.034604} {\bibfield  {journal} {\bibinfo
  {journal} {Phys. Rev. C}\ }\textbf {\bibinfo {volume} {84}},\ \bibinfo
  {pages} {034604} (\bibinfo {year} {2011})},\ \Eprint
  {https://arxiv.org/abs/1102.0983} {arXiv:1102.0983 [hep-ex]} \BibitemShut
  {NoStop}%
\bibitem [{\citenamefont {Barton}\ \emph {et~al.}(1983)\citenamefont {Barton}
  \emph {et~al.}}]{Barton:1982dg}%
  \BibitemOpen
  \bibfield  {author} {\bibinfo {author} {\bibfnamefont {D.~S.}\ \bibnamefont
  {Barton}} \emph {et~al.},\ }\href {https://doi.org/10.1103/PhysRevD.27.2580}
  {\bibfield  {journal} {\bibinfo  {journal} {Phys. Rev. D}\ }\textbf {\bibinfo
  {volume} {27}},\ \bibinfo {pages} {2580} (\bibinfo {year}
  {1983})}\BibitemShut {NoStop}%
\bibitem [{\citenamefont {Andreopoulos}\ \emph {et~al.}(2015)\citenamefont
  {Andreopoulos}, \citenamefont {Barry}, \citenamefont {Dytman}, \citenamefont
  {Gallagher}, \citenamefont {Golan}, \citenamefont {Hatcher}, \citenamefont
  {Perdue},\ and\ \citenamefont {Yarba}}]{Andreopoulos:2015wxa}%
  \BibitemOpen
  \bibfield  {author} {\bibinfo {author} {\bibfnamefont {C.}~\bibnamefont
  {Andreopoulos}}, \bibinfo {author} {\bibfnamefont {C.}~\bibnamefont {Barry}},
  \bibinfo {author} {\bibfnamefont {S.}~\bibnamefont {Dytman}}, \bibinfo
  {author} {\bibfnamefont {H.}~\bibnamefont {Gallagher}}, \bibinfo {author}
  {\bibfnamefont {T.}~\bibnamefont {Golan}}, \bibinfo {author} {\bibfnamefont
  {R.}~\bibnamefont {Hatcher}}, \bibinfo {author} {\bibfnamefont
  {G.}~\bibnamefont {Perdue}},\ and\ \bibinfo {author} {\bibfnamefont
  {J.}~\bibnamefont {Yarba}},\ }\href@noop {} {\  (\bibinfo {year} {2015})},\
  \Eprint {https://arxiv.org/abs/1510.05494} {arXiv:1510.05494 [hep-ph]}
  \BibitemShut {NoStop}%
\bibitem [{\citenamefont {Andreopoulos}\ \emph {et~al.}(2018)\citenamefont
  {Andreopoulos} \emph {et~al.}}]{genie_manual_v3}%
  \BibitemOpen
  \bibfield  {author} {\bibinfo {author} {\bibfnamefont {C.}~\bibnamefont
  {Andreopoulos}} \emph {et~al.},\ }\href@noop {} {\bibinfo {title} {{GENIE
  Physics and User Manual, version 3.0.0}}} (\bibinfo {year} {2018}),\ \bibinfo
  {note}
  {\url{https://genie-docdb.pp.rl.ac.uk/cgi-bin/ShowDocument?docid=2&version=3}}\BibitemShut
  {NoStop}%
\bibitem [{\citenamefont {Gran}\ \emph {et~al.}(2013)\citenamefont {Gran},
  \citenamefont {Nieves}, \citenamefont {Sanchez},\ and\ \citenamefont
  {Vicente~Vacas}}]{Gran:2013kda}%
  \BibitemOpen
  \bibfield  {author} {\bibinfo {author} {\bibfnamefont {R.}~\bibnamefont
  {Gran}}, \bibinfo {author} {\bibfnamefont {J.}~\bibnamefont {Nieves}},
  \bibinfo {author} {\bibfnamefont {F.}~\bibnamefont {Sanchez}},\ and\ \bibinfo
  {author} {\bibfnamefont {M.~J.}\ \bibnamefont {Vicente~Vacas}},\ }\href
  {https://doi.org/10.1103/PhysRevD.88.113007} {\bibfield  {journal} {\bibinfo
  {journal} {Phys. Rev. D}\ }\textbf {\bibinfo {volume} {88}},\ \bibinfo
  {pages} {113007} (\bibinfo {year} {2013})},\ \Eprint
  {https://arxiv.org/abs/1307.8105} {arXiv:1307.8105 [hep-ph]} \BibitemShut
  {NoStop}%
\bibitem [{\citenamefont {Meyer}\ \emph {et~al.}(2016)\citenamefont {Meyer},
  \citenamefont {Betancourt}, \citenamefont {Gran},\ and\ \citenamefont
  {Hill}}]{Meyer:2016oeg}%
  \BibitemOpen
  \bibfield  {author} {\bibinfo {author} {\bibfnamefont {A.~S.}\ \bibnamefont
  {Meyer}}, \bibinfo {author} {\bibfnamefont {M.}~\bibnamefont {Betancourt}},
  \bibinfo {author} {\bibfnamefont {R.}~\bibnamefont {Gran}},\ and\ \bibinfo
  {author} {\bibfnamefont {R.~J.}\ \bibnamefont {Hill}},\ }\href
  {https://doi.org/10.1103/PhysRevD.93.113015} {\bibfield  {journal} {\bibinfo
  {journal} {Phys. Rev. D}\ }\textbf {\bibinfo {volume} {93}},\ \bibinfo
  {pages} {113015} (\bibinfo {year} {2016})},\ \Eprint
  {https://arxiv.org/abs/1603.03048} {arXiv:1603.03048 [hep-ph]} \BibitemShut
  {NoStop}%
\bibitem [{\citenamefont {Berger}\ and\ \citenamefont
  {Sehgal}(2007)}]{Berger:2007rq}%
  \BibitemOpen
  \bibfield  {author} {\bibinfo {author} {\bibfnamefont {C.}~\bibnamefont
  {Berger}}\ and\ \bibinfo {author} {\bibfnamefont {L.~M.}\ \bibnamefont
  {Sehgal}},\ }\href {https://doi.org/10.1103/PhysRevD.76.113004} {\bibfield
  {journal} {\bibinfo  {journal} {Phys. Rev. D}\ }\textbf {\bibinfo {volume}
  {76}},\ \bibinfo {pages} {113004} (\bibinfo {year} {2007})},\ \Eprint
  {https://arxiv.org/abs/0709.4378} {arXiv:0709.4378 [hep-ph]} \BibitemShut
  {NoStop}%
\bibitem [{\citenamefont {Rein}\ and\ \citenamefont
  {Sehgal}(1981)}]{Rein:1980wg}%
  \BibitemOpen
  \bibfield  {author} {\bibinfo {author} {\bibfnamefont {D.}~\bibnamefont
  {Rein}}\ and\ \bibinfo {author} {\bibfnamefont {L.~M.}\ \bibnamefont
  {Sehgal}},\ }\href {https://doi.org/10.1016/0003-4916(81)90242-6} {\bibfield
  {journal} {\bibinfo  {journal} {Annals Phys.}\ }\textbf {\bibinfo {volume}
  {133}},\ \bibinfo {pages} {79} (\bibinfo {year} {1981})}\BibitemShut
  {NoStop}%
\bibitem [{\citenamefont {Bodek}\ and\ \citenamefont
  {Yang}(2003)}]{Bodek:2002ps}%
  \BibitemOpen
  \bibfield  {author} {\bibinfo {author} {\bibfnamefont {A.}~\bibnamefont
  {Bodek}}\ and\ \bibinfo {author} {\bibfnamefont {U.~K.}\ \bibnamefont
  {Yang}},\ }\href {https://doi.org/10.1088/0954-3899/29/8/369} {\bibfield
  {journal} {\bibinfo  {journal} {J. Phys. G}\ }\textbf {\bibinfo {volume}
  {29}},\ \bibinfo {pages} {1899} (\bibinfo {year} {2003})},\ \Eprint
  {https://arxiv.org/abs/hep-ex/0210024} {arXiv:hep-ex/0210024} \BibitemShut
  {NoStop}%
\bibitem [{\citenamefont {Yang}\ and\ \citenamefont
  {Bodek}(1999)}]{Yang:1998zb}%
  \BibitemOpen
  \bibfield  {author} {\bibinfo {author} {\bibfnamefont {U.-K.}\ \bibnamefont
  {Yang}}\ and\ \bibinfo {author} {\bibfnamefont {A.}~\bibnamefont {Bodek}},\
  }\href {https://doi.org/10.1103/PhysRevLett.82.2467} {\bibfield  {journal}
  {\bibinfo  {journal} {Phys. Rev. Lett.}\ }\textbf {\bibinfo {volume} {82}},\
  \bibinfo {pages} {2467} (\bibinfo {year} {1999})},\ \Eprint
  {https://arxiv.org/abs/hep-ph/9809480} {arXiv:hep-ph/9809480} \BibitemShut
  {NoStop}%
\bibitem [{\citenamefont {Georgi}\ and\ \citenamefont
  {Politzer}(1976)}]{Georgi:1976ve}%
  \BibitemOpen
  \bibfield  {author} {\bibinfo {author} {\bibfnamefont {H.}~\bibnamefont
  {Georgi}}\ and\ \bibinfo {author} {\bibfnamefont {H.~D.}\ \bibnamefont
  {Politzer}},\ }\href {https://doi.org/10.1103/PhysRevD.14.1829} {\bibfield
  {journal} {\bibinfo  {journal} {Phys. Rev. D}\ }\textbf {\bibinfo {volume}
  {14}},\ \bibinfo {pages} {1829} (\bibinfo {year} {1976})}\BibitemShut
  {NoStop}%
\bibitem [{\citenamefont {Barbieri}\ \emph
  {et~al.}(1976{\natexlab{a}})\citenamefont {Barbieri}, \citenamefont {Ellis},
  \citenamefont {Gaillard},\ and\ \citenamefont {Ross}}]{Barbieri:1976bj}%
  \BibitemOpen
  \bibfield  {author} {\bibinfo {author} {\bibfnamefont {R.}~\bibnamefont
  {Barbieri}}, \bibinfo {author} {\bibfnamefont {J.~R.}\ \bibnamefont {Ellis}},
  \bibinfo {author} {\bibfnamefont {M.~K.}\ \bibnamefont {Gaillard}},\ and\
  \bibinfo {author} {\bibfnamefont {G.~G.}\ \bibnamefont {Ross}},\ }\href
  {https://doi.org/10.1016/0370-2693(76)90323-3} {\bibfield  {journal}
  {\bibinfo  {journal} {Phys. Lett. B}\ }\textbf {\bibinfo {volume} {64}},\
  \bibinfo {pages} {171} (\bibinfo {year} {1976}{\natexlab{a}})}\BibitemShut
  {NoStop}%
\bibitem [{\citenamefont {Barbieri}\ \emph
  {et~al.}(1976{\natexlab{b}})\citenamefont {Barbieri}, \citenamefont {Ellis},
  \citenamefont {Gaillard},\ and\ \citenamefont {Ross}}]{Barbieri:1976rd}%
  \BibitemOpen
  \bibfield  {author} {\bibinfo {author} {\bibfnamefont {R.}~\bibnamefont
  {Barbieri}}, \bibinfo {author} {\bibfnamefont {J.~R.}\ \bibnamefont {Ellis}},
  \bibinfo {author} {\bibfnamefont {M.~K.}\ \bibnamefont {Gaillard}},\ and\
  \bibinfo {author} {\bibfnamefont {G.~G.}\ \bibnamefont {Ross}},\ }\href
  {https://doi.org/10.1016/0550-3213(76)90563-0} {\bibfield  {journal}
  {\bibinfo  {journal} {Nucl. Phys. B}\ }\textbf {\bibinfo {volume} {117}},\
  \bibinfo {pages} {50} (\bibinfo {year} {1976}{\natexlab{b}})}\BibitemShut
  {NoStop}%
\bibitem [{\citenamefont {Yang}\ \emph {et~al.}(2009)\citenamefont {Yang},
  \citenamefont {Andreopoulos}, \citenamefont {Gallagher}, \citenamefont
  {Hoffmann},\ and\ \citenamefont {Kehayias}}]{Yang:2009zx}%
  \BibitemOpen
  \bibfield  {author} {\bibinfo {author} {\bibfnamefont {T.}~\bibnamefont
  {Yang}}, \bibinfo {author} {\bibfnamefont {C.}~\bibnamefont {Andreopoulos}},
  \bibinfo {author} {\bibfnamefont {H.}~\bibnamefont {Gallagher}}, \bibinfo
  {author} {\bibfnamefont {K.}~\bibnamefont {Hoffmann}},\ and\ \bibinfo
  {author} {\bibfnamefont {P.}~\bibnamefont {Kehayias}},\ }\href
  {https://doi.org/10.1140/epjc/s10052-009-1094-z} {\bibfield  {journal}
  {\bibinfo  {journal} {Eur. Phys. J. C}\ }\textbf {\bibinfo {volume} {63}},\
  \bibinfo {pages} {1} (\bibinfo {year} {2009})},\ \Eprint
  {https://arxiv.org/abs/0904.4043} {arXiv:0904.4043 [hep-ph]} \BibitemShut
  {NoStop}%
\bibitem [{\citenamefont {Salcedo}\ \emph {et~al.}(1988)\citenamefont
  {Salcedo}, \citenamefont {Oset}, \citenamefont {Vicente-Vacas},\ and\
  \citenamefont {Garcia-Recio}}]{Salcedo:1987md}%
  \BibitemOpen
  \bibfield  {author} {\bibinfo {author} {\bibfnamefont {L.~L.}\ \bibnamefont
  {Salcedo}}, \bibinfo {author} {\bibfnamefont {E.}~\bibnamefont {Oset}},
  \bibinfo {author} {\bibfnamefont {M.~J.}\ \bibnamefont {Vicente-Vacas}},\
  and\ \bibinfo {author} {\bibfnamefont {C.}~\bibnamefont {Garcia-Recio}},\
  }\href {https://doi.org/10.1016/0375-9474(88)90310-7} {\bibfield  {journal}
  {\bibinfo  {journal} {Nucl. Phys. A}\ }\textbf {\bibinfo {volume} {484}},\
  \bibinfo {pages} {557} (\bibinfo {year} {1988})}\BibitemShut {NoStop}%
\bibitem [{\citenamefont {Pinzon~Guerra}\ \emph {et~al.}(2019)\citenamefont
  {Pinzon~Guerra} \emph {et~al.}}]{PinzonGuerra:2018rju}%
  \BibitemOpen
  \bibfield  {author} {\bibinfo {author} {\bibfnamefont {E.~S.}\ \bibnamefont
  {Pinzon~Guerra}} \emph {et~al.},\ }\href
  {https://doi.org/10.1103/PhysRevD.99.052007} {\bibfield  {journal} {\bibinfo
  {journal} {Phys. Rev. D}\ }\textbf {\bibinfo {volume} {99}},\ \bibinfo
  {pages} {052007} (\bibinfo {year} {2019})},\ \Eprint
  {https://arxiv.org/abs/1812.06912} {arXiv:1812.06912 [hep-ex]} \BibitemShut
  {NoStop}%
\bibitem [{\citenamefont {Acero}\ \emph {et~al.}(2024)\citenamefont {Acero}
  \emph {et~al.}}]{NOvA:2023iam}%
  \BibitemOpen
  \bibfield  {author} {\bibinfo {author} {\bibfnamefont {M.~A.}\ \bibnamefont
  {Acero}} \emph {et~al.} (\bibinfo {collaboration} {NOvA}),\ }\href
  {https://doi.org/10.1103/PhysRevD.110.012005} {\bibfield  {journal} {\bibinfo
   {journal} {Phys. Rev. D}\ }\textbf {\bibinfo {volume} {110}},\ \bibinfo
  {pages} {012005} (\bibinfo {year} {2024})},\ \Eprint
  {https://arxiv.org/abs/2311.07835} {arXiv:2311.07835 [hep-ex]} \BibitemShut
  {NoStop}%
\bibitem [{\citenamefont {Martinez~Casales}(2023)}]{MartinezCasales:2023bkf}%
  \BibitemOpen
  \bibfield  {author} {\bibinfo {author} {\bibfnamefont {M.}~\bibnamefont
  {Martinez~Casales}},\ }\emph {\bibinfo {title} {{Constraining neutrino
  interaction uncertainties for oscillation measurements in the NOvA experiment
  using Near Detector data}}},\ \href
  {https://doi.org/10.31274/td-20240617-320} {Ph.D. thesis},\ \bibinfo
  {school} {Iowa State U. (main), Iowa State U.} (\bibinfo {year}
  {2023})\BibitemShut {NoStop}%
\bibitem [{\citenamefont {Aurisano}\ \emph {et~al.}(2015)\citenamefont
  {Aurisano}, \citenamefont {Backhouse}, \citenamefont {Hatcher}, \citenamefont
  {Mayer}, \citenamefont {Musser}, \citenamefont {Patterson}, \citenamefont
  {Schroeter},\ and\ \citenamefont {Sousa}}]{Aurisano:2015oxj}%
  \BibitemOpen
  \bibfield  {author} {\bibinfo {author} {\bibfnamefont {A.}~\bibnamefont
  {Aurisano}}, \bibinfo {author} {\bibfnamefont {C.}~\bibnamefont {Backhouse}},
  \bibinfo {author} {\bibfnamefont {R.}~\bibnamefont {Hatcher}}, \bibinfo
  {author} {\bibfnamefont {N.}~\bibnamefont {Mayer}}, \bibinfo {author}
  {\bibfnamefont {J.}~\bibnamefont {Musser}}, \bibinfo {author} {\bibfnamefont
  {R.}~\bibnamefont {Patterson}}, \bibinfo {author} {\bibfnamefont
  {R.}~\bibnamefont {Schroeter}},\ and\ \bibinfo {author} {\bibfnamefont
  {A.}~\bibnamefont {Sousa}} (\bibinfo {collaboration} {NOvA}),\ }\href
  {https://doi.org/10.1088/1742-6596/664/7/072002} {\bibfield  {journal}
  {\bibinfo  {journal} {J. Phys. Conf. Ser.}\ }\textbf {\bibinfo {volume}
  {664}},\ \bibinfo {pages} {072002} (\bibinfo {year} {2015})}\BibitemShut
  {NoStop}%
\bibitem [{\citenamefont {Anfimov}\ \emph {et~al.}(2020)\citenamefont
  {Anfimov}, \citenamefont {Antoshkin}, \citenamefont {Aurisano}, \citenamefont
  {Samoylov},\ and\ \citenamefont {Sotnikov}}]{Anfimov:2020okt}%
  \BibitemOpen
  \bibfield  {author} {\bibinfo {author} {\bibfnamefont {N.}~\bibnamefont
  {Anfimov}}, \bibinfo {author} {\bibfnamefont {A.}~\bibnamefont {Antoshkin}},
  \bibinfo {author} {\bibfnamefont {A.}~\bibnamefont {Aurisano}}, \bibinfo
  {author} {\bibfnamefont {O.}~\bibnamefont {Samoylov}},\ and\ \bibinfo
  {author} {\bibfnamefont {A.}~\bibnamefont {Sotnikov}},\ }\href
  {https://doi.org/10.1088/1748-0221/15/06/C06066} {\bibfield  {journal}
  {\bibinfo  {journal} {JINST}\ }\textbf {\bibinfo {volume} {15}}\bibinfo
  {number} { (06)},\ \bibinfo {pages} {C06066}}\BibitemShut {NoStop}%
\bibitem [{\citenamefont {Kalman}(1960)}]{Kalman:1960mft}%
  \BibitemOpen
\bibfield  {number} {  }\bibfield  {author} {\bibinfo {author} {\bibfnamefont
  {R.~E.}\ \bibnamefont {Kalman}},\ }\href {https://doi.org/10.1115/1.3662552}
  {\bibfield  {journal} {\bibinfo  {journal} {J. Fluids Eng.}\ }\textbf
  {\bibinfo {volume} {82}},\ \bibinfo {pages} {35} (\bibinfo {year}
  {1960})}\BibitemShut {NoStop}%
\bibitem [{\citenamefont {Fruhwirth}(1987)}]{Fruhwirth:1987fm}%
  \BibitemOpen
  \bibfield  {author} {\bibinfo {author} {\bibfnamefont {R.}~\bibnamefont
  {Fruhwirth}},\ }\href {https://doi.org/10.1016/0168-9002(87)90887-4}
  {\bibfield  {journal} {\bibinfo  {journal} {Nucl. Instrum. Meth. A}\ }\textbf
  {\bibinfo {volume} {262}},\ \bibinfo {pages} {444} (\bibinfo {year}
  {1987})}\BibitemShut {NoStop}%
\bibitem [{\citenamefont {Acero}\ \emph
  {et~al.}(2023{\natexlab{b}})\citenamefont {Acero} \emph
  {et~al.}}]{NOvA:2021eqi}%
  \BibitemOpen
  \bibfield  {author} {\bibinfo {author} {\bibfnamefont {M.~A.}\ \bibnamefont
  {Acero}} \emph {et~al.} (\bibinfo {collaboration} {NOvA}),\ }\href
  {https://doi.org/10.1103/PhysRevD.107.052011} {\bibfield  {journal} {\bibinfo
   {journal} {Phys. Rev. D}\ }\textbf {\bibinfo {volume} {107}},\ \bibinfo
  {pages} {052011} (\bibinfo {year} {2023}{\natexlab{b}})},\ \Eprint
  {https://arxiv.org/abs/2109.12220} {arXiv:2109.12220 [hep-ex]} \BibitemShut
  {NoStop}%
\bibitem [{\citenamefont {McGivern}\ \emph {et~al.}(2016)\citenamefont
  {McGivern} \emph {et~al.}}]{MINERvA:2016sfc}%
  \BibitemOpen
  \bibfield  {author} {\bibinfo {author} {\bibfnamefont {C.~L.}\ \bibnamefont
  {McGivern}} \emph {et~al.} (\bibinfo {collaboration} {MINERvA}),\ }\href
  {https://doi.org/10.1103/PhysRevD.94.052005} {\bibfield  {journal} {\bibinfo
  {journal} {Phys. Rev. D}\ }\textbf {\bibinfo {volume} {94}},\ \bibinfo
  {pages} {052005} (\bibinfo {year} {2016})},\ \Eprint
  {https://arxiv.org/abs/1606.07127} {arXiv:1606.07127 [hep-ex]} \BibitemShut
  {NoStop}%
\bibitem [{\citenamefont {D'Agostini}(1995)}]{DAgostini:1994fjx}%
  \BibitemOpen
  \bibfield  {author} {\bibinfo {author} {\bibfnamefont {G.}~\bibnamefont
  {D'Agostini}},\ }\href {https://doi.org/10.1016/0168-9002(95)00274-X}
  {\bibfield  {journal} {\bibinfo  {journal} {Nucl. Instrum. Meth. A}\ }\textbf
  {\bibinfo {volume} {362}},\ \bibinfo {pages} {487} (\bibinfo {year}
  {1995})}\BibitemShut {NoStop}%
\bibitem [{Roo()}]{RooUnfold}%
  \BibitemOpen
  \href@noop {} {\bibinfo {title} {Roounfold}},\ \bibinfo {howpublished}
  {\url{https://gitlab.cern.ch/RooUnfold/RooUnfold}},\ \bibinfo {note}
  {accessed: September 2025}\BibitemShut {NoStop}%
\bibitem [{\citenamefont {Brenner}\ \emph {et~al.}(2020)\citenamefont
  {Brenner}, \citenamefont {Balasubramanian}, \citenamefont {Burgard},
  \citenamefont {Verkerke}, \citenamefont {Cowan}, \citenamefont
  {Verschuuren},\ and\ \citenamefont {Croft}}]{Brenner:2019lmf}%
  \BibitemOpen
  \bibfield  {author} {\bibinfo {author} {\bibfnamefont {L.}~\bibnamefont
  {Brenner}}, \bibinfo {author} {\bibfnamefont {R.}~\bibnamefont
  {Balasubramanian}}, \bibinfo {author} {\bibfnamefont {C.}~\bibnamefont
  {Burgard}}, \bibinfo {author} {\bibfnamefont {W.}~\bibnamefont {Verkerke}},
  \bibinfo {author} {\bibfnamefont {G.}~\bibnamefont {Cowan}}, \bibinfo
  {author} {\bibfnamefont {P.}~\bibnamefont {Verschuuren}},\ and\ \bibinfo
  {author} {\bibfnamefont {V.}~\bibnamefont {Croft}},\ }\href
  {https://doi.org/10.1142/S0217751X20501456} {\bibfield  {journal} {\bibinfo
  {journal} {Int. J. Mod. Phys. A}\ }\textbf {\bibinfo {volume} {35}},\
  \bibinfo {pages} {2050145} (\bibinfo {year} {2020})},\ \Eprint
  {https://arxiv.org/abs/1910.14654} {arXiv:1910.14654 [physics.data-an]}
  \BibitemShut {NoStop}%
\bibitem [{\citenamefont {Hocker}\ and\ \citenamefont
  {Kartvelishvili}(1996)}]{Hocker:1995kb}%
  \BibitemOpen
  \bibfield  {author} {\bibinfo {author} {\bibfnamefont {A.}~\bibnamefont
  {Hocker}}\ and\ \bibinfo {author} {\bibfnamefont {V.}~\bibnamefont
  {Kartvelishvili}},\ }\href {https://doi.org/10.1016/0168-9002(95)01478-0}
  {\bibfield  {journal} {\bibinfo  {journal} {Nucl. Instrum. Meth. A}\ }\textbf
  {\bibinfo {volume} {372}},\ \bibinfo {pages} {469} (\bibinfo {year}
  {1996})},\ \Eprint {https://arxiv.org/abs/hep-ph/9509307}
  {arXiv:hep-ph/9509307} \BibitemShut {NoStop}%
\bibitem [{\citenamefont {Birks}(1951)}]{Birk}%
  \BibitemOpen
  \bibfield  {author} {\bibinfo {author} {\bibfnamefont {J.~B.}\ \bibnamefont
  {Birks}},\ }\href {https://doi.org/10.1088/0370-1298/64/10/303} {\bibfield
  {journal} {\bibinfo  {journal} {Proceedings of the Physical Society. Section
  A}\ }\textbf {\bibinfo {volume} {64}},\ \bibinfo {pages} {874} (\bibinfo
  {year} {1951})}\BibitemShut {NoStop}%
\bibitem [{\citenamefont {Chou}(1952)}]{Chou}%
  \BibitemOpen
  \bibfield  {author} {\bibinfo {author} {\bibfnamefont {C.~N.}\ \bibnamefont
  {Chou}},\ }\href {https://doi.org/10.1103/PhysRev.87.904} {\bibfield
  {journal} {\bibinfo  {journal} {Phys. Rev.}\ }\textbf {\bibinfo {volume}
  {87}},\ \bibinfo {pages} {904} (\bibinfo {year} {1952})}\BibitemShut
  {NoStop}%
\bibitem [{\citenamefont {Kohley}\ \emph {et~al.}(2012)\citenamefont {Kohley}
  \emph {et~al.}}]{Kohley:2012awa}%
  \BibitemOpen
  \bibfield  {author} {\bibinfo {author} {\bibfnamefont {Z.}~\bibnamefont
  {Kohley}} \emph {et~al.},\ }\href
  {https://doi.org/10.1016/j.nima.2012.04.060} {\bibfield  {journal} {\bibinfo
  {journal} {Nucl. Instrum. Meth. A}\ }\textbf {\bibinfo {volume} {682}},\
  \bibinfo {pages} {59} (\bibinfo {year} {2012})}\BibitemShut {NoStop}%
\bibitem [{\citenamefont {Roeder}(2008)}]{Roeder2008}%
  \BibitemOpen
  \bibfield  {author} {\bibinfo {author} {\bibfnamefont {B.}~\bibnamefont
  {Roeder}},\ }\href@noop {} {\emph {\bibinfo {title} {Development and
  validation of neutron detection simulations for EURISOL}}},\ \bibinfo {type}
  {Tech. Rep.}\ (\bibinfo  {institution} {EURISOL Design Study},\ \bibinfo
  {year} {2008})\ \bibinfo {note} {report}\BibitemShut {NoStop}%
\bibitem [{\citenamefont {Golan}\ \emph
  {et~al.}(2012{\natexlab{a}})\citenamefont {Golan}, \citenamefont {Sobczyk},\
  and\ \citenamefont {Zmuda}}]{Golan:2012rfa}%
  \BibitemOpen
  \bibfield  {author} {\bibinfo {author} {\bibfnamefont {T.}~\bibnamefont
  {Golan}}, \bibinfo {author} {\bibfnamefont {J.~T.}\ \bibnamefont {Sobczyk}},\
  and\ \bibinfo {author} {\bibfnamefont {J.}~\bibnamefont {Zmuda}},\ }\href
  {https://doi.org/10.1016/j.nuclphysbps.2012.09.136} {\bibfield  {journal}
  {\bibinfo  {journal} {Nucl. Phys. B Proc. Suppl.}\ }\textbf {\bibinfo
  {volume} {229-232}},\ \bibinfo {pages} {499} (\bibinfo {year}
  {2012}{\natexlab{a}})}\BibitemShut {NoStop}%
\bibitem [{\citenamefont {Golan}\ \emph
  {et~al.}(2012{\natexlab{b}})\citenamefont {Golan}, \citenamefont {Juszczak},\
  and\ \citenamefont {Sobczyk}}]{Golan:2012wx}%
  \BibitemOpen
  \bibfield  {author} {\bibinfo {author} {\bibfnamefont {T.}~\bibnamefont
  {Golan}}, \bibinfo {author} {\bibfnamefont {C.}~\bibnamefont {Juszczak}},\
  and\ \bibinfo {author} {\bibfnamefont {J.~T.}\ \bibnamefont {Sobczyk}},\
  }\href {https://doi.org/10.1103/PhysRevC.86.015505} {\bibfield  {journal}
  {\bibinfo  {journal} {Phys. Rev. C}\ }\textbf {\bibinfo {volume} {86}},\
  \bibinfo {pages} {015505} (\bibinfo {year} {2012}{\natexlab{b}})},\ \Eprint
  {https://arxiv.org/abs/1202.4197} {arXiv:1202.4197 [nucl-th]} \BibitemShut
  {NoStop}%
\bibitem [{\citenamefont {Hayato}\ and\ \citenamefont
  {Pickering}(2021)}]{Hayato:2021heg}%
  \BibitemOpen
  \bibfield  {author} {\bibinfo {author} {\bibfnamefont {Y.}~\bibnamefont
  {Hayato}}\ and\ \bibinfo {author} {\bibfnamefont {L.}~\bibnamefont
  {Pickering}},\ }\href {https://doi.org/10.1140/epjs/s11734-021-00287-7}
  {\bibfield  {journal} {\bibinfo  {journal} {Eur. Phys. J. ST}\ }\textbf
  {\bibinfo {volume} {230}},\ \bibinfo {pages} {4469} (\bibinfo {year}
  {2021})},\ \Eprint {https://arxiv.org/abs/2106.15809} {arXiv:2106.15809
  [hep-ph]} \BibitemShut {NoStop}%
\bibitem [{\citenamefont {{NOvA Collaboration}}()}]{nova_data_release}%
  \BibitemOpen
  \bibfield  {author} {\bibinfo {author} {\bibnamefont {{NOvA
  Collaboration}}},\ }\href@noop {} {\bibinfo {title} {{Data Release}}},\
  \bibinfo {howpublished}
  {\url{https://novaexperiment.fnal.gov/data-releases/}}\BibitemShut {NoStop}%
\end{thebibliography}%

\end{document}